\documentclass[journal]{IEEEtran}
\usepackage{bm}
\usepackage{amsfonts}
\usepackage{subfigure}
\usepackage{graphicx}
\usepackage{epstopdf}
\usepackage{float}

\usepackage{rotating}
\usepackage{algorithm,algorithmicx}
\usepackage[noend]{algpseudocode}
\usepackage{longtable}
\usepackage{multirow}
\usepackage{booktabs}
\usepackage{amsmath} 
\usepackage{array}
\ifCLASSINFOpdf
\else
\fi
\begin{document}
	%
	\title{Multi-scale Dynamic Graph Convolutional Network for Hyperspectral Image Classification}
	%
	%
	%
	
	\author{Sheng~Wan,
		Chen~Gong,~\IEEEmembership{Member,~IEEE},
		Ping~Zhong,~\IEEEmembership{Senior~Member,~IEEE},
		Bo~Du,~\IEEEmembership{Senior~Member,~IEEE},
		Lefei~Zhang,~\IEEEmembership{Member,~IEEE},
		and~Jian~Yang,~\IEEEmembership{Member,~IEEE}
		
		\thanks{S. Wan, C. Gong, and J. Yang are with the PCA Lab, the Key Laboratory of Intelligent Perception and Systems for High-Dimensional Information of Ministry of Education, the Jiangsu Key Laboratory of Image and Video Understanding for Social Security, and the School of Computer Science and Engineering, Nanjing University of Science and Technology, Nanjing, 210094, P.R. China. (e-mail: wansheng315@hotmail.com; chen.gong@njust.edu.cn; csjyang@njust.edu.cn).}
		\thanks{P. Zhong is with the National Key Laboratory of Science and Technology on ATR, National University of Defense Technology, Changsha 410073, China (e-mail: zhongping@nudt.edu.cn).}
		\thanks{B. Du, and L. Zhang are with the State Key Laboratory of Software Engineering, School of Computer, Wuhan University, Wuhan 430079, China (e-mail: gunspace@163.com; zhanglefei@whu.edu.cn).}
		\thanks{\emph{Corresponding~authors:~Chen~Gong~and~Jian~Yang.}}
	}
	%
	%

	\markboth{}%
	{Shell \MakeLowercase{\textit{et al.}}: Bare Demo of IEEEtran.cls for IEEE Journals}
	%



	\maketitle
	
	\begin{abstract}
		Convolutional Neural Network (CNN) has demonstrated impressive ability to represent hyperspectral images and to achieve promising results in hyperspectral image classification. However, traditional CNN models can only operate convolution on regular square image regions with fixed size and weights, so they cannot universally adapt to the distinct local regions with various object distributions and geometric appearances. Therefore, their classification performances are still to be improved, especially in class boundaries. To alleviate this shortcoming, we consider employing the recently proposed Graph Convolutional Network (GCN) for hyperspectral image classification, as it can conduct the convolution on arbitrarily structured non-Euclidean data and is applicable to the irregular image regions represented by graph topological information. Different from the commonly used GCN models which work on a fixed graph, we enable the graph to be dynamically updated along with the graph convolution process, so that these two steps can be benefited from each other to gradually produce the discriminative embedded features as well as a refined graph. Moreover, to comprehensively deploy the multi-scale information inherited by hyperspectral images, we establish multiple input graphs with different neighborhood scales to extensively exploit the diversified spectral-spatial correlations at multiple scales. Therefore, our method is termed `Multi-scale Dynamic Graph Convolutional Network' (MDGCN). The experimental results on three typical benchmark datasets firmly demonstrate the superiority of the proposed MDGCN to other state-of-the-art methods in both qualitative and quantitative aspects.
		
	\end{abstract}

	%
	\IEEEpeerreviewmaketitle

	\section{Introduction}
	%
	%
	%
	%
	
	\IEEEPARstart{T}{he} rapid development of optics and photonics has significantly advanced hyperspectral techniques. As a result, hyperspectral images, which consist of hundreds of contiguous bands and contain large amounts of useful information, can be easily acquired \cite{Chen2015Spectral, Zhong2019Multiple}. Over the past few decades, hyperspectral image classification has played an important role in various fields, such as military target detection, vegetation monitoring, and disaster prevention and control.
	
	Up to now, diverse kinds of approaches have been proposed for classifying the pixels of a hyperspectral image into certain land-cover categories. The early-staged methods are mainly based on conventional pattern recognition methods, such as nearest neighbor classifier and linear classifier. Among these conventional methods, $K$-nearest neighbor \cite{Li2010Local} has been widely used due to its simplicity in both theory and practice. Support Vector Machine (SVM) \cite{Kuo2010Spatial} also performs robustly and satisfactorily with high-dimensional hyperspectral data. In addition to these, graph-based methods \cite{Shi2013Supervised}, extreme learning machine \cite{Wei2015Local}, sparse representation-based classifier \cite{Yi2011Hyperspectral}, and many other methods have been further employed to promote the performance of hyperspectral image classification. Nevertheless, it is difficult to distinguish different land-cover categories accurately by only using the spectral information \cite{Hang2016Matrix}. With the observation that spatially neighboring pixels usually carry correlated information within a smooth spatial domain, many researchers have resorted to spectral-spatial classification methods and several models have been proposed to exploit such local continuity \cite{Zhang2018Simultaneous, Zhong2017Discriminant}. For example, Markov Random Field (MRF)-based models \cite{Wang2005A} have been widely used for deploying spatial information and have achieved great popularity. In MRF-based models, spatial information is usually regarded as a priori before optimizing an energy function via posteriori maximization. Meanwhile, morphological profiles-based methods \cite{Fauvel2008Spectral, Song2014Remotely} have also been proposed to effectively combine spatial and spectral information.

	However, the aforementioned methods are all based on the handcrafted spectral-spatial features \cite{Zhang2018Diverse} which heavily depend on professional expertise and are quite empirical. To address this defect, deep learning \cite{Liu2016Deep, Fan2014Saliency, Zhong2017Learning, Mou2017Deep} has been extensively employed for hyperspectral image classification and has attracted increasing attention for its strong representation ability. The main reason is that deep learning methods can automatically obtain abstract high-level representations by gradually aggregating the low-level features, by which the complicated feature engineering can be avoided \cite{Yang2018Hyperspectral}. The first attempt to use deep learning methods for hyperspectral image classification was made by Chen \emph{et} \emph{al.} \cite{Chen2014Deep}, where the stacked autoencoder was built for high-level feature extraction. Subsequently, Mou \emph{et} \emph{al.} \cite{Mou2017Deep} first employed Recurrent Neural Network (RNN) for hyperspectral image classification. Besides, Ma \emph{et} \emph{al.} \cite{Ma2015Hyperspectral} attempted to learn the spectral-spatial features via a deep learning architecture by fine-tuning the network via a supervised strategy. Recently, Convolutional Neural Network (CNN) has emerged as a powerful tool for hyperspectral image classification \cite{Chen2016Deep, Hao2018Two, Zhao2016Spectral}. For instance, Jia \emph{et} \emph{al.} \cite{Jia2016Convolutional} employed CNN to extract spectral features and achieved superior performance to SVM. In addition, Hu \emph{et} \emph{al.} \cite{Hu2015Deep} proposed a five-layer 1-D CNN to classify hyperspectral images directly in spectral domain. In these methods, the convolution operation is mainly applied to spectral domain while the spatial details are largely neglected. Another set of deep learning approaches perform hyperspectral image classification by incorporating spectral-spatial information. For example, in \cite{Makantasis2015Deep}, Makantasis \emph{et} \emph{al.} encoded spectral-spatial information with a CNN and conducted classification with a multi-layer perceptron. Besides, Zhang \emph{et} \emph{al.} \cite{Zhang2017Spectral} proposed a multi-dimensional CNN to automatically extract hierarchical spectral features and spatial features. Furthermore, Lee \emph{et} \emph{al.} \cite{Lee2017Going} designed a novel contextual deep CNN, which is able to optimally explore contextual interactions by exploiting local spectral-spatial relationship among spatially neighboring pixels. Specifically, the joint exploitation of spectral-spatial information is obtained by a multi-scale convolutional filter bank. Although the existing CNN-based methods have achieved good performance to some extent, they still suffer from some drawbacks. To be specific, conventional CNN models only conduct the convolution on the regular square regions, so they cannot adaptively capture the geometric variations of different object regions in a hyperspectral image. Besides, the weights of each convolution kernel are identical when convolving all image patches. As a result, the information of class boundaries may be lost during the feature abstraction process and misclassifications will probably happen due to the inflexible convolution kernel. In other words, the convolution kernels with fixed shape, size, and weights are not adaptive to all the regions in a hyperspectral image. Apart from that, CNN-based methods often take a long training time because of the large number of parameters.
	
	Consequently, in this paper, we propose to utilize the recently proposed Graph Convolutional Network (GCN) \cite{Defferrard2016Convolutional, Kipf2016Semi} for hyperspectral image classification. GCN operates on a graph and is able to aggregate and transform feature information from the neighbors of every graph node. Consequently, the convolution operation of GCN is adaptively governed by the neighborhood structure of a graph and thus GCN can be applicable to the non-Euclidean irregular data based on the predefined graph. Besides, both node features and local graph structure can be encoded by the learned hidden layers, so GCN is able to exhaustively exploit the image features and flexibly preserve the class boundaries. 
	
	Nevertheless, the direct use of traditional GCN for hyperspectral image classification is still inadequate. Since hyperspectral data is often contaminated by noise, the initial input graph may not be accurate. Specifically, the edge weight of pairwise pixels may not represent their intrinsic similarity, which makes the input graph less than optimal. Furthermore, traditional GCN can only use the spectral features of image pixels without incorporating the spatial context which is actually of great significance in hyperspectral images. Additionally, the computational complexity of traditional GCN will be unacceptable when the number of pixels gets too large. To tackle these difficulties in applying GCN to hyperspectral image classification, we propose a new type of GCN called `Multi-scale Dynamic GCN' (MDGCN). Instead of utilizing a predefined fixed graph for convolution, we design a dynamic graph convolution operation, by which the similarity measures among pixels can be updated by fusing current feature embeddings. Consequently, the graph can be gradually refined during the convolution process of GCN, which will in turn make the feature embeddings more accurate. The processes of graph updating and feature embedding alternate, which work collaboratively to yield faithful graph structure and promising classification results. To take the multi-scale cues into consideration, we construct multiple graphs with different neighborhood scales so that the spatial information at different scales can be fully exploited \cite{He2017Multi}. Different from commonly used GCN models which utilize only one fixed graph, the multi-scale design enables MDGCN to extract spectral-spatial features with varied receptive fields, by which the comprehensive contextual information from different levels can be incorporated. Moreover, due to the large number of pixels brought by the high spatial resolution of hyperspectral images, the computational complexity of network training can be extremely high. To mitigate this problem, we group the raw pixels into a certain amount of homogenous superpixels and treat each superpixel as a graph node. As a result, the number of nodes in each graph will be significantly reduced, which also helps to accelerate the subsequent convolution process. 
	
	To sum up, the main contributions of the proposed MDGCN are as follows: First, we propose a novel dynamic graph convolution operation, which can reduce the impact of a bad predefined graph. Secondly, multi-scale graph convolution is utilized to extensively exploit the spatial information and acquire better feature representation. Thirdly, the superpixel technique is involved in our proposed MDGCN framework, which significantly reduces the complexity of model training. Finally, the experimental results on three typical hyperspectral image datasets show that MDGCN achieves state-of-the-art performance when compared with the existing methods.
	
	\section{Related Works}
	\label{Relatedworks}
	
	In this section, we review some representative works on hyperspectral image classification and GCN, as they are related to this work.

	\subsection{Hyperspectral Image Classification}
	
	As a traditional yet important remote sensing technique, hyperspectral image classification has been intensively investigated and many related methods have been proposed, such as Bayesian methods \cite{8390931}, random forest \cite{709601}, and kernel methods \cite{Maji2008Classification}. Particularly, SVM has shown impressive classification performance with limited labeled examples \cite{Mercier2003Support}. However, SVM independently treats every pixel (i.e., example) and fails to exploit the correlations among different image pixels. To address this limitation, spatial information is introduced. For instance, by directly incorporating spatial information into kernel design, Camps-Valls \emph{et} \emph{al.} \cite{Camps2006Composite} used SVM with composite kernels for hyperspectral image classification. Besides, filtering-based methods have also been applied to spectral-spatial classification. In \cite{Lin2016A}, Lin \emph{et} \emph{al.} designed a three-dimensional filtering with a Gaussian kernel and its derivative for spectral-spatial information extraction. After that, they also proposed a discriminative low-rank Gabor filtering for spectral-spatial information extraction \cite{Lin2017Discriminative}. Additionally, MRF has been commonly used to exploit spatial context for hyperspectral image classification with the assumption that spatially neighboring pixels are more likely to take the same label \cite{5997308}. However, when neighboring pixels are highly correlated, the standard neighbor determination approaches will degrade the MRF models due to the insufficient contained pixels \cite{Fauvel2013Advances}. Therefore, instead of modeling the joint distribution of spatially neighboring pixels, conditional random field directly models the class posterior probability given the hyperspectral image and has achieved encouraging performance \cite{Zhang2012Simplified, Zhong2014A}.
	
	The aforementioned methods simply employ various manually-extracted spectral-spatial features to represent the pixels, which highly depends on experts' experience and is not general. In contrast, deep learning-based methods \cite{Ma2015Hyperspectral, Zhao2015On}, which can generate features automatically, have recently attracted increasing attention in hyperspectral image classification. The first attempt can be found in \cite{Chen2014Deep}, where stacked autoencoder was utilized for high-level feature extraction. Subsequently, Li \emph{et} \emph{al.} \cite{Tong2015Classification} used restricted Boltzmann machine and deep belief network for hyperspectral image feature extraction and pixel classification, by which the information contained in the original data can be well retained. Meanwhile, RNN model has been applied to hyperspectral image classification \cite{Shi2018Multi}. In \cite{Shi2018Multi}, Shi \emph{et} \emph{al.} exploited multi-scale spectral-spatial features via hierarchical RNN, which can learn the spatial dependency of non-adjacent image patches in a two-dimension spatial domain. Among these deep learning methods, CNN, which needs fewer parameters than fully-connected networks with the same number of hidden layers, has drawn great attention for its breakthrough in hyperspectral image classification. For example, in \cite{Hu2015Deep, Ghamisi2017A}, CNN was used to extract the spectral features, which performs better than SVM. Nonetheless, excavating spatial information is of great importance in hyperspectral image classification and many CNN-based methods have done explorations on this aspect. For instance, Yang \emph{et} \emph{al.} \cite{Yang2016Hyperspectral} proposed a two-channel deep CNN to jointly learn spectral-spatial features from hyperspectral images, where the channels are used for learning spectral and spatial features, respectively. Besides, in \cite{Yue2015Spectral}, Yue \emph{et} \emph{al.} projected hyperspectral data to several principal components before adopting CNN to extract spectral-spatial features. In the recent work of Li \emph{et} \emph{al.} \cite{Li2016Hyperspectral}, deep CNN is used to learn pixel-pair features, and the classification results of pixels in different pairs from the neighborhood are then fused. Additionally, Zhang \emph{et} \emph{al.} \cite{Zhang2018Diverse} proposed a deep CNN model based on diverse regions, which employs different local or global regions inputs to learn joint representation of each pixel. Although CNN-based hyperspectral image classification methods can extract spectra-spatial features automatically, the effectiveness of the obtained features is still restricted by some issues. For example, they simply apply the fixed convolution kernels to different regions in a hyperspectral image, which does not consider the geometric appearance of various local regions and may result in undesirable misclassifications.
	
	\subsection{Graph Convolutional Network}
	
	The concept of neural network for graph data was first proposed by Gori \emph{et} \emph{al.} \cite{Gori2005A}, of which the advantage over CNN and RNN is that it can work on the graph-structured non-Euclidean data. Specifically, the Graph Neural Network (GNN) can collectively aggregate the node features in a graph and properly embed the entire graph in a new discriminative space. Subsequently, Scarselli \emph{et} \emph{al.} \cite{Franco2009The} made GNN trainable by a supervised learning algorithm for practical data. However, their algorithm is computationally expensive and runs inefficiently on large-scale graphs. Therefore, Bruna \emph{et} \emph{al.} \cite{Bruna2014Spectral} developed the operation of `graph convolution' based on spectral property, which convolves on the neighborhood of every graph node and produces a node-level output. After that, many extensions of graph convolution have been investigated and achieved advanced results \cite{Dai2016Discriminative, Monti2017Geometric}. For instance, Hamilton \emph{et} \emph{al.} \cite{Hamilton2017Inductive} presented an inductive framework called `GraphSAGE', which leverages node features to effectively generate node embeddings for previously unseen data. Apart from this, Defferrard \emph{et} \emph{al.} \cite{Defferrard2016Convolutional} proposed a formulation of CNNs in the context of spectral graph theory. Based on their work, Kipf and Welling \cite{Kipf2016Semi} proposed a fast approximation localized convolution, which makes the GCN model able to encode both graph structure and node features. In their work, GCN was simplified by a first-order approximation of graph spectral convolution, which leads to more efficient filtering operations.
	
	With the rapid development of graph convolution theories, GCN has been widely applied to various applications, such as recommender systems \cite{Ying2018Graph} and semantic segmentation \cite{Qi20173D}. Besides, to our best knowledge, GCN has been deployed for hyperspectral image classification in only one prior work \cite{8474300}. However, \cite{8474300} only utilizes a fixed graph during the node convolution process, so the intrinsic relationship among the pixels cannot be precisely reflected. Moreover, the neighborhood size in their method is also fixed and thus the spectral-spatial information in different local regions cannot be flexibly captured. To cope with these issues, we propose a novel dynamic multi-scale GCN which dynamically updates the graphs and fuses multi-scale spectral-spatial information for hyperspectral image classification. As a result, the accurate node embeddings can be acquired, which ensures satisfactory classification performance.
	
	\begin{figure*}[!t]
		\centering
		\centering
		\includegraphics[width=\linewidth]{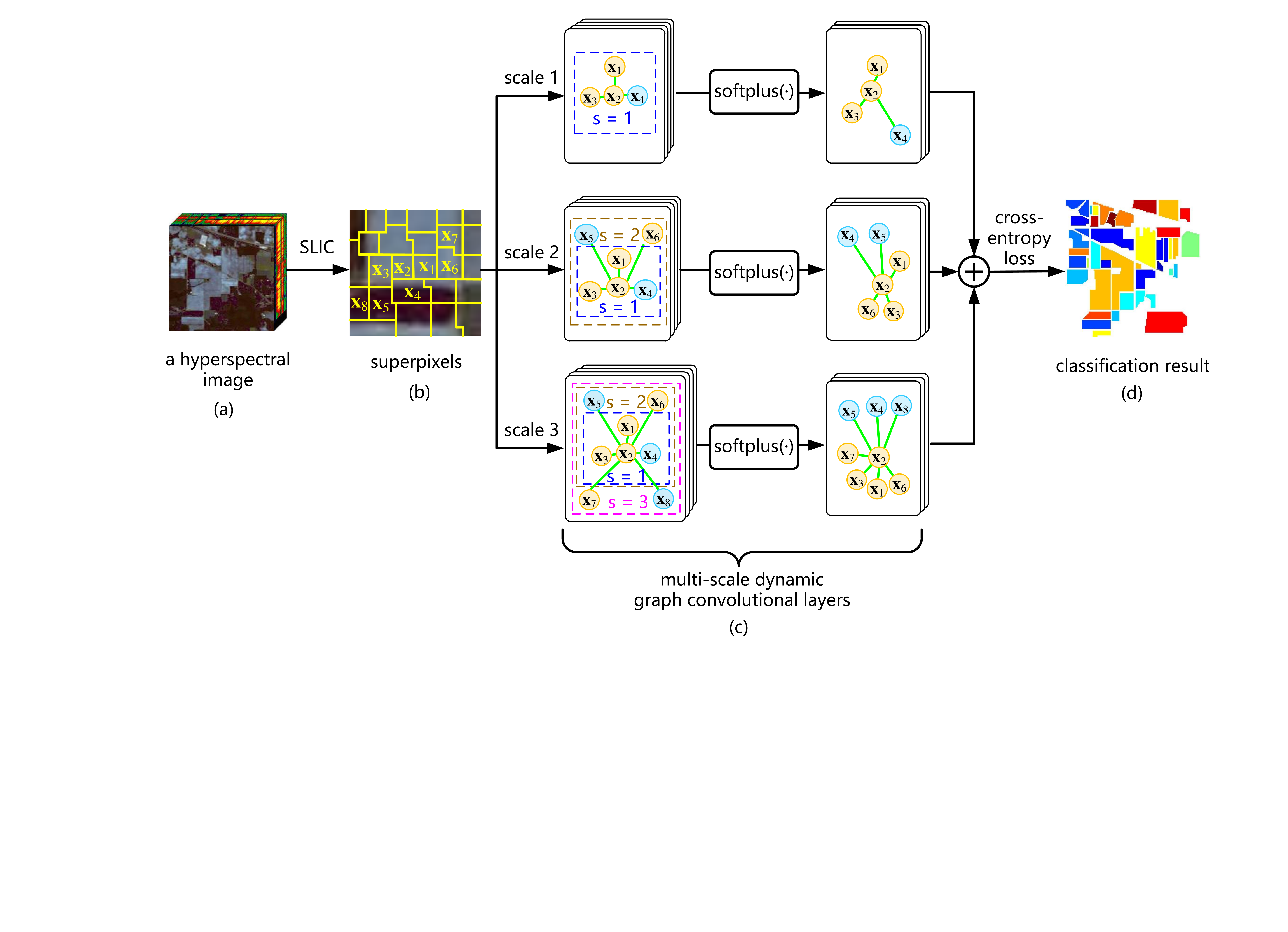}\hspace{0pt}
		\caption{The framework of our algorithm. (a) is the original hyperspectral image. (b) shows the superpixels segmented by SLIC algorithm \cite{Radhakrishna2012SLIC}, where a local region of the hyperspectral image is exhibited which contains eight superpixels $\mathbf{x}_1, \mathbf{x}_2,\cdots, \mathbf{x}_8$. In (c), the circles and green lines represent the graph nodes and edges, respectively, where different colors of the nodes represent different land-cover types. Specifically, at each scale, the edge weight is updated gradually along with the convolution on graph nodes, so that the graph can be dynamically refined. Here two dynamic graph convolutional layers are employed for each scale, where softplus \cite{7280459} is utilized as the activation function. In (d), the classification result is acquired by integrating the multi-scale outputs, and the cross-entropy loss is used to penalize the label difference between the output and the seed superpixels.}
		\label{Overview}
	\end{figure*}
	
	\section{The Proposed Method}
	\label{Proposedmethod}

	This section details our proposed MDGCN model (see Fig.~\ref{Overview}). When an input hyperspectral image is given, it is pre-processed by the Simple Linear Iterative Clustering (SLIC) algorithm \cite{Radhakrishna2012SLIC} to be segmented into several homogeneous superpixels. Then, graphs are constructed over these superpixels at different spatial scales. After that, the convolutions are conducted on these graphs, which simultaneously aggregates multi-scale spectral-spatial features and also gradually refine the input graphs. The superpixels potentially belonging to the same class will be ideally clustered together in the embedding space. Finally, the classification result is produced by the well-trained network. Next we detail the critical steps of our MDGCN by explaining the superpixel segmentation (Section \ref{subsection_sp}), presenting the GCN backbone (Section \ref{subsection_gcn}), elaborating the dynamic graph evolution (Section \ref{subsection_dgcn}), and describing the multi-scale manipulation (Section \ref{subsection_multi}).

	\subsection{Superpixel Segmentation}   
	\label{subsection_sp}
	
	A hyperspectral image usually contains hundreds of thousands of pixels, which may result in unacceptable computational complexity for the subsequent graph convolution and classification. To address this problem, we adopt a segmentation algorithm named SLIC \cite{Radhakrishna2012SLIC} to segment the entire image into a small amount of compact superpixels, and each superpixel represents a homogeneous image region with strong spectral-spatial similarity. Concretely, the SLIC algorithm starts from an initial grid on the image and then creates segmentation through iteratively growing the local clusters using a $k$-means algorithm. When the segmentation is finished, each superpixel is treated as a graph node instead of the pixel in the input image, therefore the amount of graph nodes can be significantly reduced, and the computational efficiency can be improved. Here the feature of each node (i.e., superpixel) is the average spectral signatures of the pixels involved in the corresponding superpixel. Another advantage for implementing the superpixel segmentation is that the generated superpixels also help to preserve the local structural information of a hyperspectral image, as nearby pixels with high spatial consistency have a large probability to belong to the same land-cover type (i.e., label).
	
	\subsection{Graph Convolutional Network}
	\label{subsection_gcn}
	
	GCN \cite{Kipf2016Semi} is a multi-layer neural network which operates directly on a graph and generates node embeddings by gradually fusing the features in the neighborhood. Different from traditional CNN which only applies to data represented by regular grids, GCN is able to operate on the data with arbitrary non-Euclidean structure. Formally, an undirected graph is defined as $\mathcal{G}=(\mathcal{V}, \mathcal{E})$, where $\mathcal{V}$ and $\mathcal{E}$ are the sets of nodes and edges, respectively. The notation $\mathbf{A}$ denotes the adjacency matrix of $\mathcal{G}$ which indicates whether each pair of nodes is connected and can be calculated as
	\begin{equation}
	\label{InitialAdjacencyMatirx}
	{\mathbf{A}_{ij}} = \left\{ {\begin{array}{*{20}{l}}
		{{e^{ - \gamma {{\left\| {{\mathbf{x}_i} - {\mathbf{x}_j}} \right\|}^2}}}}\,\;{\rm{if}}\,\;{\mathbf{x}}_{i} \in N(\mathbf{x}_j)\,\;{\rm{or}}\,\;{\mathbf{x}}_{j} \in N(\mathbf{x}_i)\\
		0 \;\;\;\;\;\;\;\;\;\;\;\;\;\;\;\;\;\;\rm{otherwise}
		\end{array}} \right.,
	\end{equation}
	where the parameter $\gamma$ is empirically set to 0.2 in the experiments, $\mathbf{x}_i$ represents a superpixel and $N(\mathbf{x}_j)$ is the set of neighbors of the example $\mathbf{x}_j$. 
	
	To conduct node embeddings for $\mathcal{G}$, spectral filtering on graphs is defined, which can be expressed as a signal $\mathbf{x}$ filtered by $ g_{\bm{\theta}}=\rm{diag}(\bm{\theta}) $ in the Fourier domain, namely
	\begin{equation}
	\label{FilterFourier}
	{g_{\bm{\theta}} }\star\mathbf{x} = \mathbf{U}{g_{\bm{\theta}} }{\mathbf{U}^{\top}}\mathbf{x},
	\end{equation}
	where $\mathbf{U}$ is the matrix composed of the eigenvectors of the normalized graph Laplacian $\mathbf{L}=\mathbf{I}-{\mathbf{D}^{ - \frac{1}{2}}}\mathbf{A}{\mathbf{D}^{ - \frac{1}{2}}}=\mathbf{U}\mathbf{\Lambda}\mathbf{U}^{\top}$. Here $\mathbf{\Lambda}$ is a diagonal matrix containing the eigenvalues of $\mathbf{L}$, $\mathbf{D}$ is the degree matrix ${\mathbf{D}_{ii}} = \sum\nolimits_j {{\mathbf{A}_{ij}}}$, and $\mathbf{I}$ denotes the identity matrix with proper size throughout this paper. Then, we can understand $g_{\bm{\theta}}$ as a function of the eigenvalues of $\mathbf{L}$, i.e., $g_{\bm{\theta}}(\mathbf{\Lambda})$. To reduce the computational consumption of eigenvector decomposition in Eq.~\eqref{FilterFourier}, Hammond \emph{et} \emph{al.} \cite{Hammond2009Wavelets} approximated $g_{\bm{\theta}}(\mathbf{\Lambda})$ by a truncated expansion in terms of Chebyshev polynomials $T_k(\mathbf{x})$ up to $K^{\rm{th}}$-order, which is
	\begin{equation}
	{g_{\bm{\theta} '}}(\mathbf{\Lambda} ) \approx \sum\limits_{k = 0}^K {\bm{\theta} {'_k}{T_k}(\widetilde{ \mathbf{\Lambda}} )},
	\end{equation}
	where $\bm{\theta} {'}$ is a vector of Chebyshev coefficients; $\widetilde{\mathbf{\Lambda}}=\frac{2}{{{\lambda _{\max }}}}\mathbf{\Lambda}  - {\mathbf{I}}$ with $\lambda _{\max }$ being the largest eigenvalue of $\mathbf{L}$. According to \cite{Hammond2009Wavelets}, the Chebyshev polynomials are defined as ${T_k}(\mathbf{x}) = 2\mathbf{x}{T_{k - 1}}(\mathbf{x}) - {T_{k - 2}}(\mathbf{x})$ with $T_{0}(\mathbf{x})=1$ and $T_{1}(\mathbf{x})=\mathbf{x}$. Therefore, the convolution of a signal $\mathbf{x}$ by the filter $g_{\bm{\theta} '}$ can be written as
	\begin{equation}
	\label{ConvoX}
	{g_{\bm{\theta}{'}}  }\star\mathbf{x} \approx \sum\limits_{k = 0}^K {\bm{\theta} {'_k}{T_k}(\widetilde{\mathbf{L}} )\mathbf{x}},
	\end{equation}
	where $\widetilde{\mathbf{L}}=\frac{2}{{{\lambda _{\max }}}}\mathbf{L}  - {\mathbf{I}}$ denotes the scaled Laplacian matrix. Eq.~\eqref{ConvoX} can be easily verified by using the fact $(\mathbf{U}\mathbf{\Lambda}\mathbf{U}^{\top})^k=\mathbf{U}\mathbf{\Lambda}^k\mathbf{U}^{\top}$. It can be observed that this expression is a $K^{\rm{th}}$-order polynomial regarding the Laplacian (i.e., $K$-localized). That is to say, the filtering only depends on the nodes which are at most $K$ steps away from the central node. In this paper, we consider the first-order neighborhood, i.e., $K=1$, and thus Eq.~\eqref{ConvoX} becomes a linear function on the graph Laplacian spectrum with respect to $\mathbf{L}$.
	
	After that, we can build a neural network based on graph convolutions by stacking multiple convolutional layers in the form of Eq.~\eqref{ConvoX}, and each layer is followed by an element-wise non-linear operation softplus($\cdot$) \cite{7280459}. In this way, we can acquire a diverse class of convolutional filter functions by stacking multiple layers with the same configuration. With the linear formulation, Kipf and Welling \cite{Kipf2016Semi} further approximated $\lambda_{\rm{max}}\approx 2$, as the neural network parameters can adapt to this change in scale during the training process. Therefore, Eq.~\eqref{ConvoX} can be simplified to
	\begin{equation}
	\label{ConvoSim1}
	{g_{\bm{\theta}{'}} }\star\mathbf{x} \approx \bm{\theta} {'_0}\mathbf{x}+\bm{\theta} {'_1}(\mathbf{L}-\mathbf{I})\mathbf{x}=\bm{\theta} {'_0}\mathbf{x}-\bm{\theta} {'_1}{\mathbf{D}^{ - \frac{1}{2}}}\mathbf{A}{\mathbf{D}^{ - \frac{1}{2}}}\mathbf{x},
	\end{equation}
	where $\bm{\theta} {'_0}$ and $\bm{\theta} {'_1}$ are two free parameters. Since reducing the number of parameters is beneficial to address overfitting, Eq.~\eqref{ConvoSim1} is converted to
	\begin{equation}
	\label{ConvoSim2}
	{g_{\bm{\theta}}}\star\mathbf{x} \approx {\bm\theta}(\mathbf{I}+\mathbf{D}^{ - \frac{1}{2}}\mathbf{A}{\mathbf{D}^{ - \frac{1}{2}}})\mathbf{x}
	\end{equation}
	by letting $\bm{\theta}=\bm{\theta} {'_0}=-\bm{\theta} {'_1}$. Since $\mathbf{I}+\mathbf{D}^{ - \frac{1}{2}}\mathbf{A}{\mathbf{D}^{ - \frac{1}{2}}}$ has the eigenvalues in the range $[0, 2]$, repeatedly applying this operator will lead to numerical instabilities and exploding/vanishing gradients in a deep neural network. To cope with this problem, Kipf and Welling \cite{Kipf2016Semi} performed the renormalization trick $\mathbf{I}+\mathbf{D}^{ - \frac{1}{2}}\mathbf{A}{\mathbf{D}^{ - \frac{1}{2}}} \to \widetilde{\mathbf{D}}^{ - \frac{1}{2}}\widetilde{\mathbf{A}}{\widetilde{\mathbf{D}}^{ - \frac{1}{2}}}$ with $\widetilde{\mathbf{A}} = \mathbf{A} + \mathbf{I}$ and $\widetilde{\mathbf{D}}_{ii} = \sum\nolimits_j {{\widetilde{\mathbf{A}}_{ij}}}$. As a result, the convolution operation of GCN model can be expressed as
	\begin{equation}
	{\mathbf{H}^{(l)}} = \sigma (\widetilde{\mathbf{A}}{\mathbf{H}^{(l-1)}}{\mathbf{W}^{(l)}}),
	\end{equation}
	where ${\mathbf{H}^{(l)}}$ is the output (namely, embedding result) of the $l^{\rm{th}}$ layer; $\sigma ( \cdot )$ represents an activation function, such as the softplus function \cite{7280459} used in this paper; and $\mathbf{W}^{(l)} $ denotes the trainable weight matrix included by the $l^{\rm{th}}$ layer.

	\subsection{Dynamic Graph Evolution}
	\label{subsection_dgcn}
	
	As mentioned in the introduction, one major disadvantage of conventional GCN is that the graph is fixed throughout the convolution process, which will degrade the final classification performance if the input graph is not accurate. To remedy this defect, in this paper, we propose a dynamic GCN in which the graph can be gradually refined during the convolution process. The main idea is to find an improved graph by fusing the information of current data embeddings and the graph used in the previous layer.
	
	We denote the adjacency matrix in the $l^{\rm{th}}$ layer by $\mathbf{A}^{(l)} $. The embedding kernel ${{\mathbf{K}}_{E}} = {{\rm{\mathbf{H}}}^{(l)}}{{\rm{\mathbf{H}}}^{(l)}}^{\top}$ encodes the pairwise similarity of the embeddings generated from the $l^{\rm{th}}$ layer. For each data point $\mathbf{x}^{(l)}$, it is reasonable to assume that $\mathbf{x}^{(l)}$ obeys Gaussian distribution with the covariance $\mathbf{A}^{(l)}$ and unknown mean $\bm{\mu}^{(l)}$, i.e., $p(\mathbf{x}^{(l)})=\mathcal{N}(\mathbf{x}^{(l)}|\bm{\mu}^{(l)},\mathbf{A}^{(l)} )$. Based on the definitions above, the fused kernel can be obtained by linearly combining $\mathbf{A}^{(l)}$ and ${\mathbf{K}}_E$, namely
	\begin{equation}
	\label{KernelFusion}
	{{\rm{\mathbf{F}}}^{(l)}} = {{\rm{\mathbf{A}}}^{(l)}} + \alpha {\mathbf{K}}_E,
	\end{equation}
	where $\alpha$ is the weight assigned to the embedding kernel ${\mathbf{K}}_E$. The operation in Eq.~\eqref{KernelFusion} actually corresponds to the addition operator: ${\mathbf{z}^{(l)}} = {\mathbf{x}^{(l)}} + \sqrt \alpha  {\mathbf{h}^{(l)}}$ with ${\mathbf{h}^{(l)}}$ being the embedding result of the data point ${\mathbf{x}^{(l)}}$ and ${\mathbf{z}^{(l)}}$ being the fused result. From Eq.~\eqref{KernelFusion}, we can see that this fusion technique utilizes the information of the embedding results encoded in ${\mathbf{K}}_E$ and also the previous adjacency matrix ${{\mathbf{A}}^{(l)}}$ to refine the graph. The advantages of such strategy are two-fold: firstly, the introduction of embedding information helps to find a more accurate graph; and secondly, the improved graph will in turn make the embeddings more discriminative. However, there are still some problems regarding the fused kernel ${\mathbf{F}^{(l)}}$. The fusion in Eq.~\eqref{KernelFusion} will lead to performance degradation if the embeddings are not sufficiently accurate to characterize the intrinsic similarity of the input data. As a result, according to \cite{Bo2012Unsupervised}, we need to re-emphasize the inherent structure among the input data carried by the initial adjacency matrix. Therefore, we do the following projection on the fused result ${\mathbf{z}^{(l)}}$ by using the initial adjacency matrix $\mathbf{A}$, which leads to
	\begin{equation}
	{\mathbf{x}^{(l+1)}} = \mathbf{A}{\mathbf{z}^{(l)}}+{\beta ^{(l)}}\bm{\varepsilon},
	\end{equation}
	where $\bm{\varepsilon}$ denotes white noise, i.e., $p(\bm{\varepsilon} )= \mathcal{N}(\bm{\varepsilon} |0,1)$; the parameter ${\beta ^{(l)}}$ is used to control the relative importance of $\bm{\varepsilon}$. With this projection, we have:
	\begin{equation}
	p(\mathbf{x}^{(l+1)}|\mathbf{z}^{(l)}) = \mathcal{N}(\mathbf{x}^{(l+1)}|\mathbf{A}{\mathbf{z}^{(l)}},{\beta ^{(l)}}\mathbf{I}),
	\end{equation}
	where $\mathbf{I}$ is an identity matrix. Therefore, the marginal distribution of $\mathbf{x}^{(l+1)}$ is
	
	\begin{equation}
	\begin{array}{l}
	p({\mathbf{x}^{(l + 1)}}) = \int {\mathcal{N}({\mathbf{z}^{(l)}}|{\bm{\mu} ^{(l)}},{{\rm{\mathbf{F}}}^{(l)}})} \mathcal{N}({\mathbf{x}^{(l + 1)}}|\mathbf{A}{\mathbf{z}^{(l)}},{\beta ^{(l)}}\mathbf{I})d{\mathbf{z}^{(l)}}\\
	\;\;\;\;\;\;\;\;\;\;\;\;\;\;\, = \mathcal{N}({{\rm{\mathbf{x}}}^{(l+1)}}|{\rm{\mathbf{A}}}{\bm{\mu} ^{(l)}},\mathbf{A}{{\rm{\mathbf{F}}}^{(l)}}{\mathbf{A}^{\top}} + {\beta ^{(l)}}\mathbf{I}).
	\end{array}
	\end{equation}
	Since $\mathbf{x}^{(l+1)}$ is Gaussian distributed with the covariance $\mathbf{A}^{(l+1)}$, the graph can be dynamically updated as
	\begin{equation}
	\label{DynamicA}
	\mathbf{A}^{(l+1)} \leftarrow \mathbf{A}{({{\rm{\mathbf{A}}}^{(l)}} + \alpha {{\rm{\mathbf{H}}}^{(l)}}{{\rm{\mathbf{H}}}^{(l)}}^{\top})}{\mathbf{A}^{\top}} + {\beta ^{(l)}}\mathbf{I}.
	\end{equation}
	To start Eq.~\eqref{DynamicA}, here $\mathbf{A}^{(1)}$ for the first graph convolutional layer is the initial adjacency matrix $\mathbf{A}$. By expanding Eq.~\eqref{DynamicA}, each element of $\mathbf{A}^{(l+1)}$ can be written as
	\begin{equation}
	\label{DynamicAExtension}
	\begin{array}{l}
	\mathbf{A}^{(l+1)}_{ij}=\sum\limits_{\mathbf{x}_k \in N({\mathbf{x}_i})} {\sum\limits_{\mathbf{x}_p \in N({\mathbf{x}_j})}} {{\mathbf{A}_{ik}}{\mathbf{A}_{jp}}(\mathbf{A}_{kp}^{(l)}}+\alpha  \langle {\mathbf{H}_{k,:}^{(l)}, \mathbf{H}_{p,:}^{(l)}} \rangle )  \\
	\;\,\;\,\;\,\;\,\;\,\;\,\;\, +\;\beta^{(l)}{\mathbf{I}_{ij}},
	\end{array}
	\end{equation}
	where $\langle {\cdot,\cdot} \rangle$ is the inner product of two vectors. By observing Eq.~\eqref{DynamicAExtension}, we can find that $\mathbf{x}_i$ is similar to $\mathbf{x}_j$ if they have many common neighbors.

	\begin{figure}[!t]
		\centering
		\centering
		\resizebox*{!}{5cm}{\includegraphics{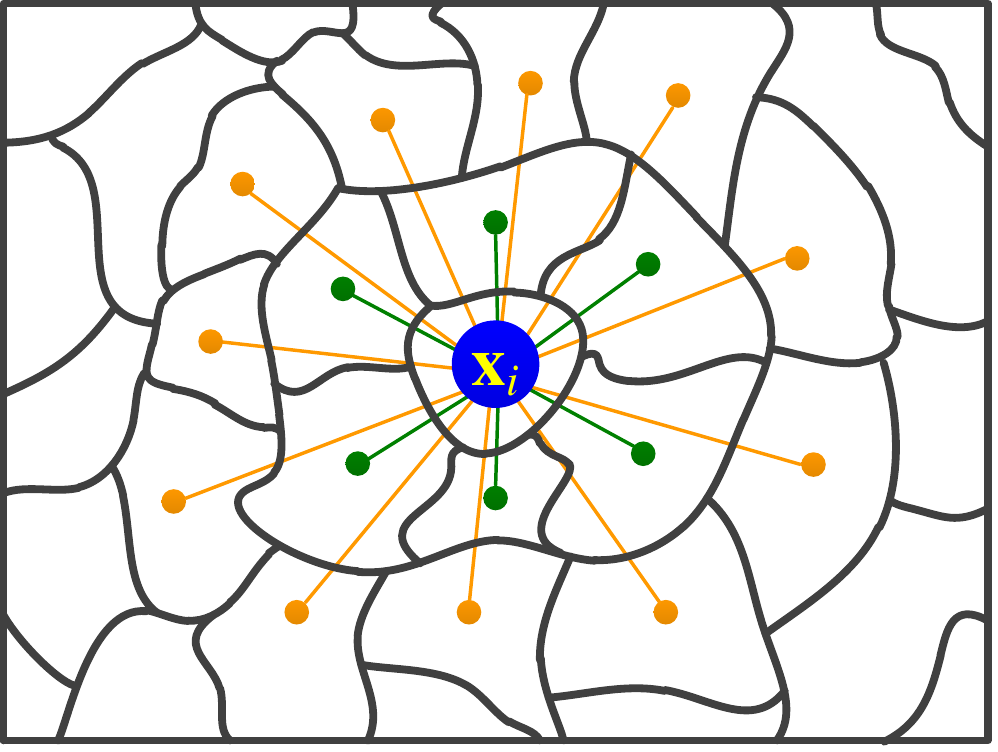}}\hspace{0pt}
		\caption{The illustration of multiple scales considered by our method. The green nodes denote the $1$-hop neighbors of $\mathbf{x}_i$, and the orange nodes together with the green nodes represent $\mathbf{x}_i$'s $2$-hop neighbors.}
		\label{OrdersOfNeighborhood}
	\end{figure}

	\subsection{Multi-Scale Manipulation}
	\label{subsection_multi}
	
	Multi-scale information has been widely demonstrated to be useful for hyperspectral image classification problems \cite{Srivastava2014Dropout, 7729625}. This is because that the objects in a hyperspectral image usually have different geometric appearances, and the contextual information revealed by different scales helps to exploit the abundant local property of image regions from diverse levels. In our method, the multi-scale spectral-spatial information is captured by constructing graphs at different neighborhood scales. Specifically, at the scale $s$, every superpixel $\mathbf{x}_{i}$ is connected to its $s$-hop neighbors. Fig.~\ref{OrdersOfNeighborhood} exhibits the $1$-hop and $2$-hop neighbors of a central example $\mathbf{x}_{i}$ to illustrate the multi-scale design. Then, the receptive field of $\mathbf{x}_{i}$ at the scale $s$ is formed as
	\begin{equation}
	{R_{s}}({\mathbf{x}_i}) = {R_{s - 1}}({\mathbf{x}_i}) \cup {R_{1}}({R_{s - 1}}({\mathbf{x}_i})),
	\end{equation}
	where ${R_{0}}({\mathbf{x}_i})=\mathbf{x}_i$ and ${R_{1}}({\mathbf{x}_i})$ is the set of $1$-hop neighbors of $\mathbf{x}_i$.  By considering both the effectiveness and efficiency, in our method, we construct the graphs at the scale $1$, $2$, and $3$, respectively. Therefore, the formulation of the graph convolutional layer is expressed as
	\begin{equation}
	\label{MultiscaleGraphConvolution}
	\begin{array}{l}
	{\mathbf{H}^{(l)}_{s}} = \sigma ( \mathbf{A}^{(l)}_{s}{\mathbf{H}^{(l-1)}_{s}}{\mathbf{W}^{(l)}_{s}})\\
	\end{array},
	\end{equation}
	where $\mathbf{A}^{(l)}_{s}, \mathbf{H}^{(l)}_{s}$, and $\mathbf{W}^{(l)}_{s}$ denote the adjacency matrix, the output matrix, and the trainable weight matrix of the $l^{\rm{th}}$ graph convolutional layer at the scale $s$. Note that the input matrix $\mathbf{H}^{(0)}$ is shared by all scales. Based on Eq.~\eqref{MultiscaleGraphConvolution}, the output of MDGCN can be obtained by
	\begin{equation}
	\label{Output}
	\begin{array}{l}
	{\mathbf{O} = \sum\nolimits_s {\mathbf{H}_{s}^{(L)}}}
	\end{array},
	\end{equation}
	where $L$ is the number of graph convolutional layers shared by all scales, and $\mathbf{O}$ is the output of MDGCN. The convolution process of MDGCN is summarized in Algorithm \ref{Algorithm1}. In our model, the cross-entropy error is adopted to penalize the difference between the network output and the labels of the original labeled examples, which is
	\begin{equation}
	\label{ETerm}
	\mathcal{L} =  - \sum\limits_{g \in {\mathbf{y}_G}} {\sum\limits_{f = 1}^C {{\mathbf{Y}_{gf}}\ln {\mathbf{O}_{gf}}} },
	\end{equation}
	where ${\mathbf{y}_{G}}$ is the set of indices corresponding to the labeled examples; $C$ denotes the number of classes and $\mathbf{Y}$ denotes the label matrix. Similar to \cite{Kipf2016Semi}, the network parameters here are learned by using full-batch gradient descent, where all superpixels are utilized to perform gradient descent in each iteration. The implementation details of our MDGCN are shown in Algorithm \ref{Algorithm2}.
	
	\begin{algorithm}[!t] 
		\caption{The Multi-scale Dynamic Convolution Process of MDGCN} 
		\label{Algorithm1} 
		\begin{algorithmic}[1] 
			\Require 
			Input matrix $\mathbf{H}^{(0)}$; number of scales $S$; number of graph convolutional layers $L$; initial adjacency matrices $\mathbf{A}^{(1)}_s$ $(1\leq s\leq S)$;
			\For {$l=1$ to $L$}
			\State Calculate the outputs of the $l^{\rm{th}}$ layer $\mathbf{H}^{(l)}_{s}$ $(1\leq s\leq S)$ according to Eq.~\eqref{MultiscaleGraphConvolution};
			\State Update the graphs $\mathbf{A}^{(l+1)}_{s}$ $(1\leq s\leq S)$ according to Eq.~\eqref{DynamicA};
			\EndFor 	
			\State \textbf{end for}	
			\State Calculate the network output according to Eq.~\eqref{Output};	
			\Ensure 
			Network output $\mathbf{O}$.
		\end{algorithmic} 
	\end{algorithm}
	
	\begin{algorithm}[!t] 
		\caption{The Proposed MDGCN for Hyperspectral Image Classification} 
		\label{Algorithm2} 
		\begin{algorithmic}[1] 
			\Require 
			Input image; number of iterations $\mathcal{T}=5000$; learning rate $\eta=0.0005$; number of scales $S=3$; number of graph convolutional layers $L=2$;
			\State Segment the whole image into superpixels via SLIC algorithm;
			\State Construct the initial adjacency matrices $\mathbf{A}^{(1)}_{s}$ $(1\leq s\leq S)$ according to Eq.~\eqref{InitialAdjacencyMatirx};
			\State // Train the MDGCN model
			\For {$t=1$ to $\mathcal{T}$}
			
			\State Conduct multi-scale dynamic convolution by Algorithm \ref{Algorithm1};
			\State Calculate the error term according to Eq.~\eqref{ETerm}, and update the weight matrices $\mathbf{W}^{(l)}_{s}$ $( 1\leq l\leq L,1\leq s\leq S)$ using full-batch gradient descent;
			
			\EndFor 
			\State \textbf{end for}

			\State Conduct label prediction by Algorithm \ref{Algorithm1};
			
			\Ensure 
			Predicted label for each superpixel.
		\end{algorithmic} 
	\end{algorithm}

	\section{Experimental Results}
	\label{Experiments}
	
	In this section, we conduct exhaustive experiments to validate the effectiveness of the proposed MDGCN method, and also provide the corresponding algorithm analyses. To be specific, we first compare MDGCN with other state-of-the-art approaches on three publicly available hyperspectral image datasets, where four metrics including per-class accuracy, overall accuracy (OA), average accuracy (AA), and kappa coefficient are adopted. Then, we demonstrate that both the multi-scale manipulation and dynamic graph design in our DMGCN are beneficial to obtaining the promising performance. After that, we validate the effectiveness of our method in dealing with the boundary regions. Finally, we compare the computational time of various methods to show the efficiency of our algorithm.

	\subsection{Datasets}
	\label{dataset}
	
	The performance of the proposed MDGCN is evaluated on three datasets, i.e., the Indian Pines, the University of Pavia, and the Kennedy Space Center, which will be introduced below.
	
	\subsubsection{Indian Pines}
	
\begin{figure}[!t]
		\centering
		
		\subfigure[]{%
			\resizebox*{3cm}{!}{\includegraphics{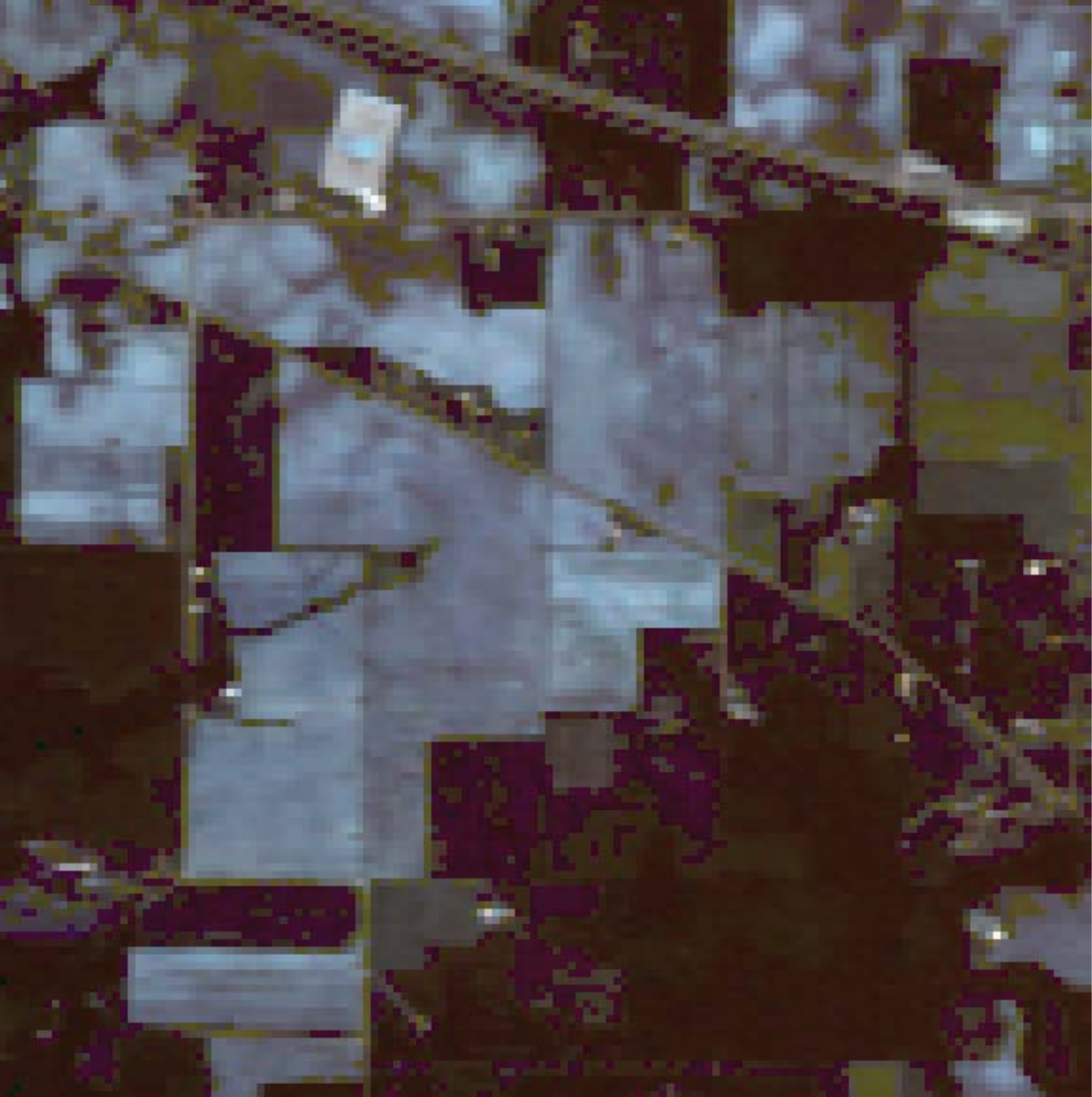}}}\hspace{0pt}	
		\subfigure[]{%
			\label{IPclassmap1}
			\resizebox*{3cm}{!}{\includegraphics{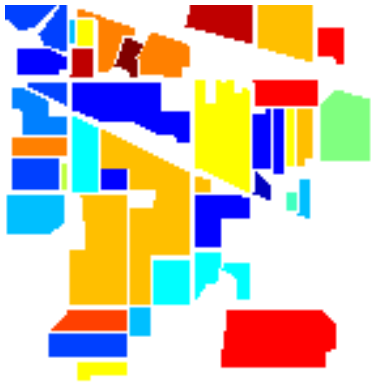}}}\hspace{0pt}

		\subfigure {%
			\resizebox*{!}{1.5cm}{\includegraphics{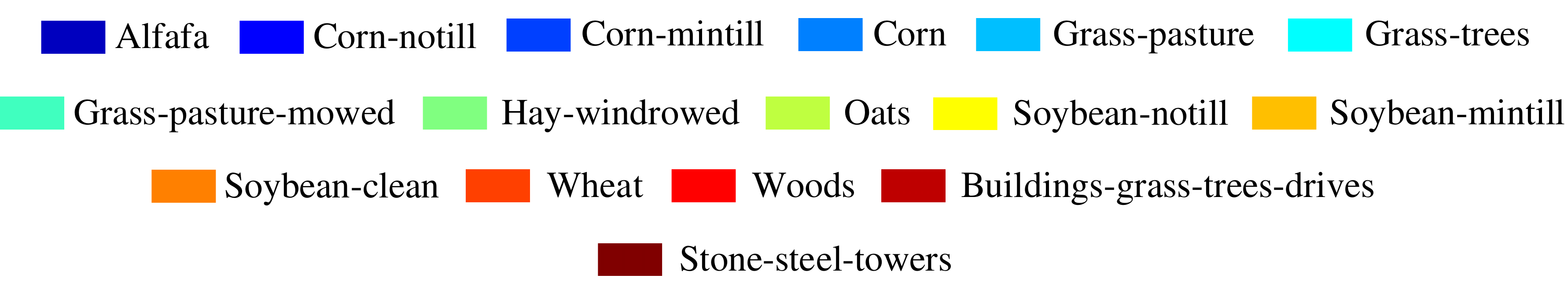}}}\hspace{1pt}
		
		\caption{Indian Pines. (a) False color image. (b) Ground-truth map.}
		\label{IPgtfc}
	\end{figure}

	The Indian Pines dataset was collected by Airborne Visible/Infrared Imaging Spectrometer sensor in 1992, which records north-western India. It consists of $145\times145$ pixels with a spatial resolution of 20 m $\times$ 20 m and has 220 spectral channels covering the range from 0.4 $\mu$m to 2.5 $\mu$m. As a usual step, 20 water absorption and noisy bands are removed, and 200 bands are reserved. The original ground truth includes 16 land-cover classes, such as `Alfalfa', `Corn-notill', and `Corn-mintill'. Fig.~\ref{IPgtfc} exhibits the false color image and ground-truth map of the Indian Pines dataset. The amounts of labeled and unlabeled pixels of various classes are listed in Table~\ref{IPnum}.

	\begin{table}[!t]
		\centering
		\caption{Numbers of Labeled and Unlabeled Pixels of All Classes in Indian Pines Dataset}
		\begin{tabular}{cccc}
			\toprule
			ID   & Class & \#Labeled  & \#Unlabeled \\
			\midrule
			1     & Alfalfa & 30    & 31 \\
			2     & Corn-notill & 30    & 1398 \\
			3     & Corn-mintill & 30    & 800 \\
			4     & Corn  & 30    & 207 \\
			5     & Grass-pasture & 30    & 453 \\
			6     & Grass-trees & 30    & 700 \\
			7     & Grass-pasture-mowed & 15    & 13 \\
			8     & Hay-windrowed & 30    & 448 \\
			9     & Oats  & 15    & 5 \\
			10    & Soybean-notill & 30    & 942 \\
			11    & Soybean-mintill & 30    & 2425 \\
			12    & Soybean-clean & 30    & 563 \\
			13    & Wheat & 30    & 175 \\
			14    & Woods & 30    & 1235 \\
			15    & Buildings-grass-trees-drives & 30    & 356 \\
			16    & Stone-steel-towers & 30    &63 \\
			\bottomrule
		\end{tabular}%
		\label{IPnum}%
	\end{table}%
	
	\subsubsection{University of Pavia}
	
\begin{figure}[!t]
		\centering
		\subfigure[]{%
			\resizebox*{3cm}{!}{\includegraphics{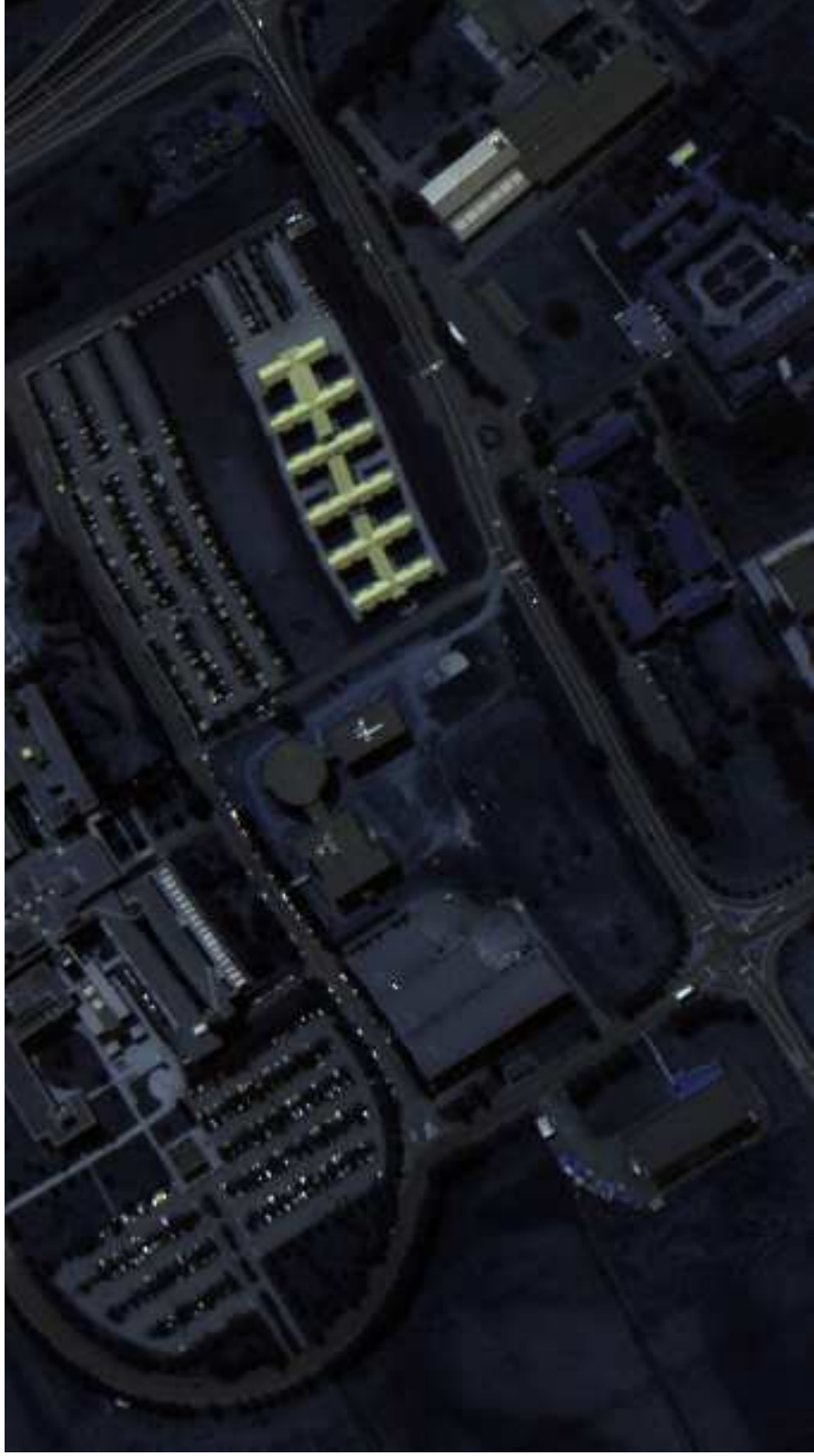}}}\hspace{6pt}
		\subfigure[]{%
			\resizebox*{3cm}{!}{\includegraphics{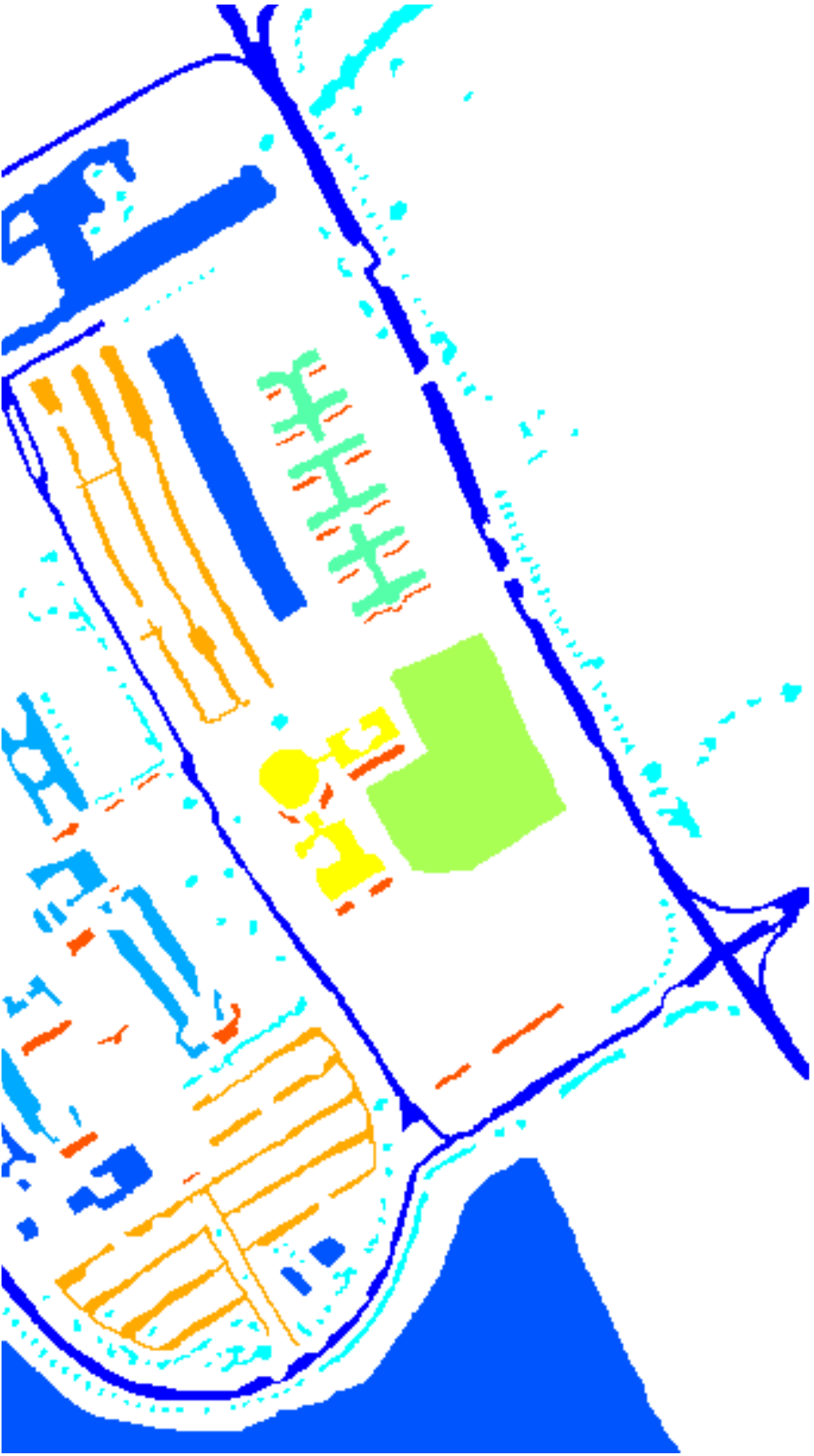}}}\hspace{6pt}

		\subfigure {%
			\resizebox*{!}{0.7cm}{\includegraphics{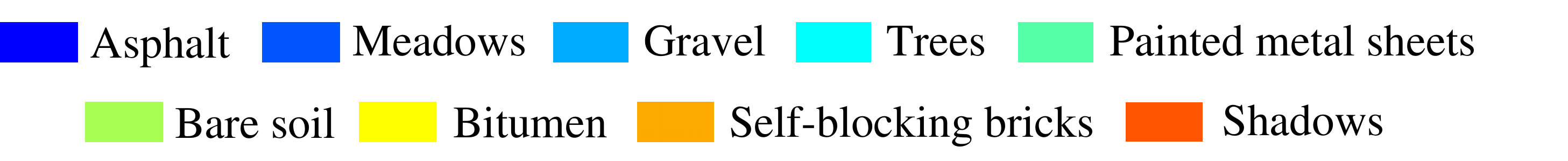}}}\hspace{0pt}	
		\caption{University of Pavia. (a) False color image. (b) Ground-truth map.} 
		\label{PUSgtfc}
	\end{figure}

	The University of Pavia dataset captured the Pavia University in Italy with the ROSIS sensor in 2001. It consists of $610\times340$ pixels with a spatial resolution of 1.3 m $\times$ 1.3 m and has 103 spectral channels in the wavelength range from 0.43 $\mu$m to 0.86 $\mu$m after removing noisy bands. This dataset includes 9 land-cover classes, such as `Asphalt', `Meadows', and `Gravel', which are shown in Fig.~\ref{PUSgtfc}. Table~\ref{PUSnum} lists the amounts of labeled and unlabeled pixels of each class.
	
	\begin{table}[!t]
		\centering
		\caption{Numbers of Labeled and Unlabeled Pixels of All Classes in University of Pavia Dataset}
		\begin{tabular}{cccc}
			\toprule
			ID   & \multicolumn{1}{c}{Class} & \#Labeled & \#Unlabeled \\
			\midrule
			1     & \multicolumn{1}{c}{Asphalt} & 30    & 6601 \\
			2     & Meadows & 30    & 18619 \\
			3     & Gravel & 30    & 2069 \\
			4     & Trees & 30    & 3034 \\
			5     & Painted metal sheets & 30    & 1315 \\
			6     & Bare soil & 30    & 4999 \\
			7     & Bitumen & 30    & 1300 \\
			8     & Self-blocking bricks & 30    & 3652 \\
			9     & Shadows & 30    & 917 \\
			\bottomrule
		\end{tabular}%
		\label{PUSnum}%
	\end{table}%
	
	\subsubsection{Kennedy Space Center}
	
\begin{figure}[!t]
		\centering
		\subfigure[]{%
			\resizebox*{3.5cm}{!}{\includegraphics{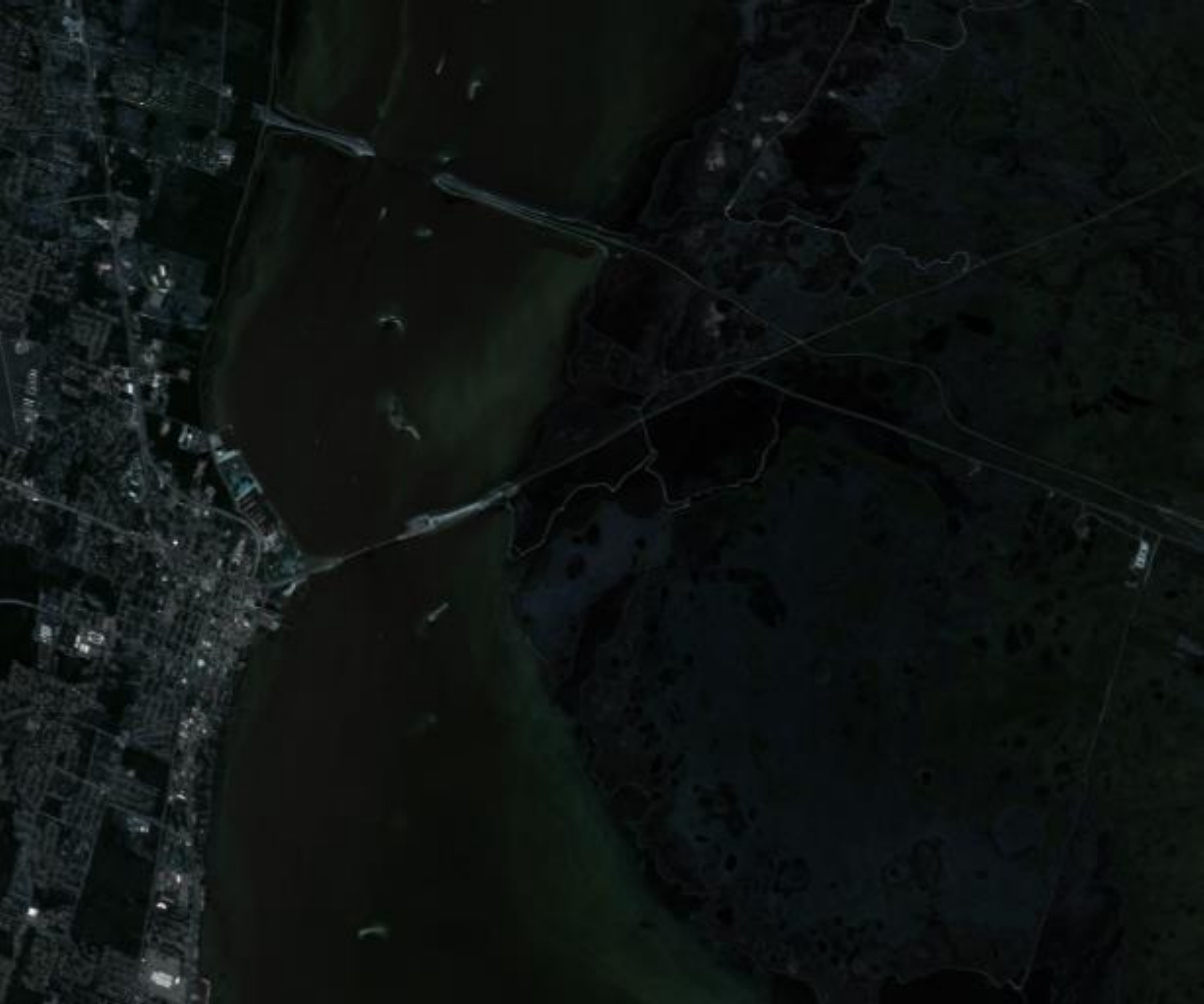}}}\hspace{0pt}
		\subfigure[]{%
			\resizebox*{3.5cm}{!}{\includegraphics{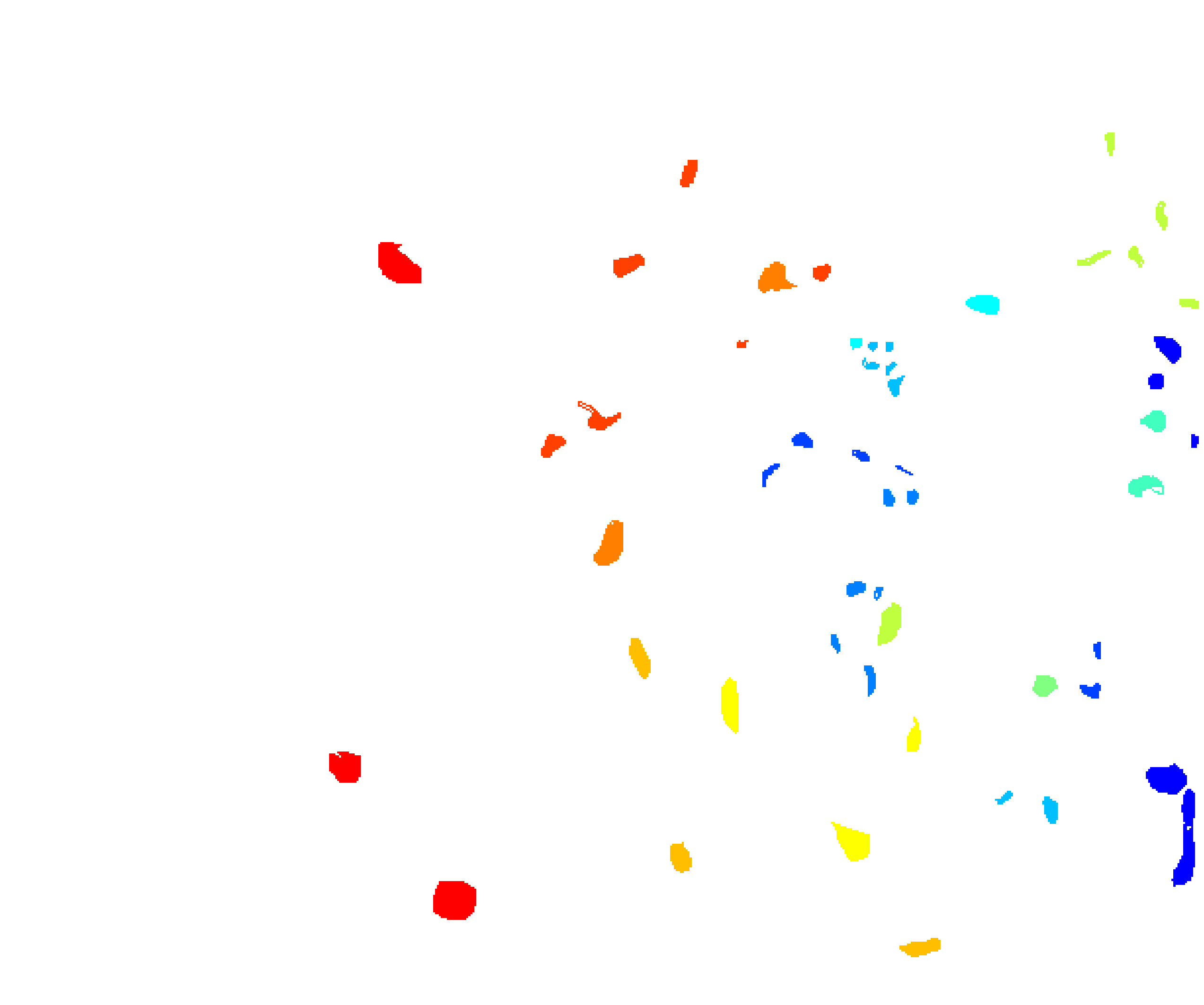}}}\hspace{0pt}

		\subfigure {%
			\resizebox*{!}{1.2cm}{\includegraphics{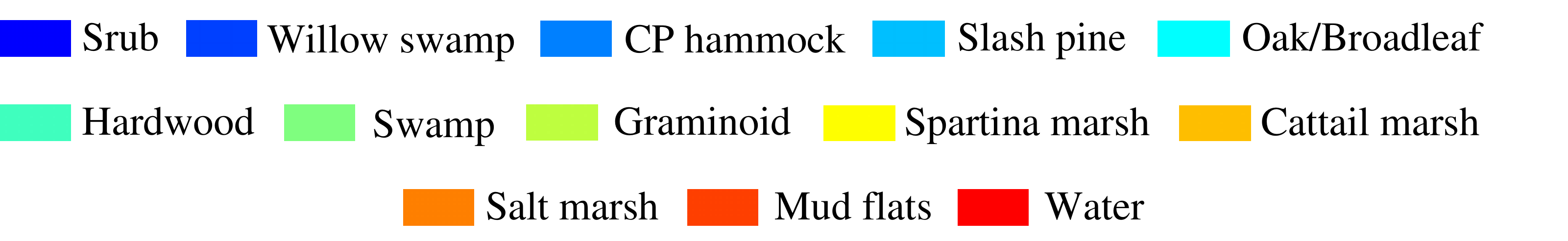}}}\hspace{0pt}		
		
		\caption{Kennedy Space Center. (a) False color image. (b) Ground-truth map.} 
		\label{KSCgtfc}
	\end{figure}

	The Kennedy Space Center dataset was taken by AVIRIS sensor over Florida with a spectral coverage ranging from 0.4 $\mu$m to 2.5 $\mu$m. This dataset contains 224 bands and 614 $\times$ 512 pixels with a spatial resolution of 18 m. After removing water absorption and noisy bands, the remaining 176 bands of the image have been preserved. The Kennedy Space Center dataset includes 13 land-cover classes, such as `Srub', `Willow swamp', and `CP hammock'. Fig.~\ref{KSCgtfc} exhibits the false color image and ground-truth map of the Kennedy Space Center dataset. The numbers of labeled and unlabeled pixels of different classes are listed in Table~\ref{KSCnum}.
	
	\begin{table}[!t]
		\centering
		\caption{Numbers of Labeled and Unlabeled Pixels of All Classes in Kennedy Space Center Dataset}
		\begin{tabular}{cccc}
			\toprule
			ID   & Class & \#Labeled  & \#Unlabeled \\
			\midrule
			1     & Srub  & 30    & 728 \\
			2     & Willow swamp & 30    & 220 \\
			3     & CP hammock & 30    & 232 \\
			4     & Slash pine & 30    & 228 \\
			5     & Oak/Broadleaf & 30    & 146 \\
			6     & Hardwood & 30    & 207 \\
			7     & Swamp & 30    & 96 \\
			8     & Graminoid & 30    & 393 \\
			9     & Spartina marsh & 30    & 469 \\
			10    & Cattail marsh & 30    & 365 \\
			11    & Salt marsh & 30    & 378 \\
			12    & Mud flats & 30    & 454 \\
			13    & Water & 30    & 836 \\
			\bottomrule
		\end{tabular}%
		\label{KSCnum}%
	\end{table}%

	\subsection{Experimental Settings}
	
	In our experiments, the proposed algorithm is implemented via TensorFlow with Adam optimizer. For all the adopted three datasets introduced in Section \ref{dataset}, usually 30 labeled pixels (i.e., examples) are randomly selected in each class for training, and only 15 labeled examples are chosen if the corresponding class has less than 30 examples. During training, 90\% of the labeled examples are used to learn the network parameters and 10\% are used as validation set to tune the hyperparameters. Meanwhile, all the unlabeled examples are used as the test set to evaluate the classification performance. The network architecture of our proposed MDGCN is kept identical for all the datasets. Specifically, three neighborhood scales, namely $s=1$, $s=2$, and $s=3$, are respectively employed for graph construction to incorporate multi-scale spectral-spatial information into our model. For each scale, we employ two graph convolutional layers with 20 hidden units, as GCN-based methods usually do not require deep structure to achieve satisfactory performance \cite{8474300, Gao:2018:LLG:3219819.3219947}. Besides, the learning rate and the number of training epochs are set to 0.0005 and 5000, respectively. 
	
	To evaluate the classification ability of our proposed method, other recent state-of-the-art hyperspectral image classification methods are also used for comparison. Specifically, we employ two CNN-based methods, i.e., Diverse Region-based deep CNN (DR-CNN) \cite{Zhang2018Diverse} and Recurrent 2D-CNN (R-2D-CNN) \cite{Yang2018Hyperspectral}, together with two GCN-based methods, i.e., Graph Convolutional Network (GCN) \cite{Kipf2016Semi} and Spectral-Spatial Graph Convolutional Network (S$^2$GCN) \cite{8474300}. Meanwhile, we compare the proposed MDGCN with three traditional hyperspectral image classification methods, namely Matrix-based Discriminant Analysis (MDA) \cite{Hang2016Matrix}, Hierarchical guidance Filtering-based ensemble classification (HiFi) \cite{7906599}, and Joint collaborative representation and SVM with Decision Fusion (JSDF) \cite{7360896}, respectively. All these methods are implemented ten times on each dataset, and the mean accuracies and standard deviations over these ten independent implementations are reported.

	\begin{table*}[!t]
		\centering
		\caption{Per-Class Accuracy, OA, AA (\%), and Kappa Coefficient Achieved by Different Methods on Indian Pines Dataset}
		\begin{tabular}{ccccccccc}
			\toprule
			ID   & GCN \cite{Kipf2016Semi} & S$^{2}$GCN \cite{8474300} & R-2D-CNN \cite{Yang2018Hyperspectral} & DR-CNN \cite{Zhang2018Diverse} & MDA \cite{Hang2016Matrix} & HiFi \cite{7906599} & JSDF \cite{7360896} & MDGCN \\
			\midrule
			1     & 95.00$\pm$2.80 & \textbf{100.00$\pm$0.00} & \textbf{100.00$\pm$0.00} & \textbf{100.00$\pm$0.00} & \textbf{100.00$\pm$0.00} & 99.38$\pm$1.92 & \textbf{100.00$\pm$0.00} & \textbf{100.00$\pm$0.00} \\
			2     & 56.71$\pm$4.42 & 84.43$\pm$2.50 & 54.94$\pm$2.23 & 80.38$\pm$1.50 & 75.08$\pm$7.35 & 87.42$\pm$4.29 & \textbf{90.75$\pm$3.19} & 80.18$\pm$0.84 \\
			3     & 51.50$\pm$2.56 & 82.87$\pm$5.53 & 73.31$\pm$4.33 & 82.21$\pm$3.53 & 81.44$\pm$5.03 & 93.39$\pm$2.81 & 77.84$\pm$3.81 & \textbf{98.26$\pm$0.00} \\
			4     & 84.64$\pm$3.16 & 93.08$\pm$1.95 & 84.06$\pm$12.98 & 99.19$\pm$0.74 & 95.29$\pm$3.02 & 97.68$\pm$2.76 & \textbf{99.86$\pm$0.33} & 98.57$\pm$0.00 \\
			5     & 83.71$\pm$3.20 & \textbf{97.13$\pm$1.34} & 87.64$\pm$0.31 & 96.47$\pm$1.10 & 91.72$\pm$3.56 & 94.33$\pm$2.71 & 87.20$\pm$2.73 & 95.14$\pm$0.33 \\
			6     & 94.03$\pm$2.11 & 97.29$\pm$1.27 & 91.21$\pm$4.34 & 98.62$\pm$1.90 & 95.46$\pm$1.76 & \textbf{98.71$\pm$1.86} & 98.54$\pm$0.28 & 97.16$\pm$0.57 \\
			7     & 92.31$\pm$0.00 & 92.31$\pm$0.00 & \textbf{100.00$\pm$0.00} & \textbf{100.00$\pm$0.00} & \textbf{100.00$\pm$0.00} & 95.00$\pm$3.36 & \textbf{100.00$\pm$0.00} & \textbf{100.00$\pm$0.00} \\
			8     & 96.61$\pm$1.86 & 99.03$\pm$0.93 & 99.11$\pm$0.95 & 99.78$\pm$0.22 & \textbf{99.80$\pm$0.28} & 99.59$\pm$0.48 & \textbf{99.80$\pm$0.31} & 98.89$\pm$0.00 \\
			9     & \textbf{100.00$\pm$0.00} & \textbf{100.00$\pm$0.00} & \textbf{100.00$\pm$0.00} & \textbf{100.00$\pm$0.00} & \textbf{100.00$\pm$0.00} & \textbf{100.00$\pm$0.00} & \textbf{100.00$\pm$0.00} & \textbf{100.00$\pm$0.00} \\
			10    & 77.47$\pm$1.24 & \textbf{93.77$\pm$3.72} & 70.81$\pm$5.11 & 90.41$\pm$1.95 & 81.95$\pm$4.68 & 91.37$\pm$3.49 & 89.99$\pm$4.24 & 90.02$\pm$1.02 \\
			11    & 56.56$\pm$1.53 & 84.98$\pm$2.82 & 56.35$\pm$1.08 & 74.46$\pm$0.37 & 69.13$\pm$7.07 & 84.33$\pm$3.55 & 76.75$\pm$5.12 & \textbf{93.35$\pm$1.47} \\
			12    & 58.29$\pm$6.58 & 80.05$\pm$5.17 & 63.06$\pm$12.81 & 91.00$\pm$3.14 & 76.58$\pm$5.12 & \textbf{95.02$\pm$3.10} & 87.10$\pm$2.82 & 93.05$\pm$2.30 \\
			13    & \textbf{100.00$\pm$0.00} & 99.43$\pm$0.00 & 98.86$\pm$1.62 & \textbf{100.00$\pm$0.00} & 99.43$\pm$0.75 & 99.29$\pm$0.25 & 99.89$\pm$0.36 & \textbf{100.00$\pm$0.00} \\
			14    & 80.03$\pm$3.93 & 96.73$\pm$0.92 & 88.74$\pm$2.58 & 91.85$\pm$3.40 & 90.92$\pm$3.21 & 98.32$\pm$0.76 & 97.21$\pm$2.78 & \textbf{99.72$\pm$0.05} \\
			15    & 69.55$\pm$6.66 & 86.80$\pm$3.42 & 87.08$\pm$2.78 & 99.44$\pm$0.28 & 91.57$\pm$5.27 & 96.71$\pm$2.06 & 99.58$\pm$0.68 & \textbf{99.72$\pm$0.00} \\
			16    & 98.41$\pm$0.00 & \textbf{100.00$\pm$0.00} & 97.62$\pm$1.12 & \textbf{100.00$\pm$0.00} & 96.63$\pm$4.91 & 99.13$\pm$0.81 & \textbf{100.00$\pm$0.00} & 95.71$\pm$0.00 \\
			\midrule
			OA    & 69.24$\pm$1.56 & 89.49$\pm$1.08 & 72.11$\pm$1.28 & 86.65$\pm$0.59 & 81.91$\pm$1.33 & 91.90$\pm$1.36 & 88.34$\pm$1.39 & \textbf{93.47$\pm$0.38} \\
			AA    & 80.93$\pm$1.71 & 92.99$\pm$1.04 & 84.55$\pm$1.79 & 93.99$\pm$0.25 & 90.31$\pm$0.71 & 95.60$\pm$0.58 & 94.03$\pm$0.55 & \textbf{96.24$\pm$0.21} \\
			Kappa & 65.27$\pm$1.80 & 88.00$\pm$1.23 & 68.66$\pm$1.46 & 84.88$\pm$0.67 & 79.54$\pm$1.46 & 90.77$\pm$1.53 & 86.80$\pm$1.55 & \textbf{92.55$\pm$0.43} \\
			\bottomrule
		\end{tabular}%
		\label{IPClassificationResults}%
	\end{table*}%
	
	\begin{figure*}[!t]
		\centering
		\subfigure[]{%
			\resizebox*{3.2cm}{!}{\includegraphics{IP_fc_cut.pdf}}}\hspace{2pt}
		\subfigure[]{%
			\label{IPClassificationMaps_gt}
			\resizebox*{3.2cm}{!}{\includegraphics{IPgt.pdf}}}\hspace{2pt}
		\subfigure[]{%
			\resizebox*{3.2cm}{!}{\includegraphics{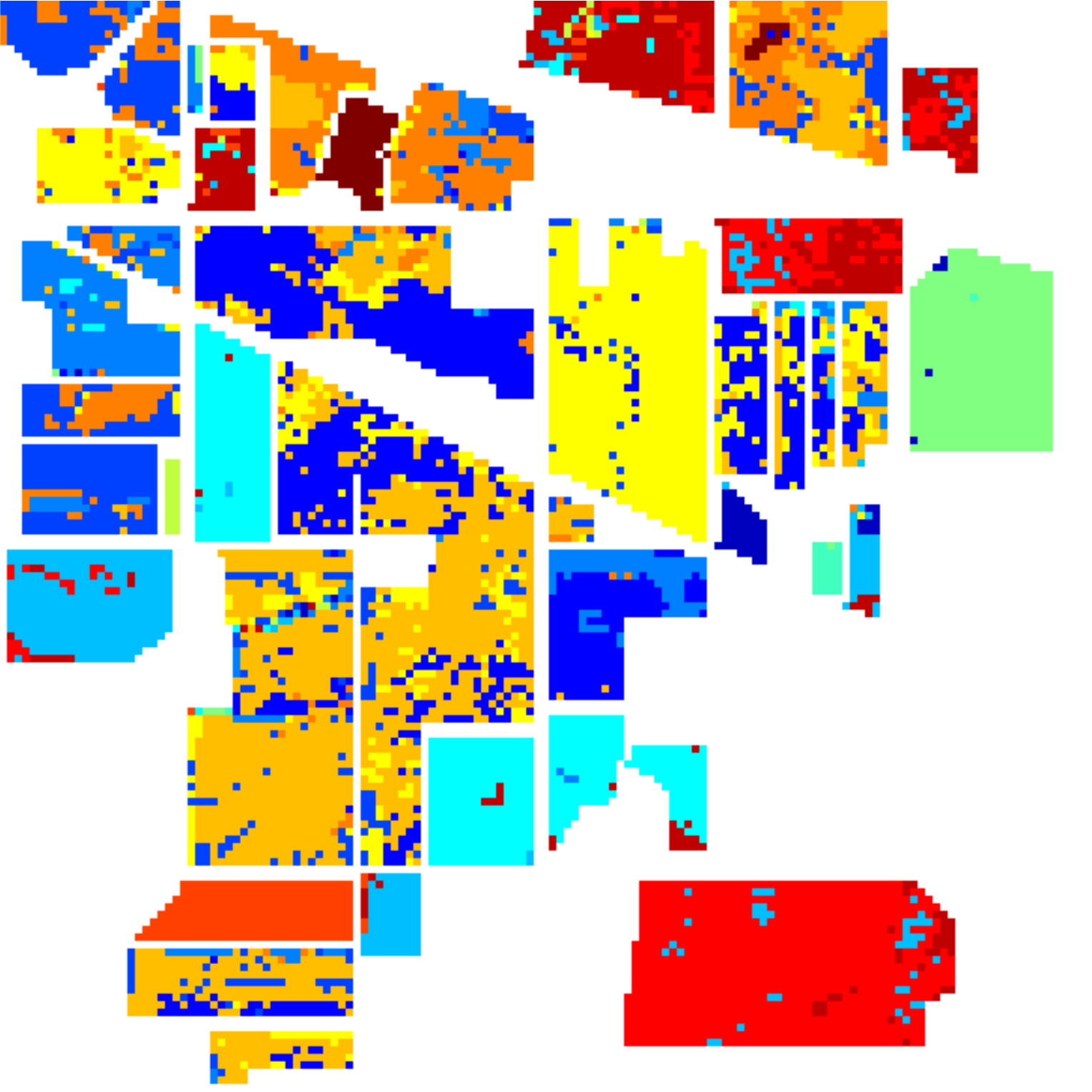}}}\hspace{2pt}	
		\subfigure[]{%
			\resizebox*{3.2cm}{!}{\includegraphics{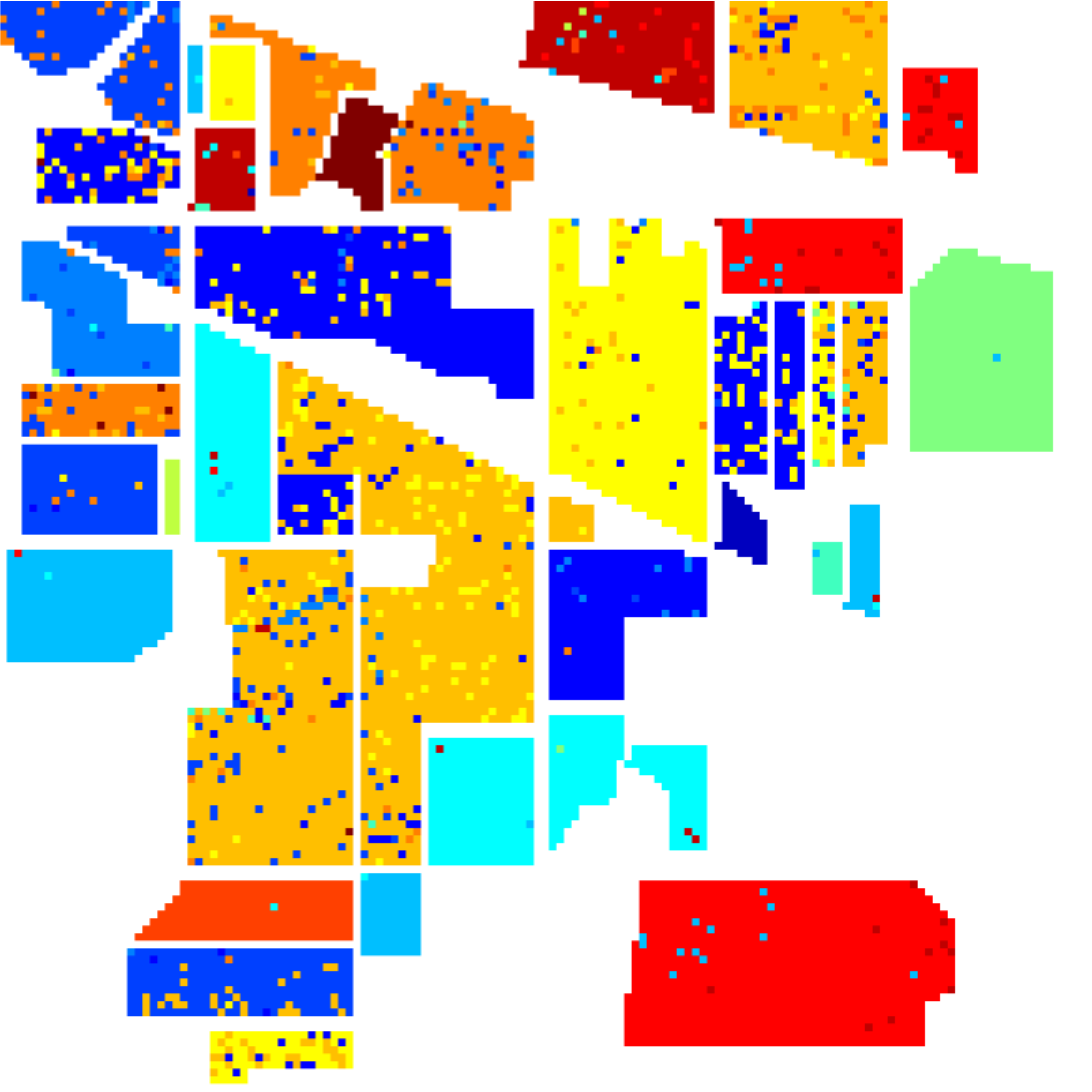}}}\hspace{2pt}	
		\subfigure[]{%
			\resizebox*{3.2cm}{!}{\includegraphics{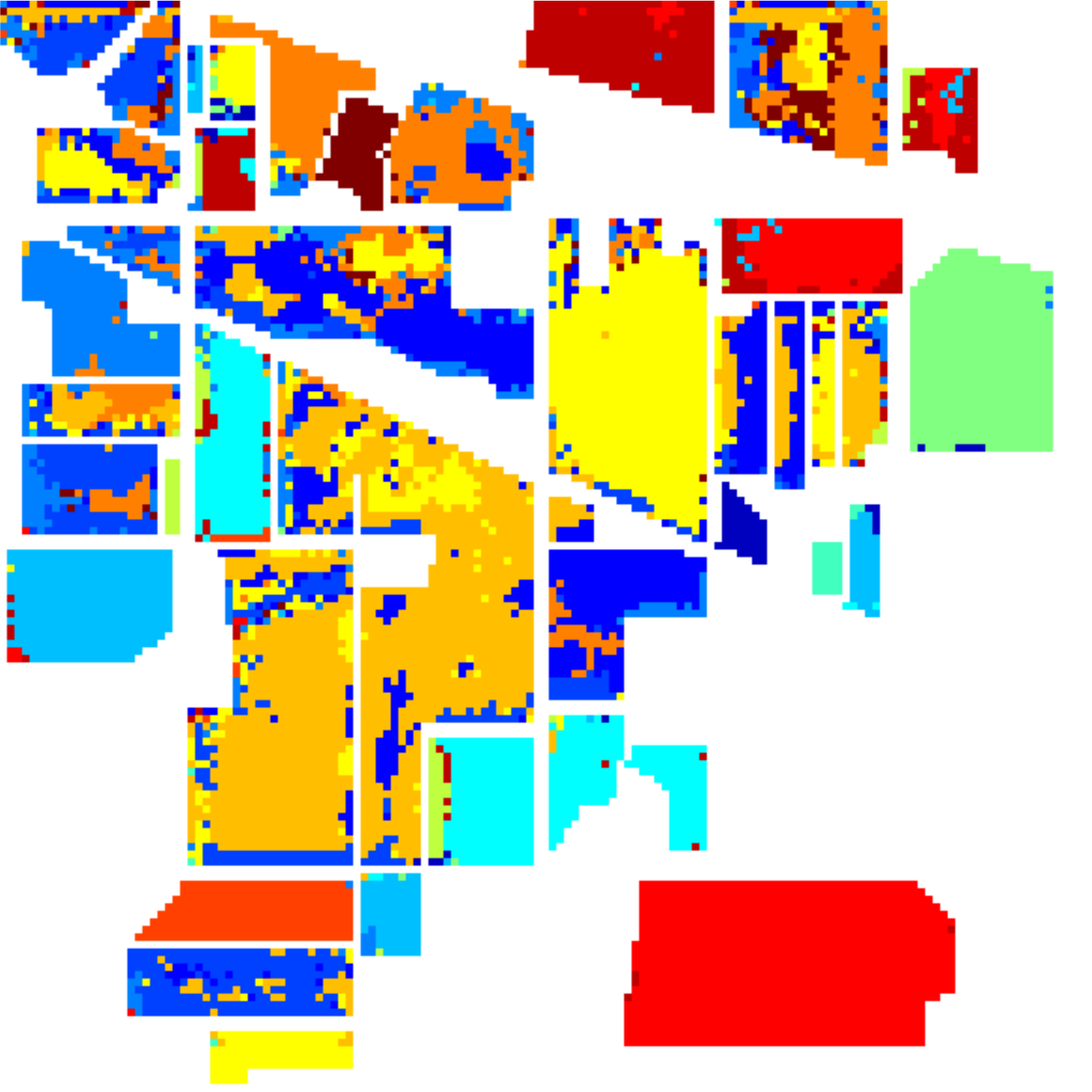}}}\hspace{2pt}	
		\subfigure[]{%
			\resizebox*{3.2cm}{!}{\includegraphics{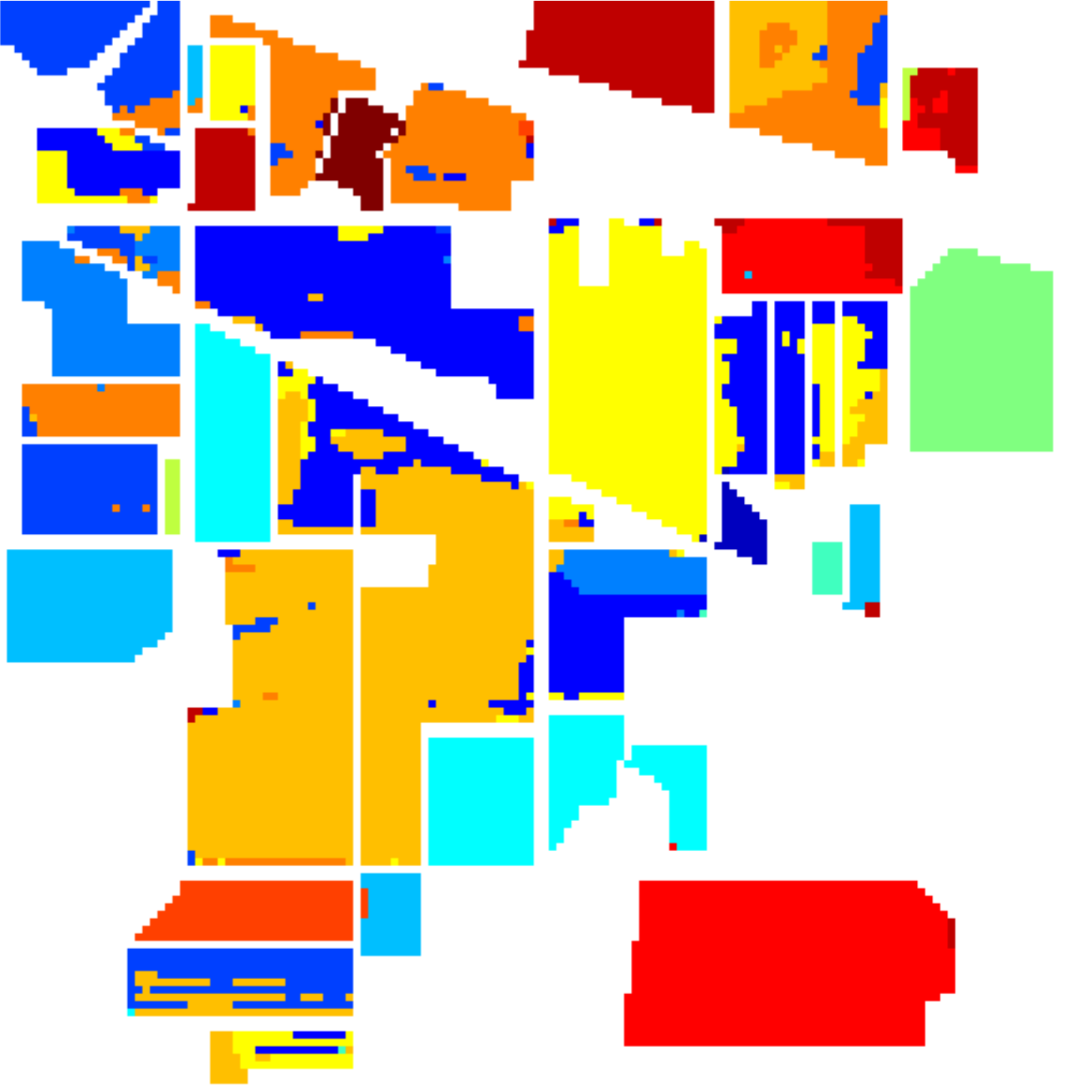}}}\hspace{2pt}	
		\subfigure[]{%
			\resizebox*{3.2cm}{!}{\includegraphics{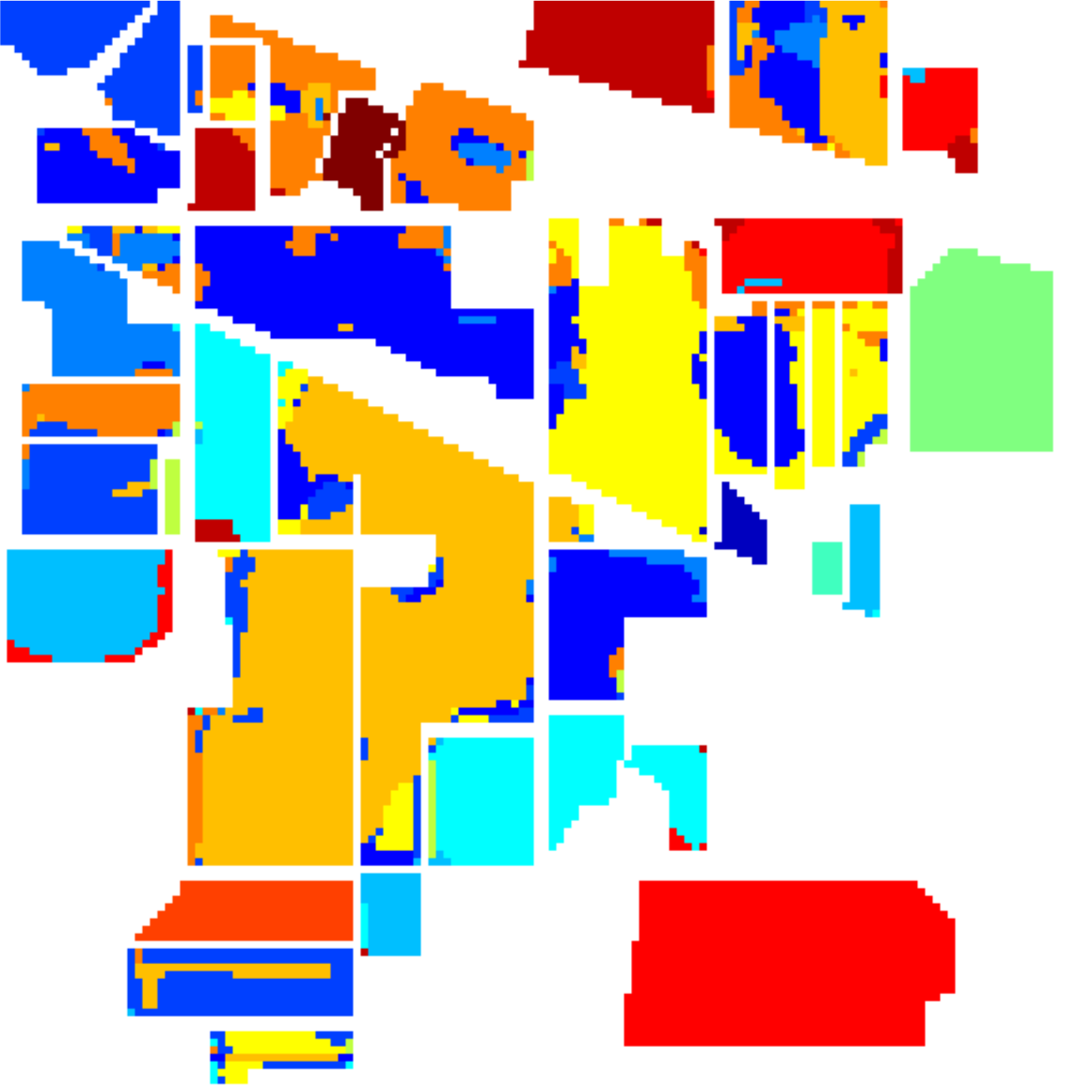}}}\hspace{2pt}
		\subfigure[]{%
			\resizebox*{3.2cm}{!}{\includegraphics{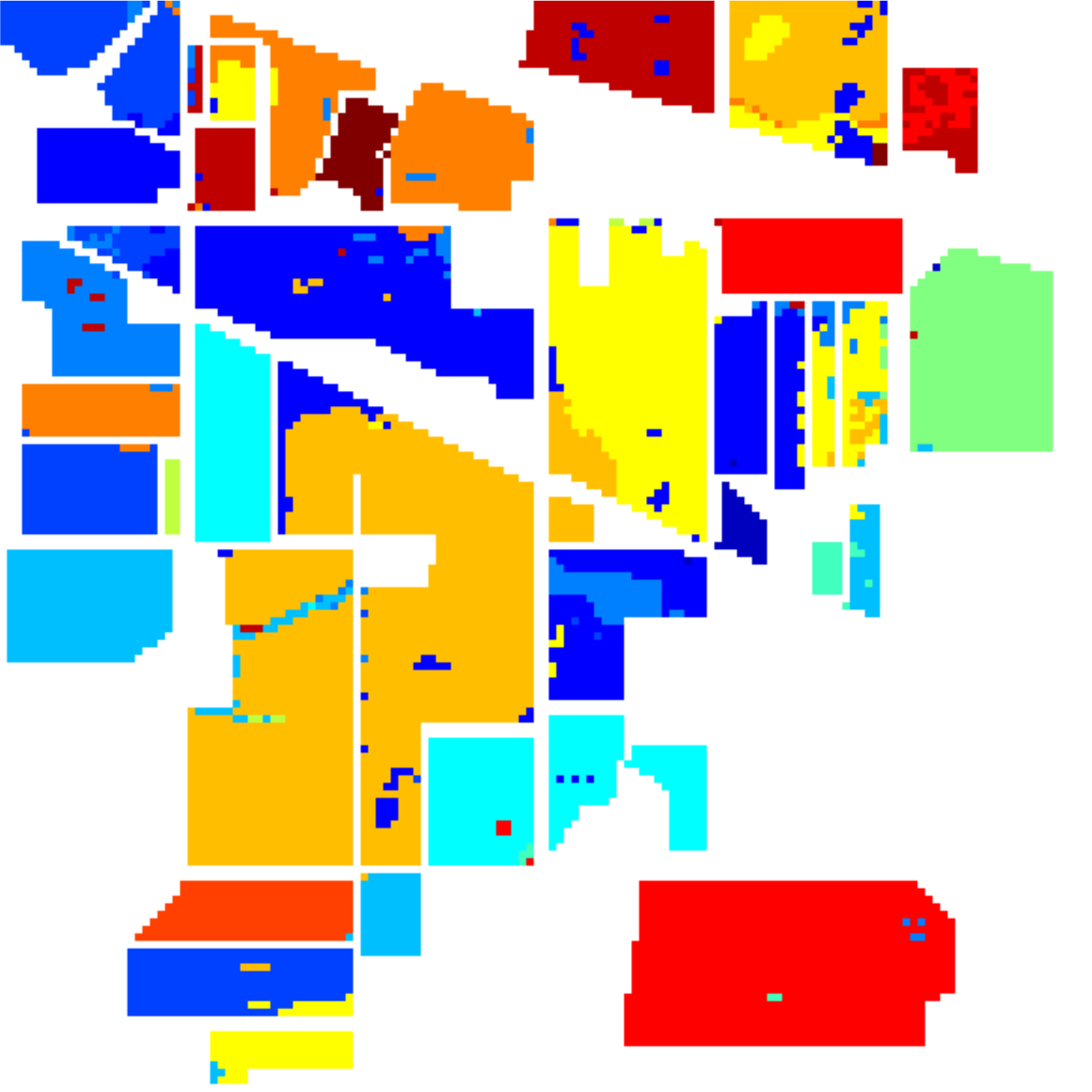}}}\hspace{2pt}			
		\subfigure[]{%
			\resizebox*{3.2cm}{!}{\includegraphics{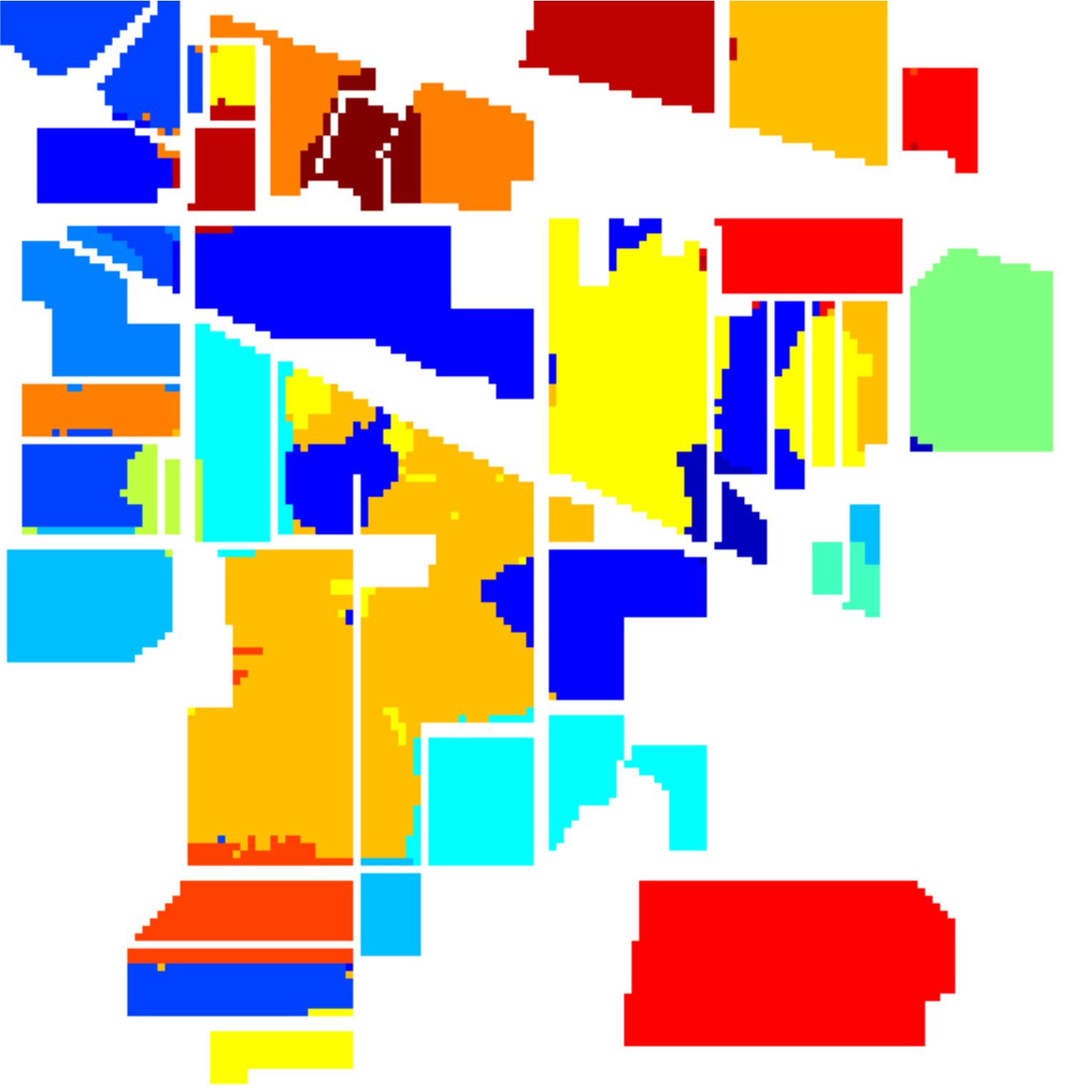}}}\hspace{2pt}	
		\subfigure[]{%
			\resizebox*{3.2cm}{!}{\includegraphics{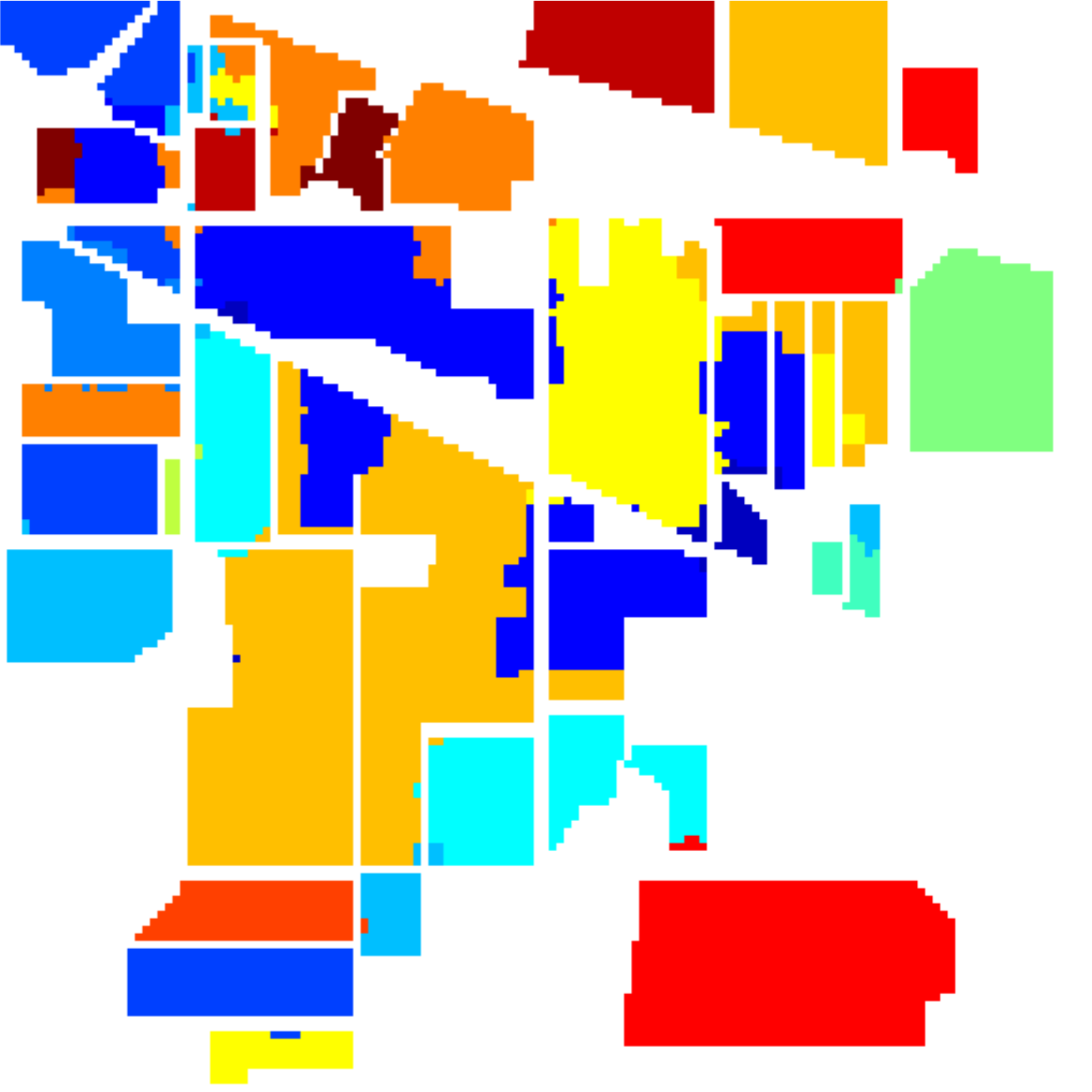}}}\hspace{2pt}	
		
		\caption{Classification maps obtained by different methods on Indian Pines dataset. (a) False color image; (b) Ground-truth map; (c) GCN; (d) S$^{2}$GCN; (e) R-2D-CNN; (f) DR-CNN; (g) MDA; (h) HiFi; (i) JSDF; (j) MDGCN.} 
		\label{IPClassificationMaps}
	\end{figure*}

	\subsection{Classification Results}
	\label{CLassificationResult}
	
	To show the effectiveness of our proposed MDGCN, here we quantitatively and qualitatively evaluate the classification performance by comparing MDGCN with the aforementioned baseline methods.
	
	\begin{table*}[!t]
		\centering
		\caption{Per-Class Accuracy, OA, AA (\%), and Kappa Coefficient Achieved by Different Methods on University of Pavia Dataset}
		\begin{tabular}{ccccccccc}
			\toprule
			ID   & GCN \cite{Kipf2016Semi} & S$^{2}$GCN \cite{8474300} & R-2D-CNN \cite{Yang2018Hyperspectral} & DR-CNN \cite{Zhang2018Diverse} & MDA \cite{Hang2016Matrix} & HiFi \cite{7906599} & JSDF \cite{7360896} & MDGCN \\
			\midrule
			1     & -     & -     & 84.96$\pm$0.56 & 92.10$\pm$3.34 & 78.84$\pm$1.99 & 77.77$\pm$4.52 & 82.40$\pm$4.07 & \textbf{93.55$\pm$0.37} \\
			2     & -     & -     & 79.99$\pm$2.29 & 96.39$\pm$3.20 & 79.96$\pm$3.41 & 95.49$\pm$2.07 & 90.76$\pm$3.74 & \textbf{99.25$\pm$0.23} \\
			3     & -     & -     & 89.49$\pm$0.17 & 84.23$\pm$0.71 & 85.58$\pm$3.46 & \textbf{94.12$\pm$3.26} & 86.71$\pm$4.14 & 92.03$\pm$0.24 \\
			4     & -     & -     & \textbf{98.12$\pm$0.65} & 95.26$\pm$0.67 & 90.90$\pm$2.29 & 82.68$\pm$4.25 & 92.88$\pm$2.16 & 83.78$\pm$1.55 \\
			5     & -     & -     & 99.85$\pm$0.11 & 97.77$\pm$0.00 & 99.93$\pm$0.08 & 97.25$\pm$1.67 & \textbf{100.00$\pm$0.00} & 99.47$\pm$0.09 \\
			6     & -     & -     & 76.79$\pm$7.40 & 90.44$\pm$2.27 & 75.52$\pm$7.11 & \textbf{99.63$\pm$0.78} & 94.30$\pm$4.55 & 95.26$\pm$0.50 \\
			7     & -     & -     & 88.69$\pm$4.57 & 89.05$\pm$1.76 & 84.28$\pm$5.11 & 97.77$\pm$2.43 & 96.62$\pm$1.37 & \textbf{98.92$\pm$1.04} \\
			8     & -     & -     & 67.54$\pm$5.67 & 78.49$\pm$1.53 & 81.82$\pm$6.92 & \textbf{95.12$\pm$2.34} & 94.69$\pm$3.74 & 94.99$\pm$1.33 \\
			9     & -     & -     & \textbf{99.84$\pm$0.08} & 96.34$\pm$0.22 & 97.50$\pm$1.48 & 83.86$\pm$3.40 & 99.56$\pm$0.36 & 81.03$\pm$0.49 \\
			\midrule
			OA    & -     & -     & 82.38$\pm$0.88 & 92.62$\pm$1.15 & 81.60$\pm$2.07 & 92.08$\pm$1.28 & 90.82$\pm$1.30 & \textbf{95.68$\pm$0.22} \\
			AA    & -     & -     & 87.25$\pm$0.68 & 91.12$\pm$0.12 & 86.04$\pm$1.67 & 91.52$\pm$0.99 & 93.10$\pm$0.65 & \textbf{93.15$\pm$0.28} \\
			Kappa & -     & -     & 77.31$\pm$0.97 & 90.27$\pm$1.44 & 76.24$\pm$2.63 & 89.60$\pm$1.65 & 88.02$\pm$1.62 & \textbf{94.25$\pm$0.29} \\
			\bottomrule
		\end{tabular}%
		\label{PUSClassificationResults}%
	\end{table*}%
	
	\begin{figure*}[!t]
		\centering
		\subfigure[]{%
			\resizebox*{3cm}{!}{\includegraphics{PUS_fc_cut.pdf}}}\hspace{10pt}
		\subfigure[]{%
			\resizebox*{3cm}{!}{\includegraphics{PUS_gt.pdf}}}\hspace{10pt}	
		\subfigure[]{%
			\resizebox*{3cm}{!}{\includegraphics{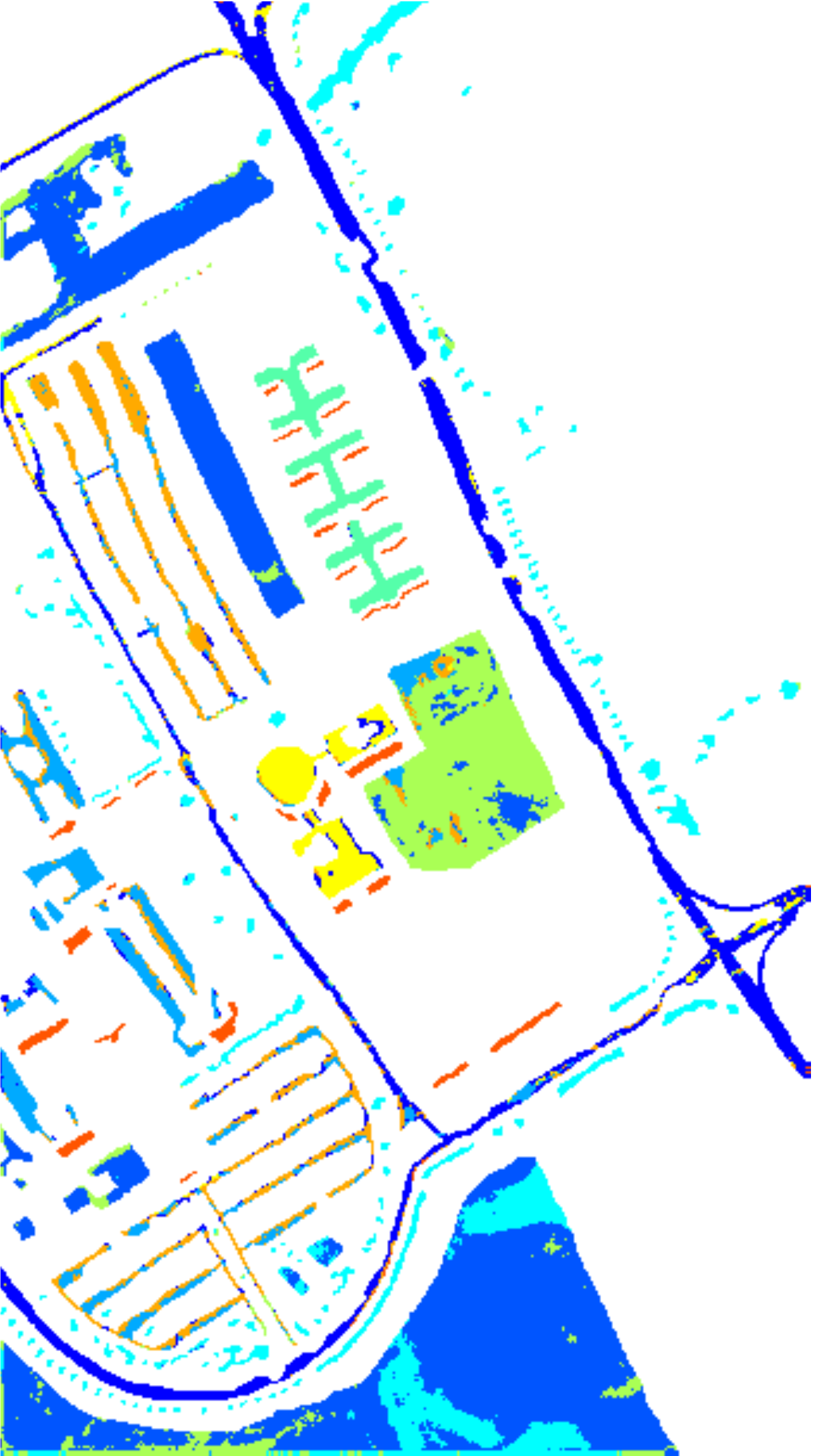}}}\hspace{10pt}	
		\subfigure[]{%
			\resizebox*{3cm}{!}{\includegraphics{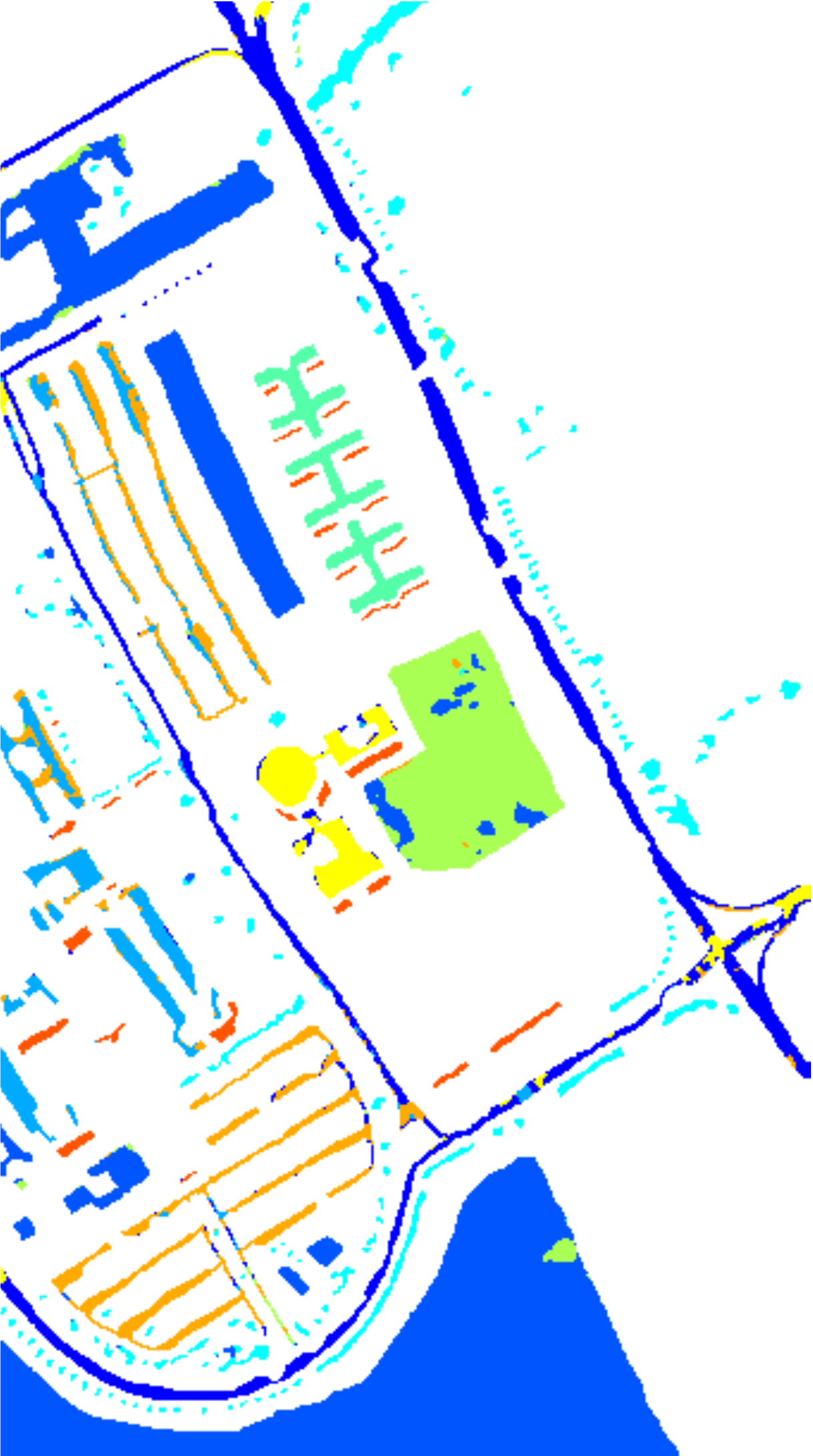}}}\hspace{10pt}
		
		\subfigure[]{%
			\resizebox*{3cm}{!}{\includegraphics{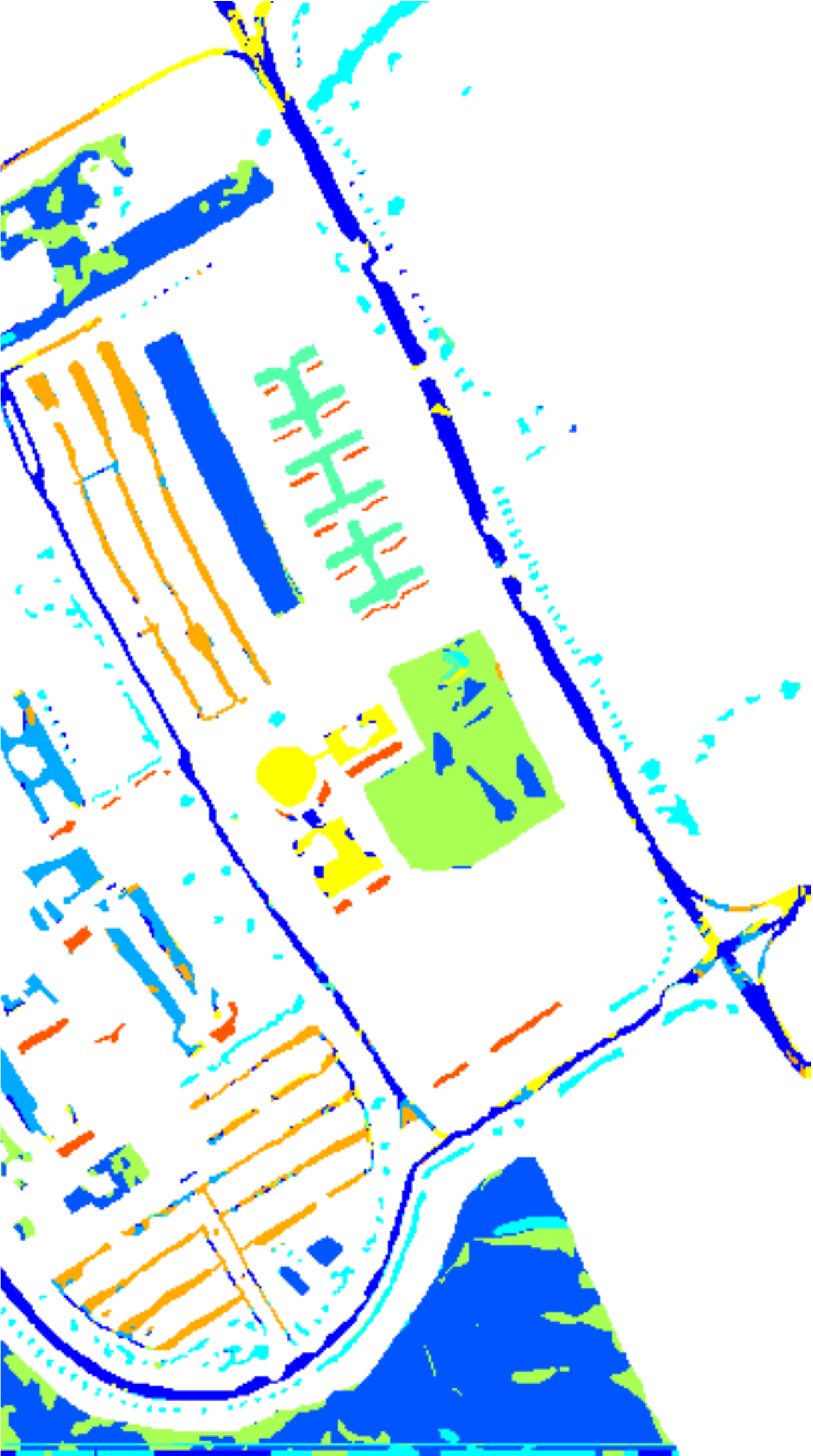}}}\hspace{10pt}
		\subfigure[]{%
			\resizebox*{3cm}{!}{\includegraphics{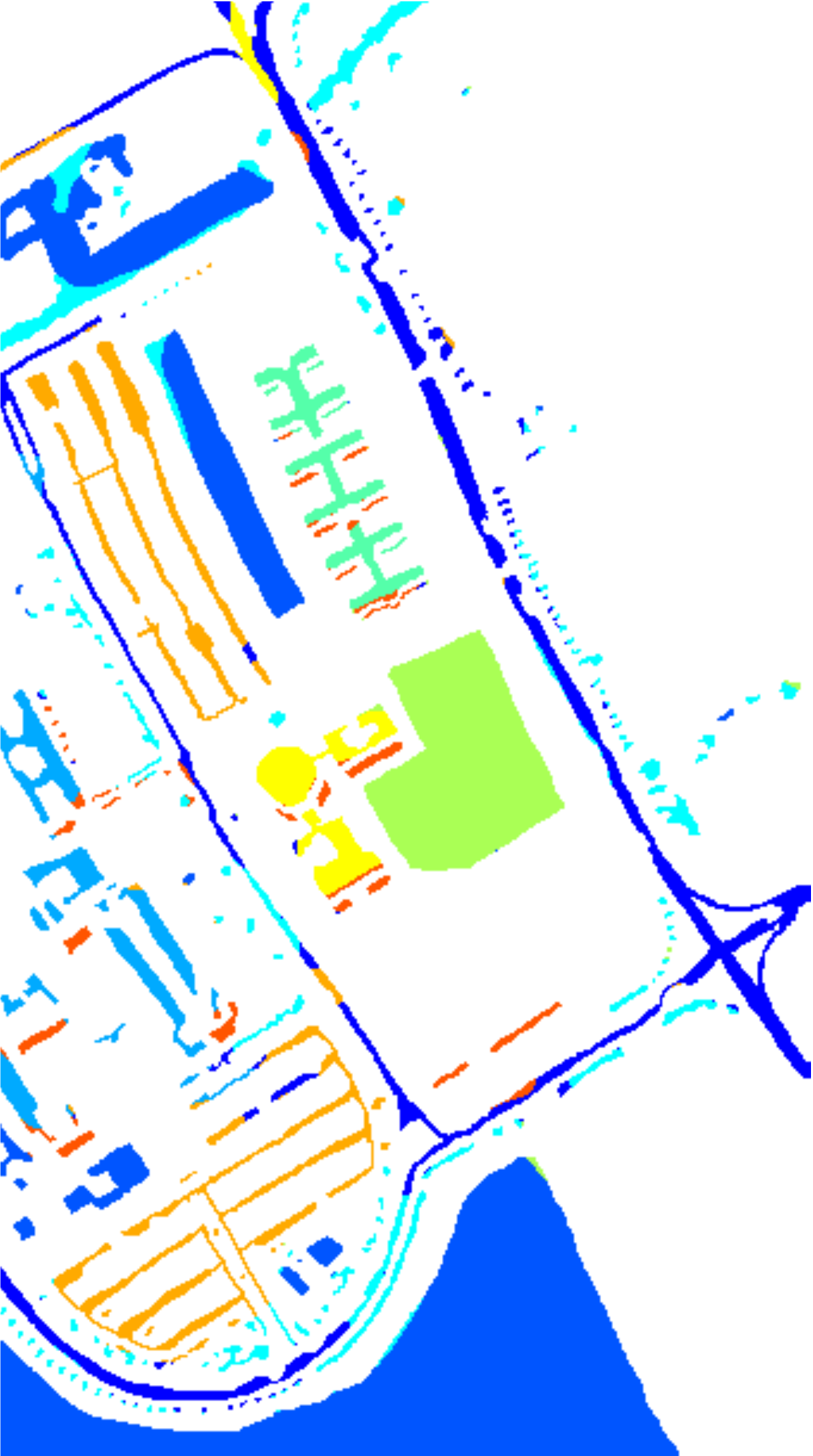}}}\hspace{10pt}			
		\subfigure[]{%
			\resizebox*{3cm}{!}{\includegraphics{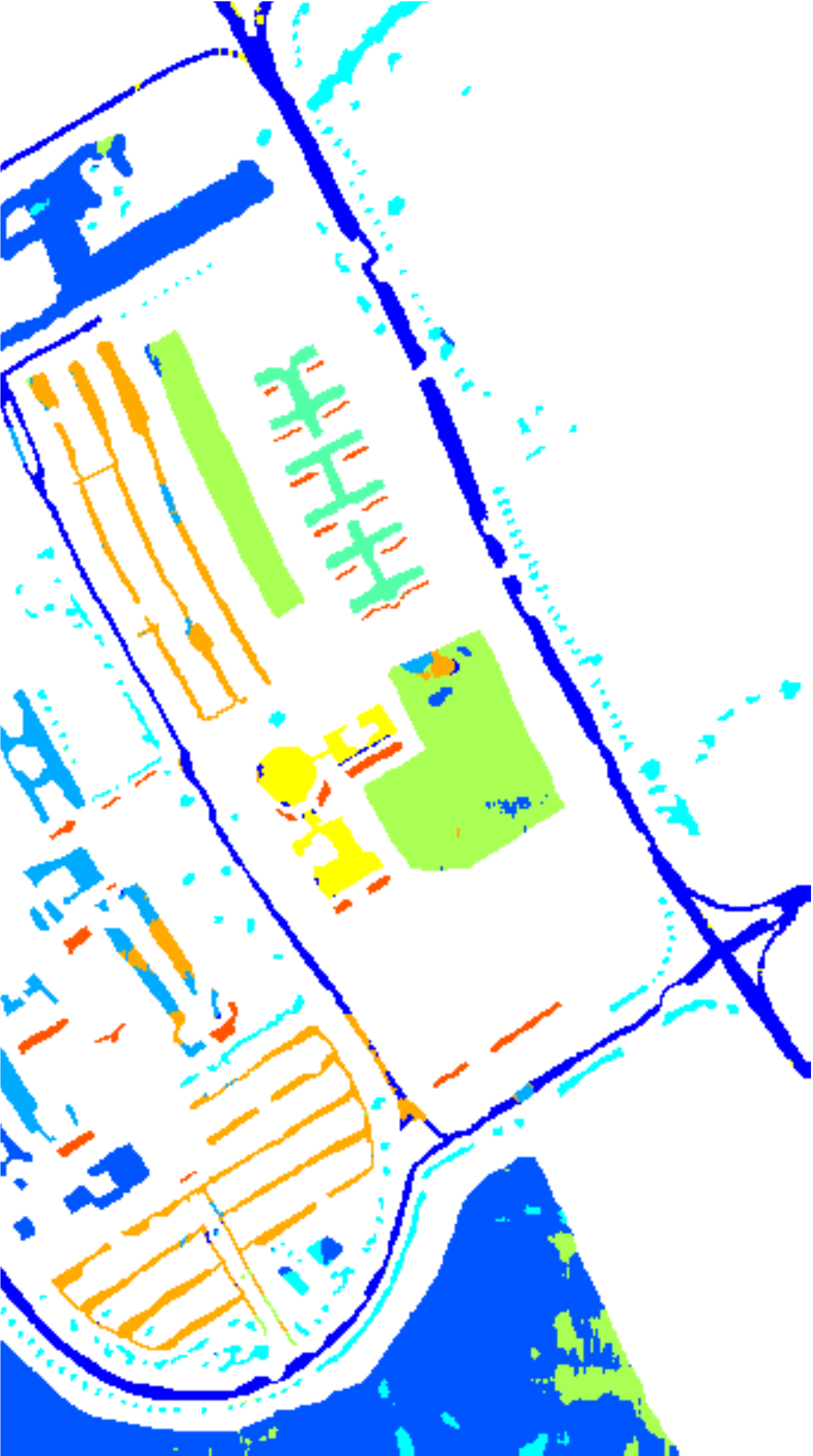}}}\hspace{10pt}	
		\subfigure[]{%
			\resizebox*{3cm}{!}{\includegraphics{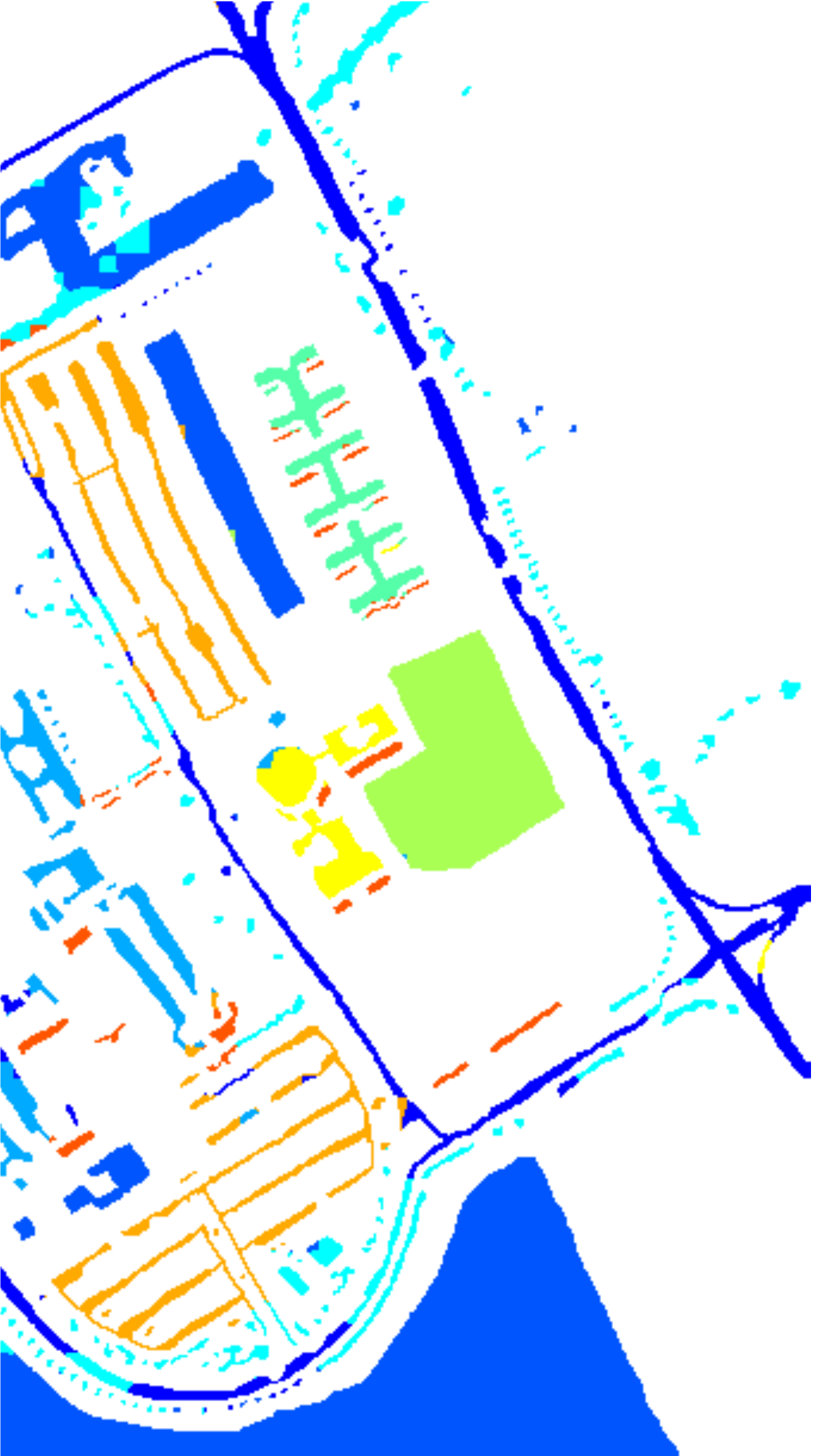}}}\hspace{10pt}	
		
		\caption{Classification maps obtained by different methods on University of Pavia dataset. (a) False color image; (b) Ground-truth map; (c) R-2D-CNN; (d) DR-CNN; (e) MDA; (f) HiFi; (g) JSDF; (h) MDGCN.} 
		\label{PUSClassificationMaps}
	\end{figure*}
	
	\subsubsection{Results on the Indian Pines Dataset} The quantitative results obtained by different methods on the Indian Pines dataset are summarized in Table~\ref{IPClassificationResults}, where the highest value in each row is highlighted in bold. We observe that the CNN-based methods including R-2D-CNN and DR-CNN achieve relatively low accuracy, which is due to the reason that they can only conduct the convolution on a regular image grid, so the specific local spatial information cannot be captured. By contrast, GCN-based methods such as S$^{2}$GCN and MDGCN are capable of adaptively aggregating the features on irregular non-Euclidean regions, so they can yield better performance than R-2D-CNN and DR-CNN. The HiFi algorithm, which combines spectral and spatial information in diverse scales, ranks in the second place. This implies that the multi-scale spectral-spatial features are quite useful to enhance the classification performance. Furthermore, we observe that the proposed MDGCN achieves the top level performance among all the methods in terms of OA, AA, and Kappa coefficient, and the standard deviations are also very small, which reflects that the proposed MDGCN is more stable and effective than the compared methods.
	
	Fig.~\ref{IPClassificationMaps} exhibits a visual comparison of the classification results generated by different methods on the Indian Pines dataset, and the ground-truth map is provided in Fig.~\ref{IPClassificationMaps_gt}. Compared with the ground-truth map, it can be seen that some pixels of `Soybean-mintill' are misclassified into `Corn-notill' in all the classification maps because these two land-cover types have similar spectral signatures. Meanwhile, due to the lack of spatial context, the classification map obtained by GCN suffers from pepper-noise-like mistakes within certain regions. Comparatively, the result of the proposed MDGCN method yields smoother visual effect and shows fewer misclassifications than other compared methods.

	\subsubsection{Results on the University of Pavia Dataset} Table~\ref{PUSClassificationResults} presents the quantitative results of different methods on the University of Pavia dataset. Note that GCN and S$^{2}$GCN are not used for comparison as they are not scalable to this large dataset. Similar to the results on the Indian Pines dataset, the results in Table~\ref{PUSClassificationResults} indicate that the proposed MDGCN is in the first place and outperforms the compared methods by a substantial margin, which again validates the strength of our proposed multi-scale dynamic graph convolution. Besides, it is also notable that DR-CNN performs better than HiFi and JSDF, which is different from the results on the Indian Pines dataset. This is mainly because that DR-CNN and our MDGCN exploit spectral-spatial information with diverse region-based inputs, which can well adapt to the hyperspectral images containing many boundary regions. Since the objects belonging to the same class in the University of Pavia dataset are often distributed in widely scattered regions, DR-CNN and MDGCN are able to achieve better performance than HiFi, JSDF, and other baseline methods. Furthermore, from the classification results in Fig.~\ref{PUSClassificationMaps}, stronger spatial correlation and fewer misclassifications can be observed in the classification map of the proposed MDGCN when compared with DR-CNN and other competitors.
	
	\begin{table*}[!t]
		\centering
		\caption{Per-Class Accuracy, OA, AA (\%), and Kappa Coefficient Achieved by Different Methods on Kennedy Space Center Dataset}
		\begin{tabular}{ccccccccc}
			\toprule
			ID   & GCN \cite{Kipf2016Semi} & S$^{2}$GCN \cite{8474300} & R-2D-CNN \cite{Yang2018Hyperspectral} & DR-CNN \cite{Zhang2018Diverse} & MDA \cite{Hang2016Matrix} & HiFi \cite{7906599} & JSDF \cite{7360896} & MDGCN \\
			\midrule
			1     & 86.91$\pm$3.46 & 95.12$\pm$0.32 & 94.71$\pm$0.16 & 98.72$\pm$0.21 & 96.88$\pm$2.22 & 97.28$\pm$1.72 & \textbf{100.00$\pm$0.00} & \textbf{100.00$\pm$0.00} \\
			2     & 83.29$\pm$3.08 & 95.15$\pm$5.15 & 79.03$\pm$0.98 & 97.97$\pm$1.36 & 97.84$\pm$2.31 & 99.66$\pm$0.89 & 92.07$\pm$1.59 & \textbf{100.00$\pm$0.00} \\
			3     & 87.57$\pm$4.31 & 96.17$\pm$0.51 & 80.24$\pm$4.11 & 97.49$\pm$2.00 & 88.45$\pm$6.24 & \textbf{100.00$\pm$0.00} & 95.13$\pm$4.01 & \textbf{100.00$\pm$0.00} \\
			4     & 24.86$\pm$12.31 & 71.17$\pm$8.58 & 42.19$\pm$5.88 & 62.46$\pm$3.94 & 78.29$\pm$6.98 & 99.03$\pm$1.12 & 59.01$\pm$8.13 & \textbf{100.00$\pm$0.00} \\
			5     & 63.36$\pm$5.47 & 97.71$\pm$2.64 & 79.39$\pm$3.33 & 94.66$\pm$2.75 & 86.76$\pm$5.06 & \textbf{100.00$\pm$0.00} & 85.34$\pm$7.82 & \textbf{100.00$\pm$0.00} \\
			6     & 61.01$\pm$4.43 & 89.95$\pm$3.48 & 77.05$\pm$7.85 & 97.65$\pm$1.05 & 94.27$\pm$3.95 & \textbf{99.21$\pm$1.00} & 86.48$\pm$3.63 & 94.91$\pm$0.25 \\
			7     & 91.20$\pm$5.63 & 98.22$\pm$3.08 & \textbf{100.00$\pm$0.00} & \textbf{100.00$\pm$0.00} & \textbf{100.00$\pm$0.00} & \textbf{100.00$\pm$0.00} & 98.93$\pm$1.51 & \textbf{100.00$\pm$0.00} \\
			8     & 78.20$\pm$6.45 & 89.11$\pm$0.58 & 98.17$\pm$0.76 & 97.42$\pm$1.77 & 94.08$\pm$3.60 & \textbf{100.00$\pm$0.00} & 94.76$\pm$3.56 & \textbf{100.00$\pm$0.00} \\
			9     & 85.39$\pm$3.96 & 99.59$\pm$0.35 & 96.67$\pm$0.62 & 99.93$\pm$0.12 & 98.79$\pm$2.94 & \textbf{100.00$\pm$0.00} & \textbf{100.00$\pm$0.00} & \textbf{100.00$\pm$0.00} \\
			10    & 84.28$\pm$4.93 & 98.04$\pm$1.08 & 98.30$\pm$1.30 & 98.84$\pm$0.56 & 98.02$\pm$3.41 & 97.78$\pm$2.60 & \textbf{100.00$\pm$0.00} & \textbf{100.00$\pm$0.00} \\
			11    & 94.68$\pm$1.95 & 99.23$\pm$0.45 & 89.03$\pm$1.16 & \textbf{100.00$\pm$0.00} & 98.62$\pm$1.53 & 99.78$\pm$0.23 & \textbf{100.00$\pm$0.00} & \textbf{100.00$\pm$0.00} \\
			12    & 82.14$\pm$2.42 & 95.63$\pm$0.24 & 94.64$\pm$0.80 & 98.94$\pm$0.73 & 94.98$\pm$1.96 & 99.97$\pm$0.08 & 95.52$\pm$2.02 & \textbf{100.00$\pm$0.00} \\
			13    & 98.99$\pm$0.67 & \textbf{100.00$\pm$0.00} & \textbf{100.00$\pm$0.00} & \textbf{100.00$\pm$0.00} & \textbf{100.00$\pm$0.00} & \textbf{100.00$\pm$0.00} & \textbf{100.00$\pm$0.00} & \textbf{100.00$\pm$0.00} \\
			\midrule
			OA    & 83.60$\pm$0.81 & 95.44$\pm$0.92 & 91.11$\pm$0.60 & 97.21$\pm$0.27 & 95.92$\pm$0.81 & 99.30$\pm$0.39 & 95.69$\pm$0.34 & \textbf{99.79$\pm$0.01} \\
			AA    & 78.60$\pm$1.01 & 94.24$\pm$1.84 & 86.88$\pm$1.03 & 95.70$\pm$0.33 & 94.38$\pm$0.96 & 99.44$\pm$0.31 & 92.87$\pm$0.63 & \textbf{99.61$\pm$0.02} \\
			Kappa & 81.70$\pm$0.90 & 94.91$\pm$1.03 & 90.06$\pm$0.66 & 0.97$\pm$0.00 & 95.44$\pm$0.91 & 99.22$\pm$0.44 & 95.17$\pm$0.37 & \textbf{1.00$\pm$0.00} \\
			\bottomrule
		\end{tabular}%
		\label{KSCClassificationResults}%
	\end{table*}%
	
	\begin{figure*}[!t]
	\centering
	\subfigure[]{%
		\resizebox*{2.5cm}{!}{\includegraphics{KSC_fc.pdf}}}\hspace{15pt}
	\subfigure[]{%
		\label{KSC_classified_gt}
		\resizebox*{2.5cm}{!}{\includegraphics{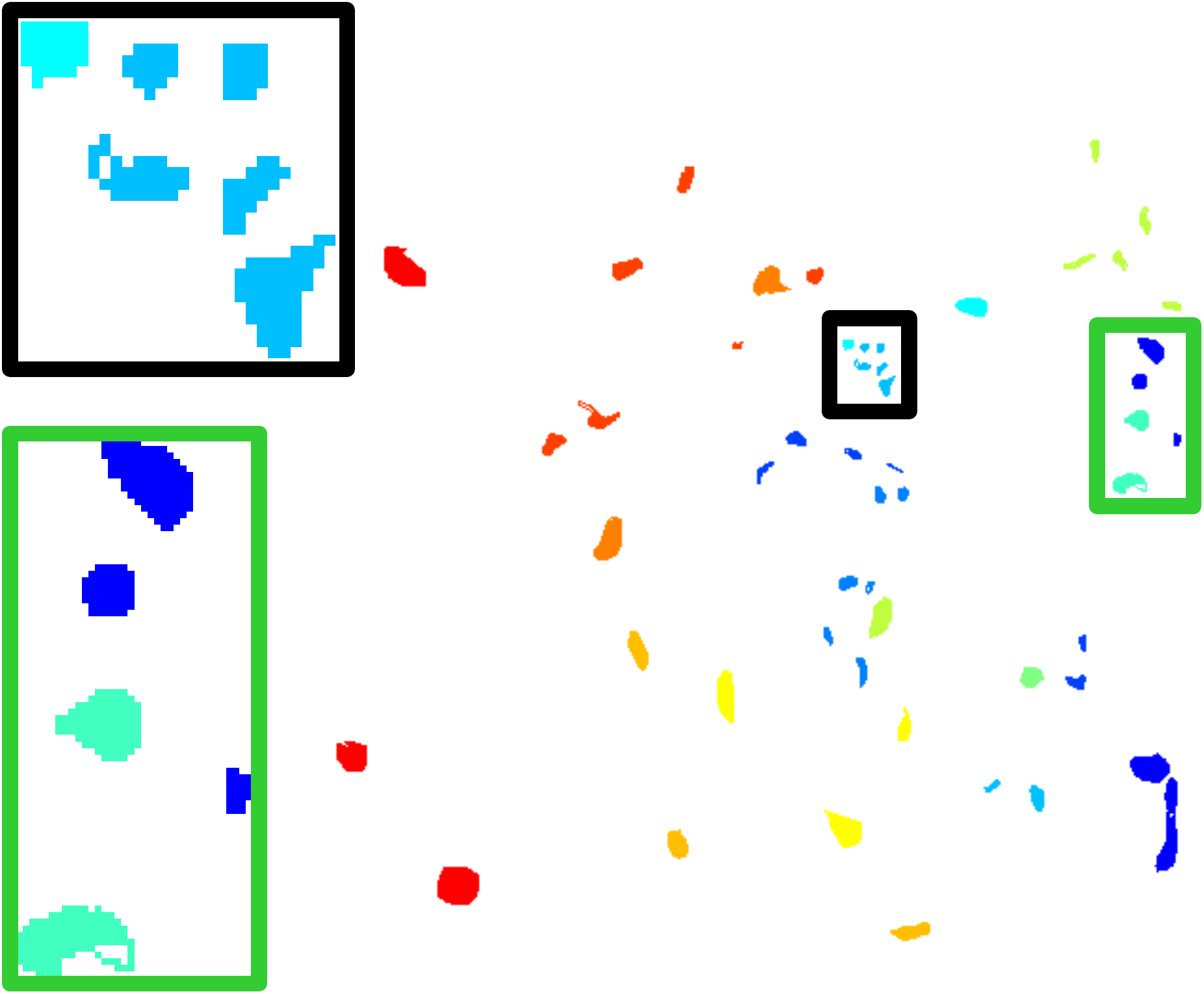}}}\hspace{15pt}
	\subfigure[]{%
		\resizebox*{2.5cm}{!}{\includegraphics{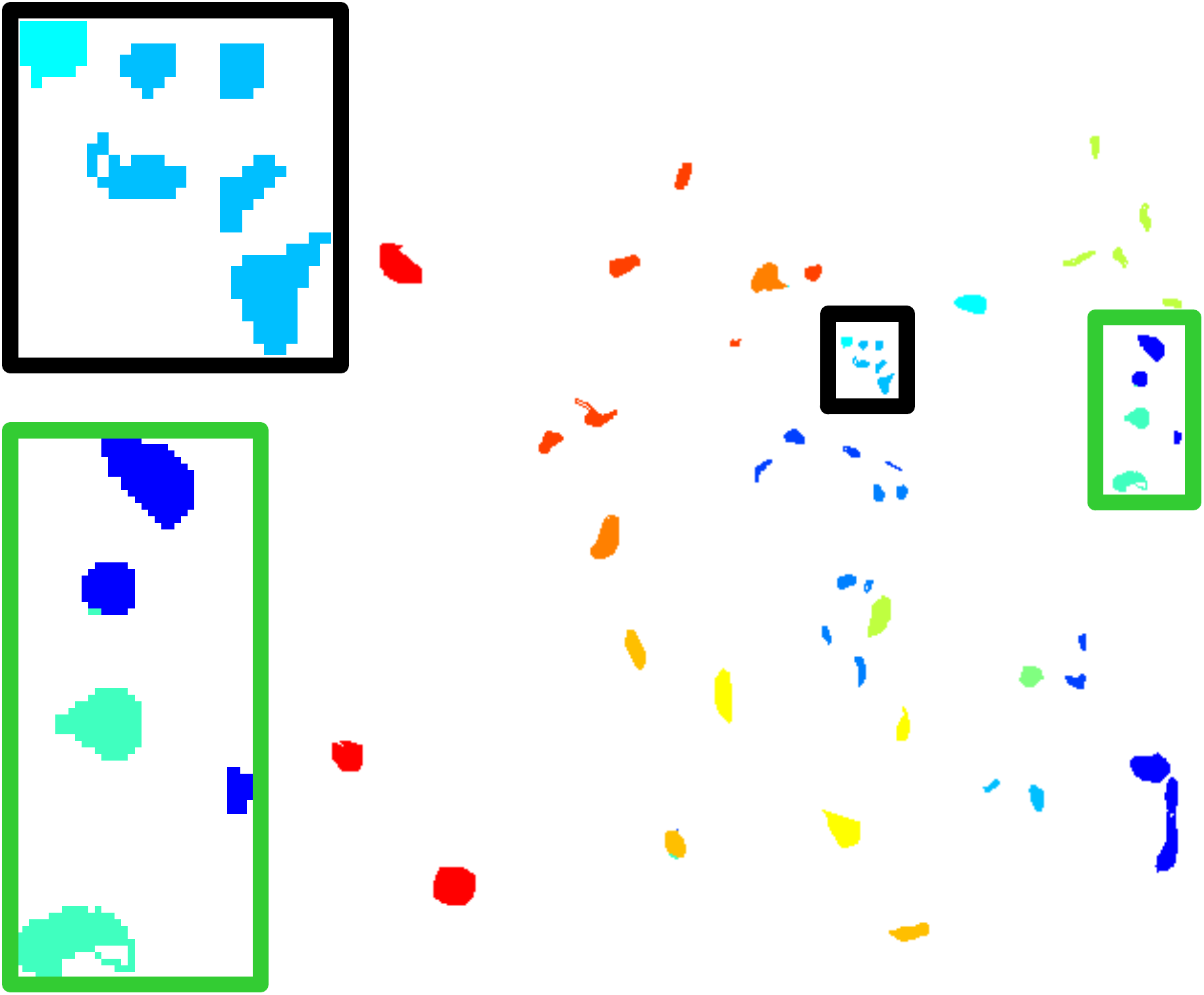}}}\hspace{15pt}	
	\subfigure[]{%
		\resizebox*{2.5cm}{!}{\includegraphics{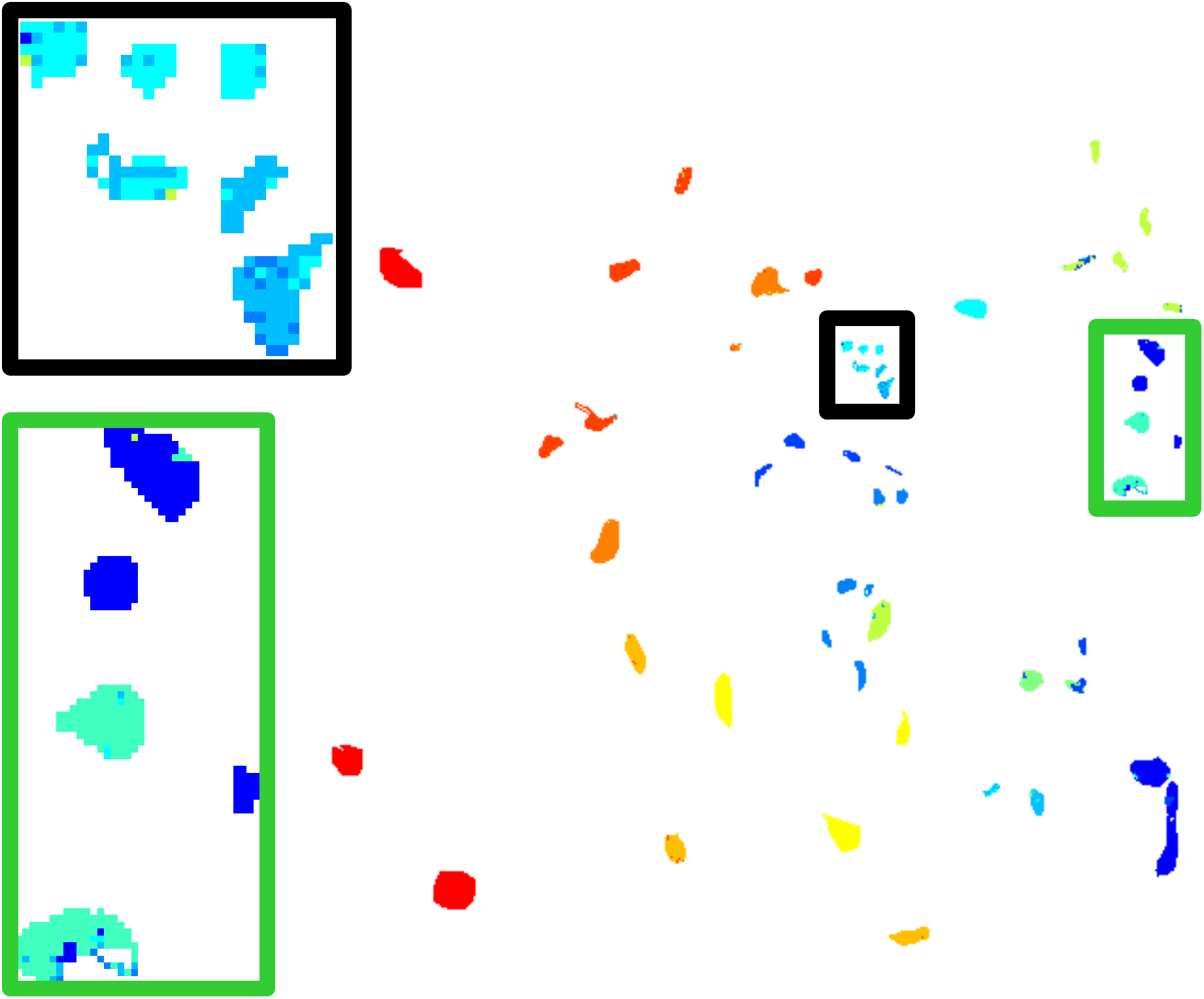}}}\hspace{15pt}	
	\subfigure[]{%
		\resizebox*{2.5cm}{!}{\includegraphics{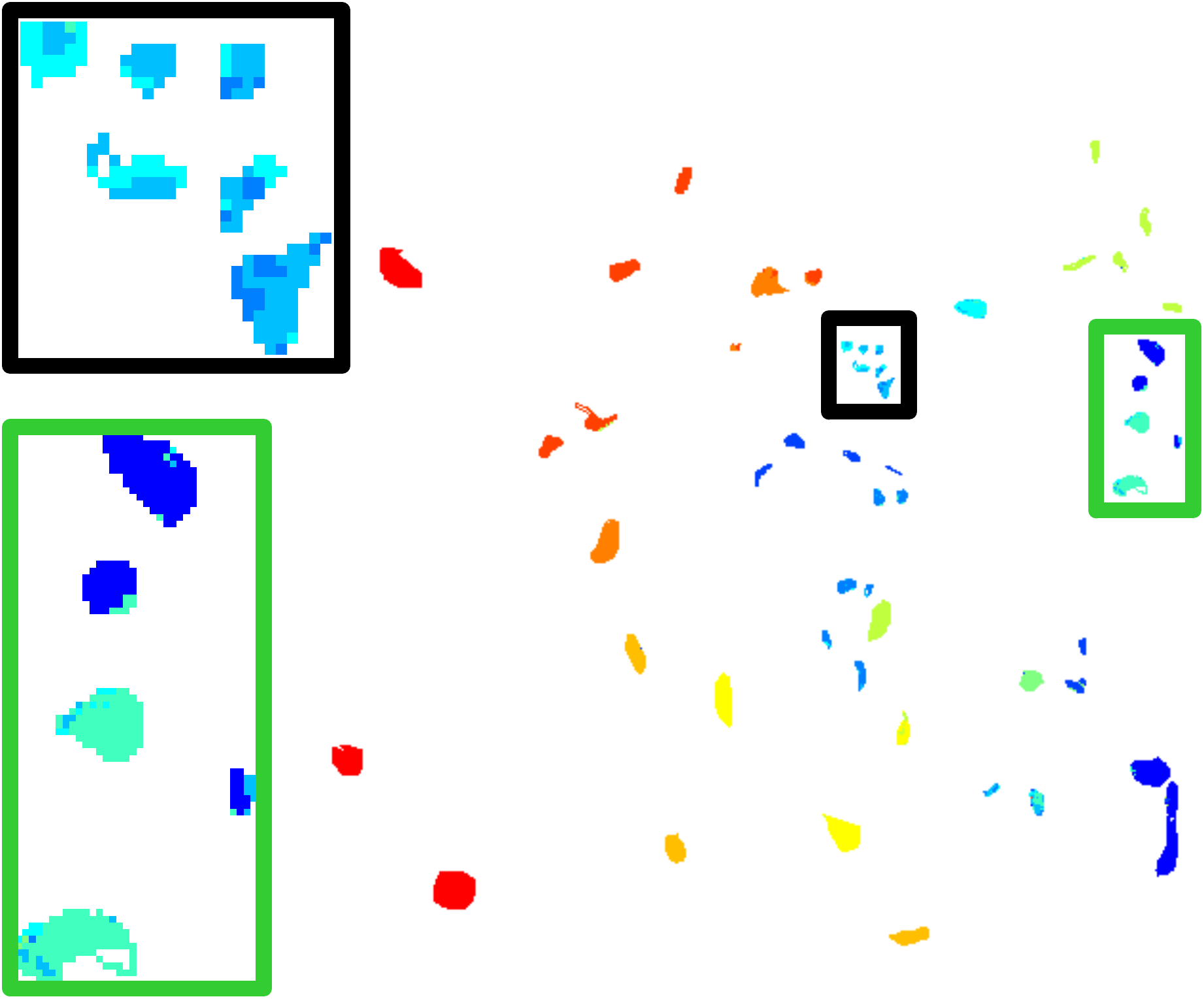}}}\hspace{16pt}	
	\subfigure[]{%
		\resizebox*{2.5cm}{!}{\includegraphics{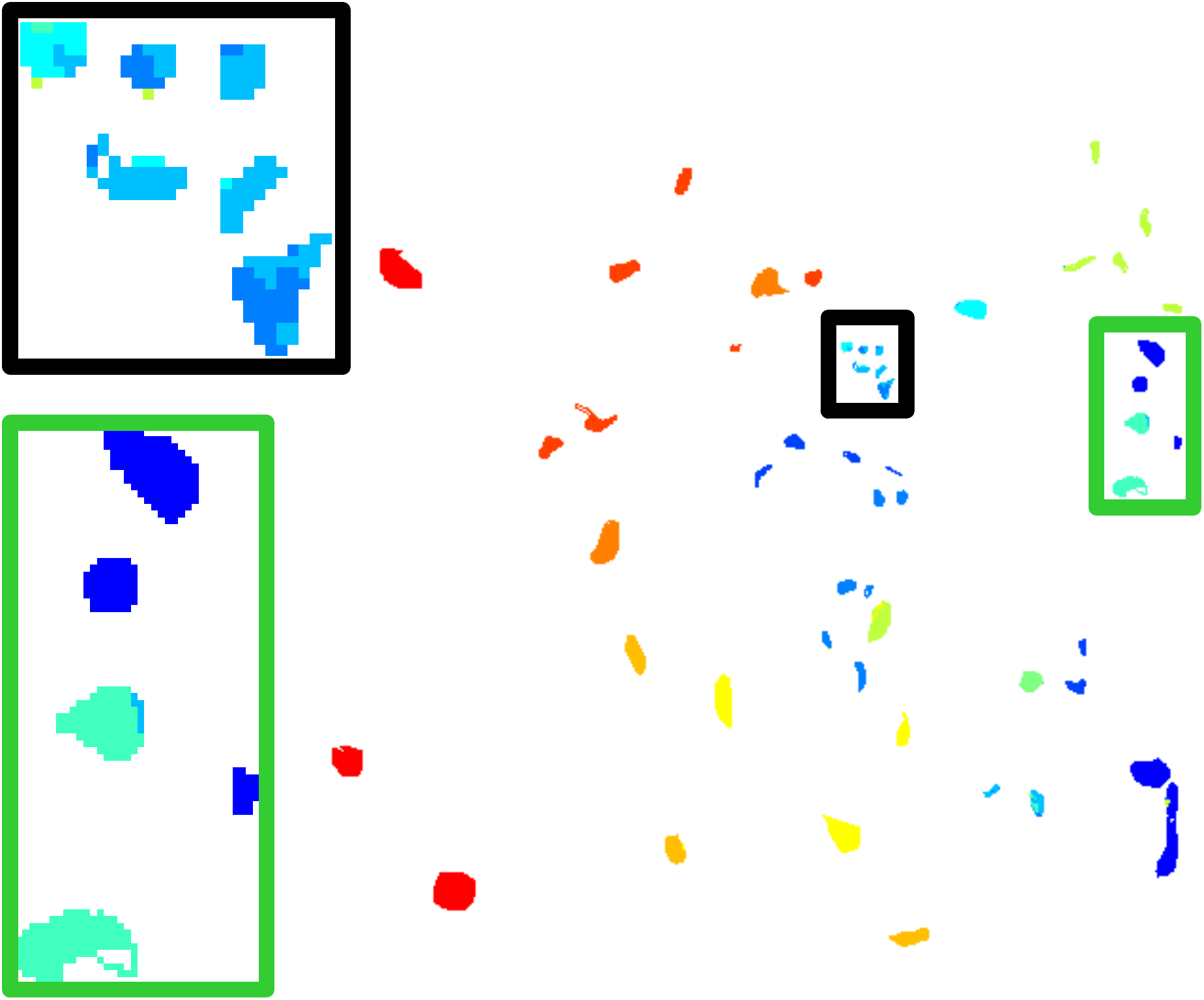}}}\hspace{15pt}	
	\subfigure[]{%
		\resizebox*{2.5cm}{!}{\includegraphics{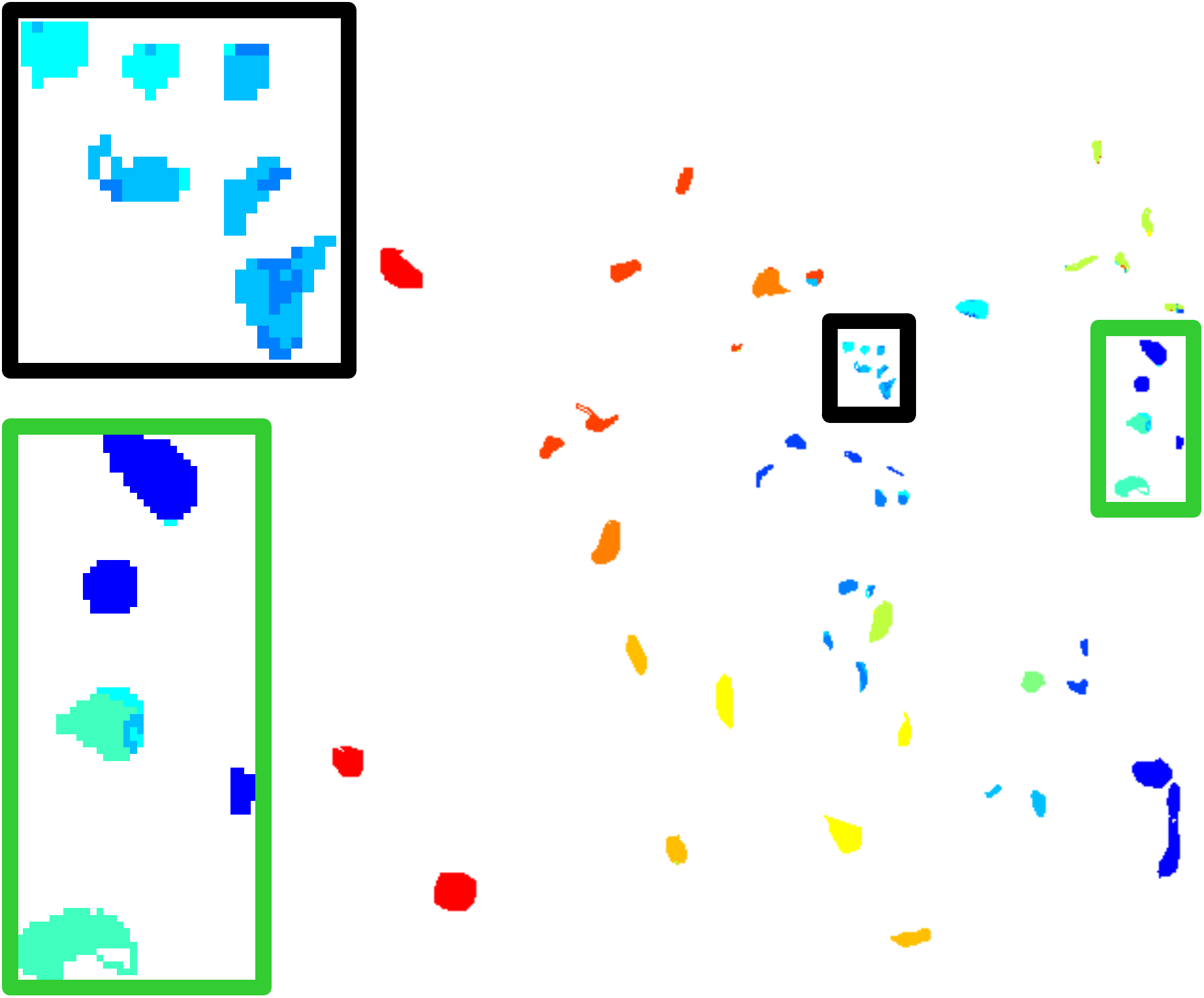}}}\hspace{15pt}
	\subfigure[]{%
		\resizebox*{2.5cm}{!}{\includegraphics{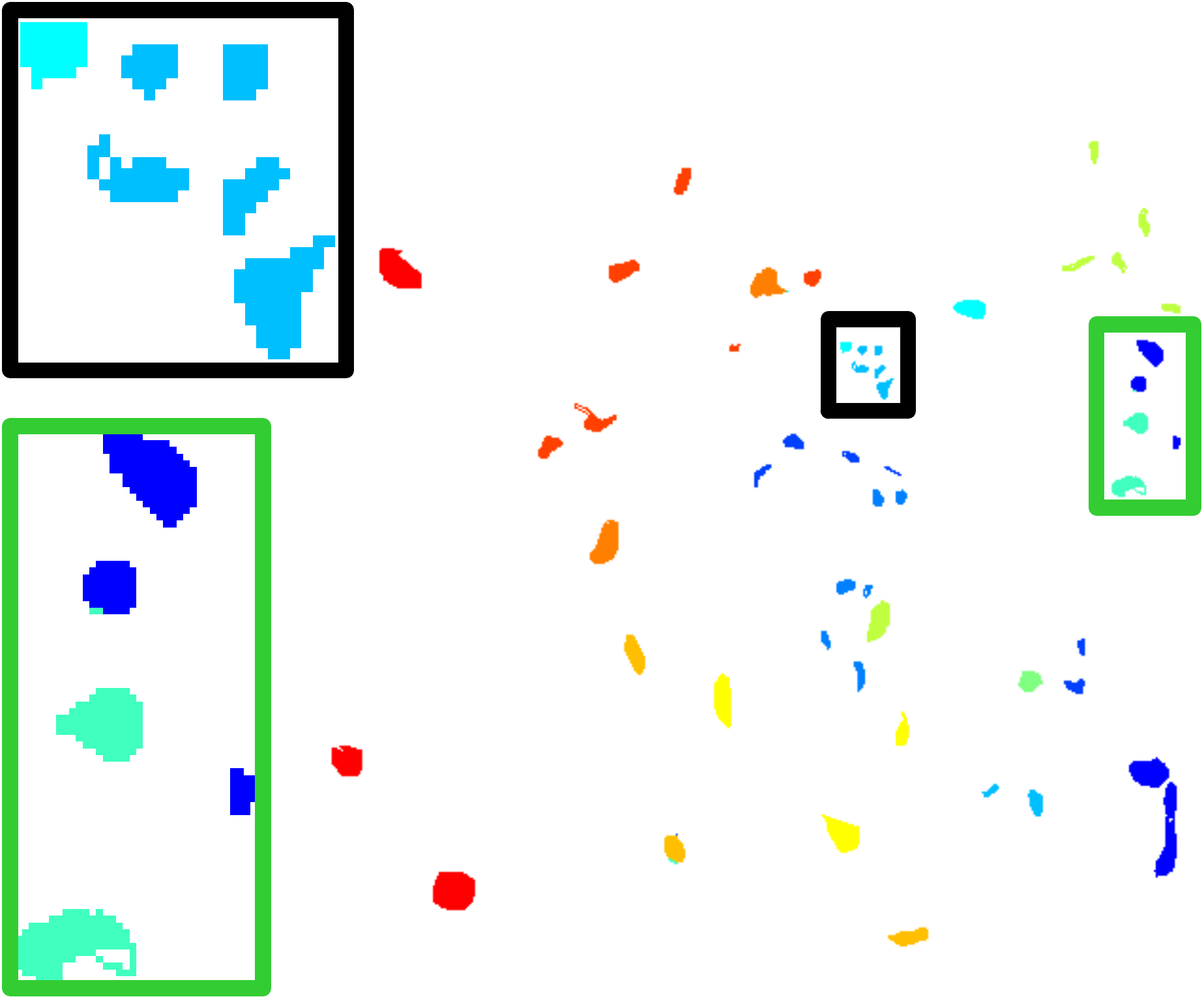}}}\hspace{15pt}			
	\subfigure[]{%
		\resizebox*{2.5cm}{!}{\includegraphics{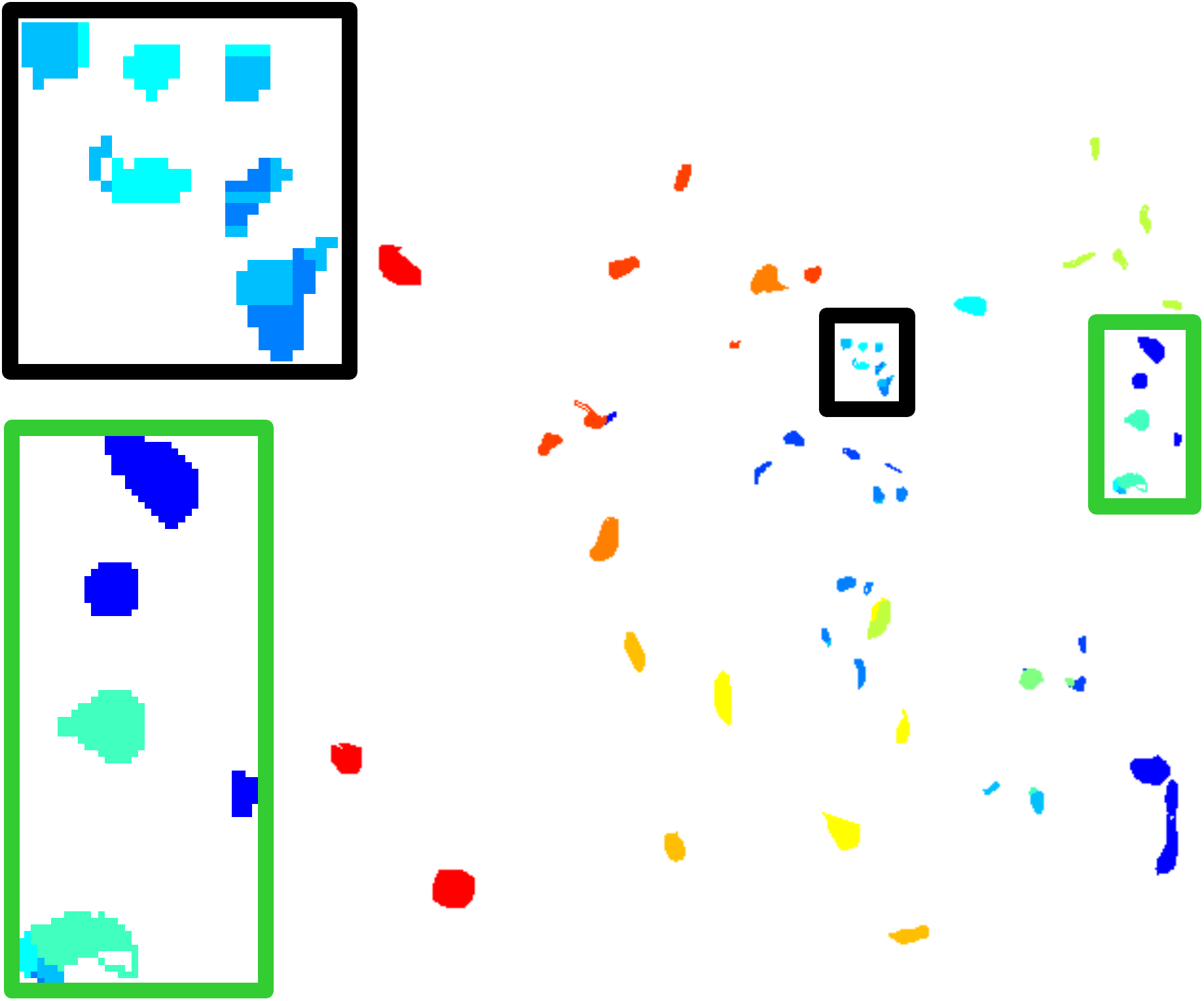}}}\hspace{15pt}	
	\subfigure[]{%
		\label{KSC_classified_MDGCN}
		\resizebox*{2.5cm}{!}{\includegraphics{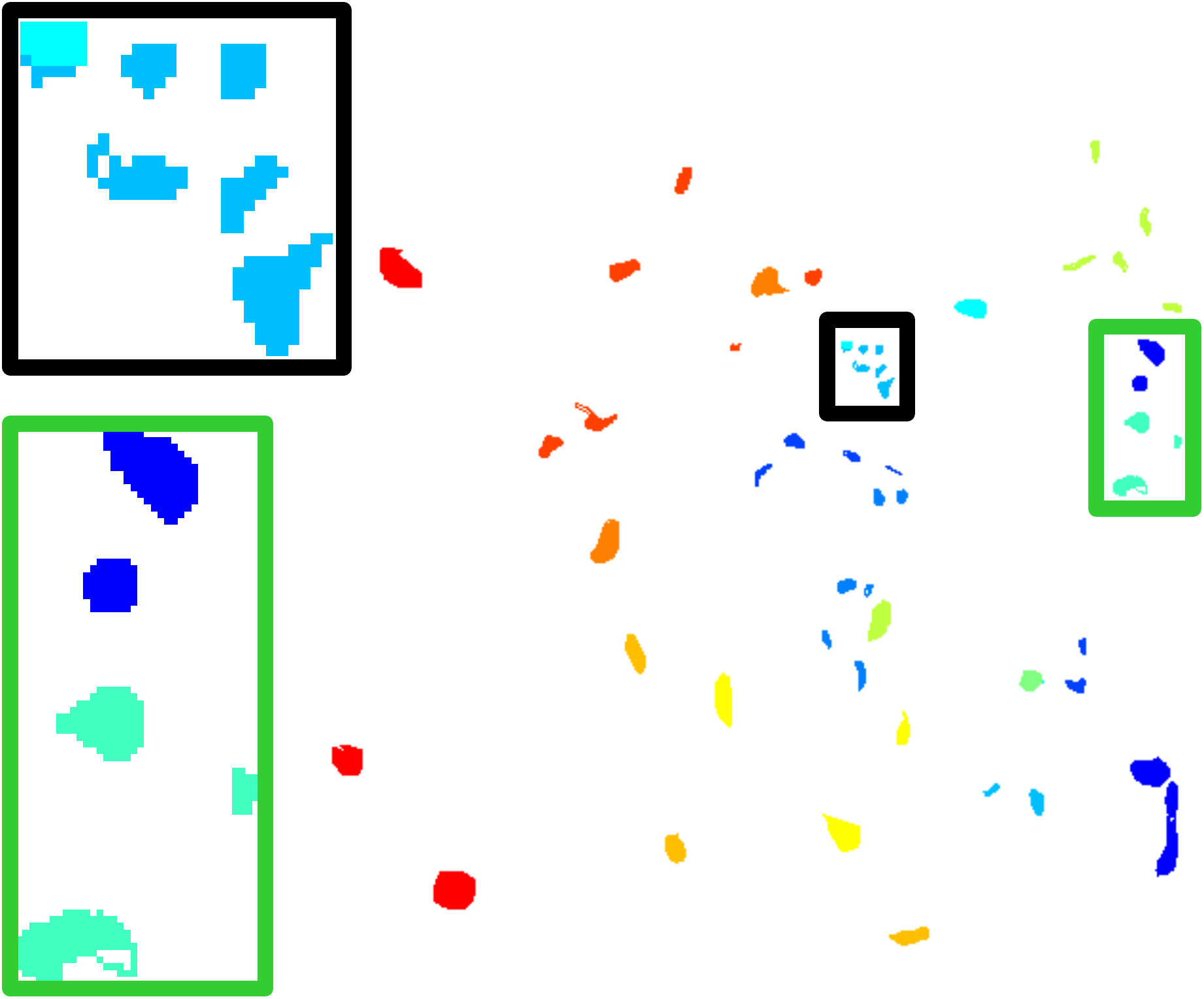}}}\hspace{0pt}	
	
	\caption{Classification maps obtained by different methods on Kennedy Space Center dataset. (a) False color image; (b) Ground-truth map; (c) GCN; (d) S$^{2}$GCN; (e) R-2D-CNN; (f) DR-CNN; (g) MDA; (h) HiFi; (i) JSDF; (j) MDGCN. In (b)-(j), zoomed-in views of the regions enclosed in black and green boxes are shown at the left side of each map.} 
	\label{KSCClassificationMaps}
\end{figure*}
	
\subsubsection{Results on the Kennedy Space Center Dataset} Table~\ref{KSCClassificationResults} presents the experimental results of different methods on the Kennedy Space Center dataset. It is apparent that the performance of all methods is better than that on the Indian Pines and the University of Pavia dataset. This could be due to that the Kennedy Space Center dataset has higher spatial resolution and contains less noise than the Indian Pines and the University of Pavia dataset, and thus is more suitable for classification. As can be noticed, HiFi algorithm achieves the highest OA among all the baseline methods. However, slight gaps can still be observed between HiFi and our MDGCN in terms of OA. For the proposed MDGCN, it is also worth noting that misclassifications only occur in the 6th class (`Hardwood'), which further demonstrates the advantage of our proposed MDGCN. Fig.~\ref{KSCClassificationMaps} visualizes the classification results of the eight different methods, where some critical regions of each classification map are enlarged for better performance comparison. We can see that our MDGCN is able to produce precise classification results on these small and difficult regions.
	
	\subsection{Impact of the Number of Labeled Examples}
	
	\begin{figure}[!t]
		\centering
		\resizebox*{8cm}{!}{\includegraphics{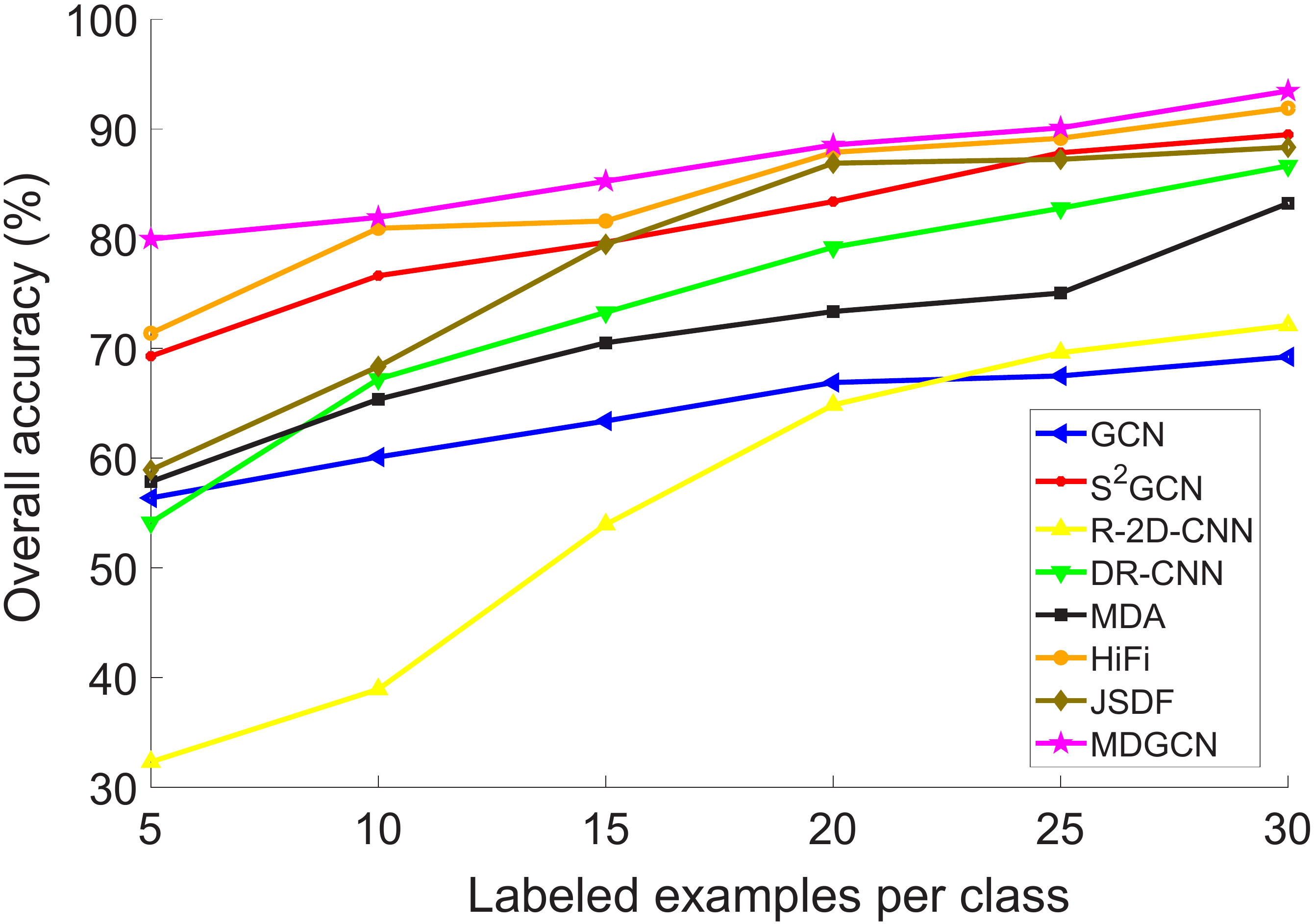}}
		\caption{Overall accuracies of various methods on Indian Pines dataset under different numbers of labeled examples per class.} 
		\label{IPclassnum}
	\end{figure}
	
	In this experiment, we investigate the classification accuracies of the proposed MDGCN and the other competitors under different numbers of labeled examples. To this end, we vary the number of labeled examples per class from 5 to 30, and report the OA gained by all the methods on the Indian Pines dataset. The results are shown in Fig.~\ref{IPclassnum}. We see that the performance of all methods can be improved by increasing the number of labeled examples. It is noteworthy that the CNN-based methods, i.e., R-2D-CNN, DR-CNN, have poor performance when labeled examples are very limited, since these methods require a large number of labeled examples for training. Besides, we can observe that the proposed MDGCN consistently yields higher OA than the other methods with the increase of the number of labeled examples. Moreover, the performance of MDGCN is more stable than the compared methods with the changed number of labeled examples. All these observations indicate the effectiveness and stability of our MDGCN method.
	
	\subsection{Ablation Study}
	
	\begin{table}[!t]
		\scriptsize
		\centering
		
		\caption{Per-Class Accuracy, OA, AA (\%), and Kappa Coefficient Achieved by Different Graph Convolution Approaches on Indian Pines Dataset}
		
		\begin{tabular}{p{10 pt}<{\centering}p{35 pt}<{\centering}p{35 pt}<{\centering}p{35 pt}<{\centering}p{35 pt}<{\centering}p{35 pt}<{\centering}}
			\toprule
			ID   & $s=1$ & $s=2$ & $s=3$ & MGCN & MDGCN \\
			\midrule
			1     & \textbf{100.00$\pm$0.00} & \textbf{100.00$\pm$0.00} & \textbf{100.00$\pm$0.00} & \textbf{100.00$\pm$0.00} & \textbf{100.00$\pm$0.00} \\
			2     & \textbf{81.15$\pm$1.06} & 76.89$\pm$4.17 & 71.22$\pm$1.35 & 78.74$\pm$2.59 & 80.18$\pm$0.84 \\
			3     & 92.81$\pm$2.21 & 96.70$\pm$1.62 & 96.02$\pm$2.81 & 94.12$\pm$1.00 & \textbf{98.26$\pm$0.00} \\
			4     & \textbf{100.00$\pm$0.00} & 98.76$\pm$0.55 & 98.47$\pm$2.37 & 98.55$\pm$0.00 & 98.57$\pm$0.00 \\
			5     & 89.62$\pm$0.31 & 93.76$\pm$2.22 & 89.51$\pm$2.54 & 90.18$\pm$0.22 & \textbf{95.14$\pm$0.33} \\
			6     & 93.36$\pm$0.91 & 96.78$\pm$0.56 & 95.21$\pm$2.27 & 94.57$\pm$1.49 & \textbf{97.16$\pm$0.57} \\
			7     & \textbf{100.00$\pm$0.00} & \textbf{100.00$\pm$0.00} & \textbf{100.00$\pm$0.00} & \textbf{100.00$\pm$0.00} & \textbf{100.00$\pm$0.00} \\
			8     & 98.66$\pm$0.32 & 98.57$\pm$1.40 & 99.81$\pm$0.46 & \textbf{100.00$\pm$0.00} & 98.89$\pm$0.00 \\
			9     & \textbf{100.00$\pm$0.00} & \textbf{100.00$\pm$0.00} & \textbf{100.00$\pm$0.00} & \textbf{100.00$\pm$0.00} & \textbf{100.00$\pm$0.00} \\
			10    & 84.50$\pm$0.30 & 88.03$\pm$2.01 & 76.26$\pm$4.30 & 85.67$\pm$2.01 & \textbf{90.02$\pm$1.02} \\
			11    & 79.59$\pm$0.93 & 91.51$\pm$1.27 & 91.70$\pm$1.46 & 90.37$\pm$0.64 & \textbf{93.35$\pm$1.47} \\
			12    & 91.21$\pm$0.38 & 91.73$\pm$3.41 & 86.35$\pm$4.17 & 90.90$\pm$3.15 & \textbf{93.05$\pm$2.30} \\
			13    & \textbf{100.00$\pm$0.00} & \textbf{100.00$\pm$0.00} & \textbf{100.00$\pm$0.00} & \textbf{100.00$\pm$0.00} & \textbf{100.00$\pm$0.00} \\
			14    & 99.55$\pm$0.06 & 99.78$\pm$0.08 & \textbf{99.87$\pm$0.07} & 99.76$\pm$0.07 & 99.72$\pm$0.05 \\
			15    & 98.31$\pm$1.99 & 99.64$\pm$0.21 & 99.63$\pm$0.15 & 98.67$\pm$1.38 & \textbf{99.72$\pm$0.00} \\
			16    & \textbf{98.41$\pm$0.00} & 96.15$\pm$1.55 & 96.83$\pm$1.74 & \textbf{98.41$\pm$0.00} & 95.71$\pm$0.00 \\
			\midrule
			OA    & 88.53$\pm$0.54 & 92.03$\pm$0.39 & 89.53$\pm$0.55 & 91.24$\pm$0.70 & \textbf{93.47$\pm$0.38} \\
			AA    & 94.20$\pm$0.28 & 95.52$\pm$0.32 & 93.81$\pm$0.30 & 95.00$\pm$0.58 & \textbf{96.24$\pm$0.21} \\
			Kappa & 86.97$\pm$0.60 & 90.90$\pm$0.44 & 88.06$\pm$0.63 & 90.00$\pm$0.80 & \textbf{92.55$\pm$0.43} \\
			\bottomrule
		\end{tabular}%
		\label{TestMulticaleDynamic}%
	\end{table}%

	As is mentioned in the introduction, our proposed MDGCN contains two critical parts for boosting the classification performance, i.e., multi-scale operation and dynamic graph convolution. Here we use the Indian Pines dataset to demonstrate the usefulness of these two operations, where the number of labeled pixels per class is kept identical to the above experiments in Section \ref{CLassificationResult}. To show the importance of multi-scale technique, we exhibit the classification results in Table~\ref{TestMulticaleDynamic} by using the dynamic graphs with three different neighborhood scales, i.e., $s=1$, $s=2$, and $s=3$. It can be observed that higher neighborhood scale does not necessarily result in better performance, since the spectral-spatial information cannot be sufficiently exploited with only a single-scale graph. Comparatively, we find that MDGCN consistently performs better than the settings of $s=1$, $s=2$, and $s=3$ in terms of OA, AA, and Kappa coefficient, which indicates the usefulness of incorporating the multi-scale spectral-spatial information into the graphs. 
	
	To show the effectiveness of dynamic graphs, Table~\ref{TestMulticaleDynamic} also lists the results acquired by only using multi-scale graph convolution network (MGCN), where the graphs are fixed throughout the classification procedure. Compared with the results of MDGCN, there is a noticeable performance drop in the OA, AA, and Kappa coefficient of MGCN, which indicates that utilizing fixed graph convolution is not ideal for accurate classification. Therefore, the dynamically updated graph in our method is useful for rendering good classification results.

	\subsection{Classification Performance in the Boundary Region}
	
	\begin{figure*}[!t]
		\centering
		\subfigure[]{%
			\label{bd_gt}
			\resizebox*{3cm}{!}{\includegraphics{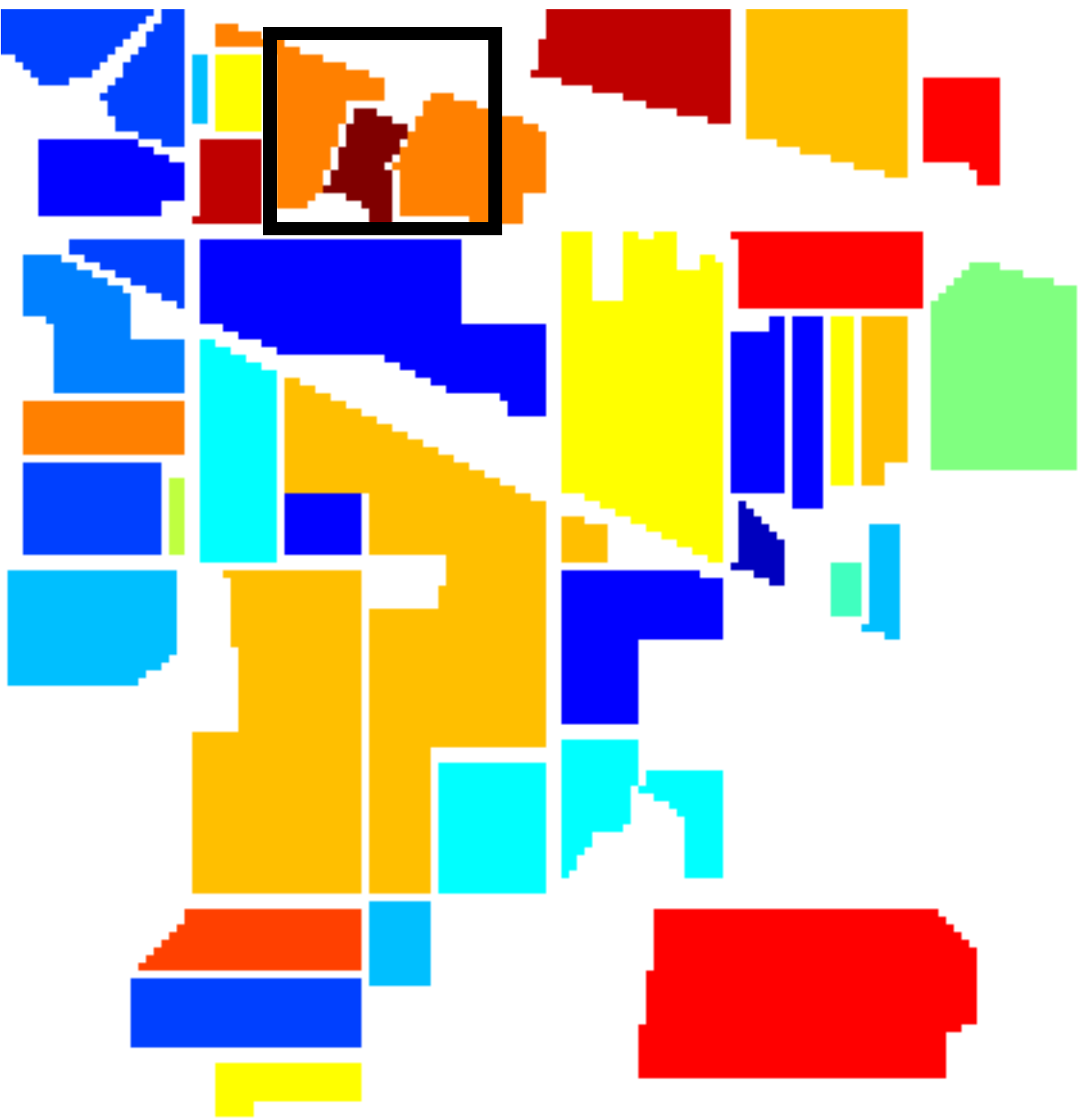}}}\hspace{2pt}
		\subfigure[]{%
			\resizebox*{3cm}{!}{\includegraphics{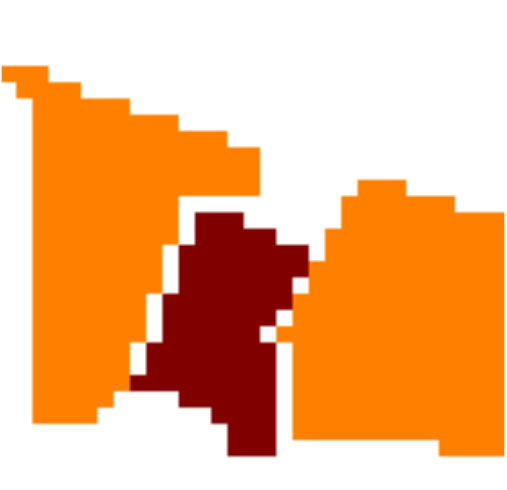}}}\hspace{2pt}
		\subfigure[]{%
			\resizebox*{3cm}{!}{\includegraphics{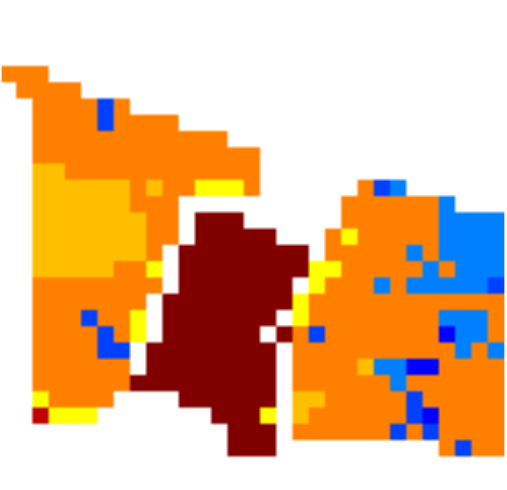}}}\hspace{2pt}	
		\subfigure[]{%
			\resizebox*{3cm}{!}{\includegraphics{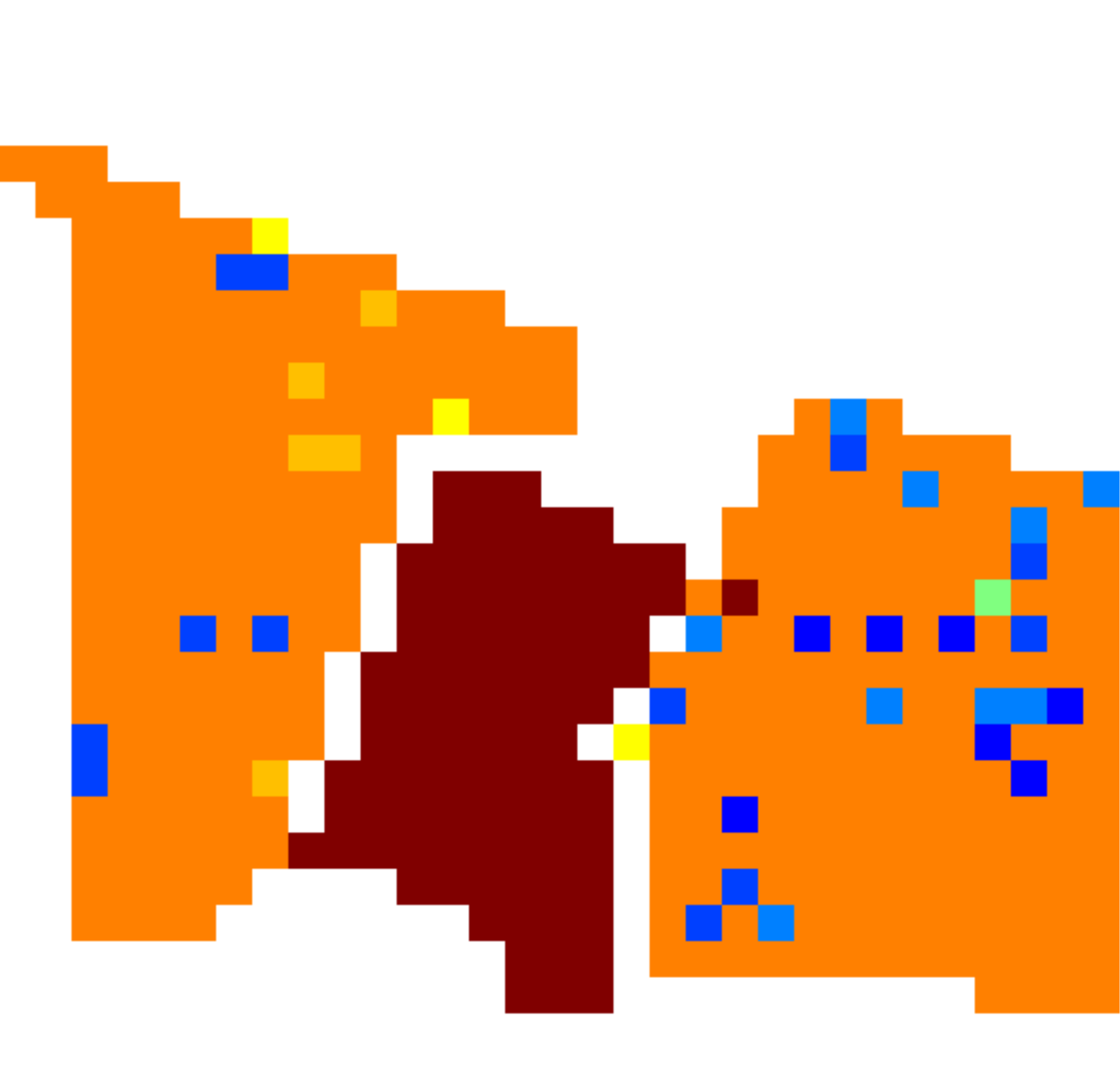}}}\hspace{2pt}	
		\subfigure[]{%
			\resizebox*{3cm}{!}{\includegraphics{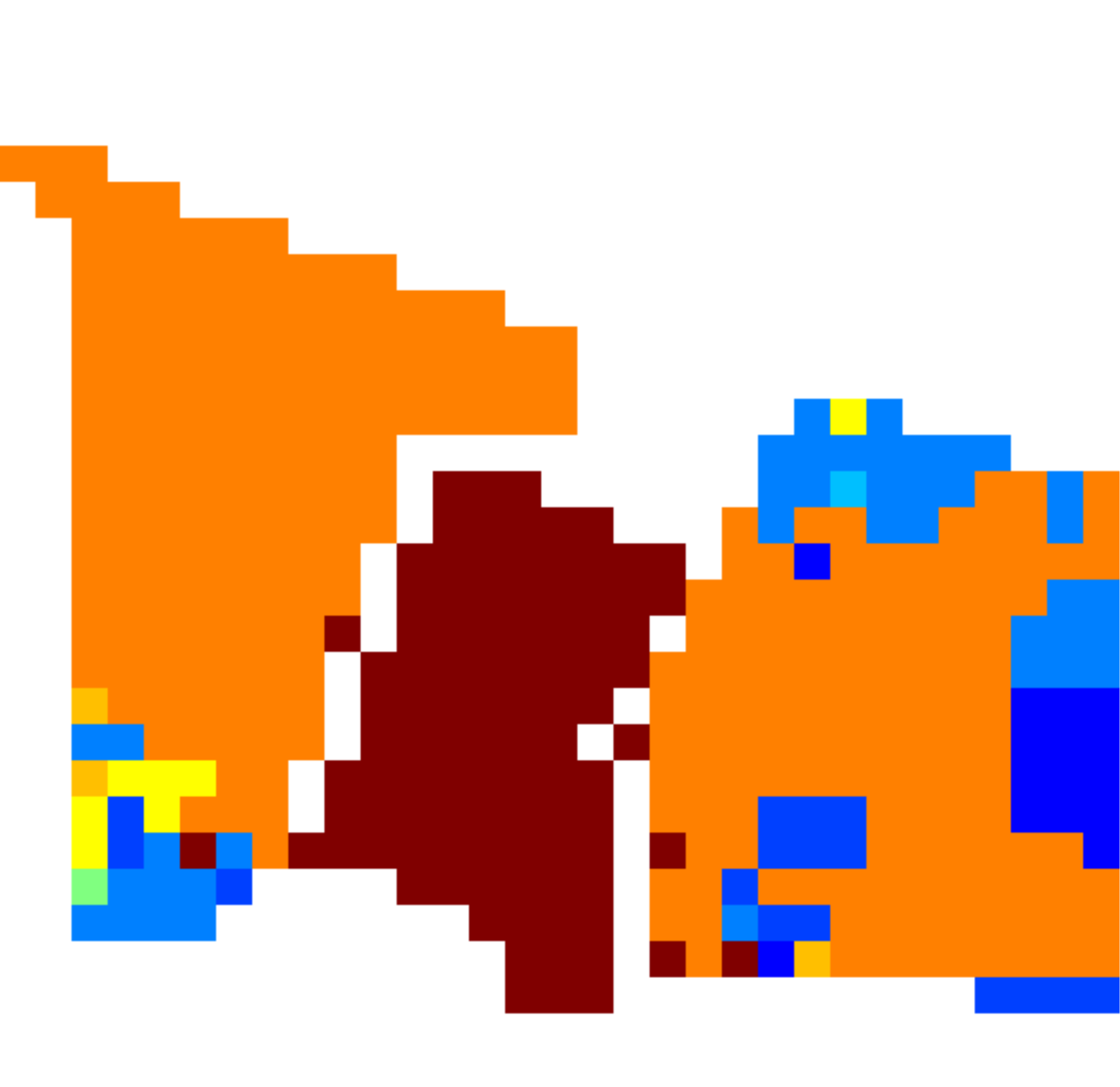}}}\hspace{2pt}	
		\subfigure[]{%
			\resizebox*{3cm}{!}{\includegraphics{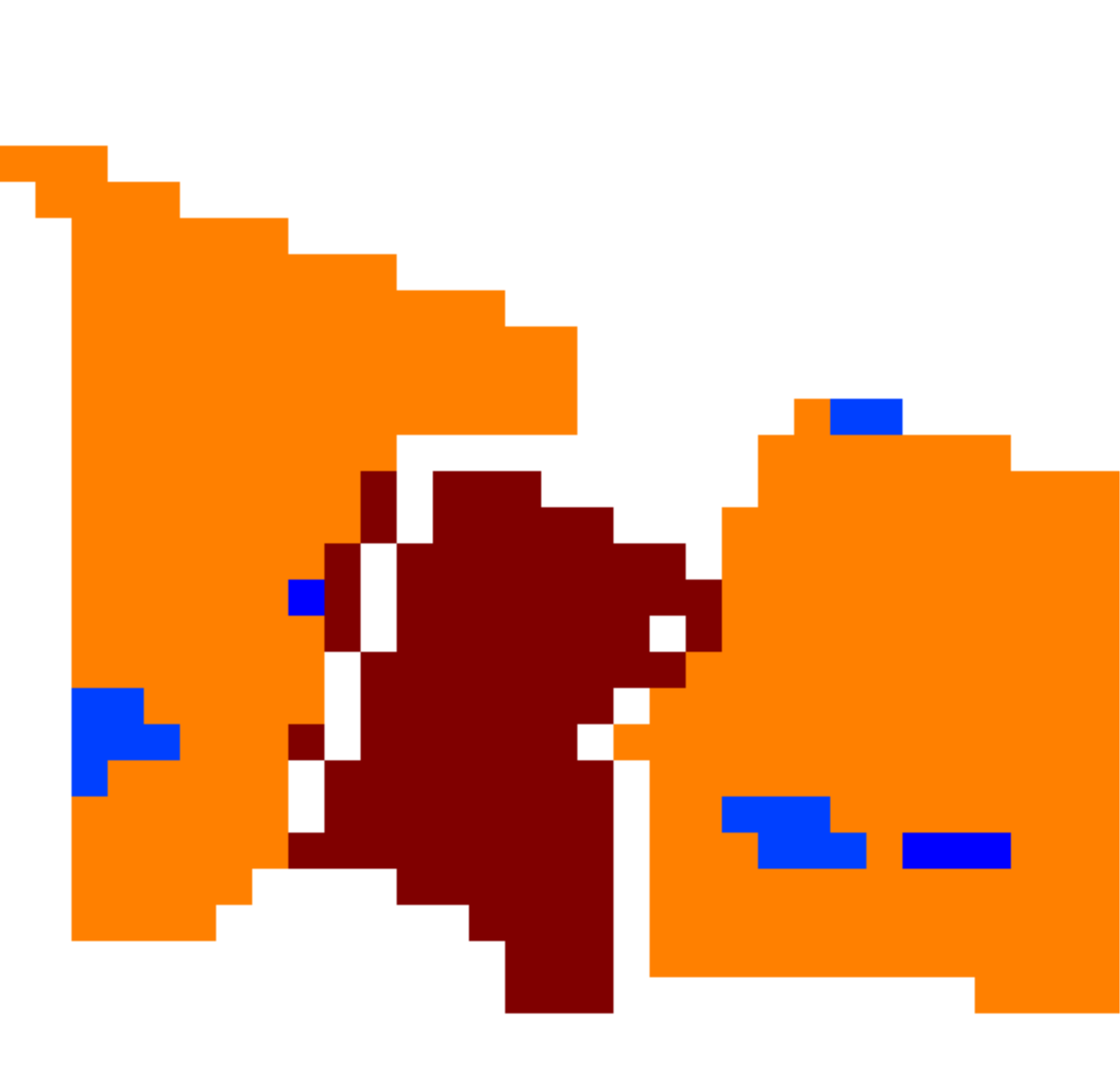}}}\hspace{2pt}	
		\subfigure[]{%
			\resizebox*{3cm}{!}{\includegraphics{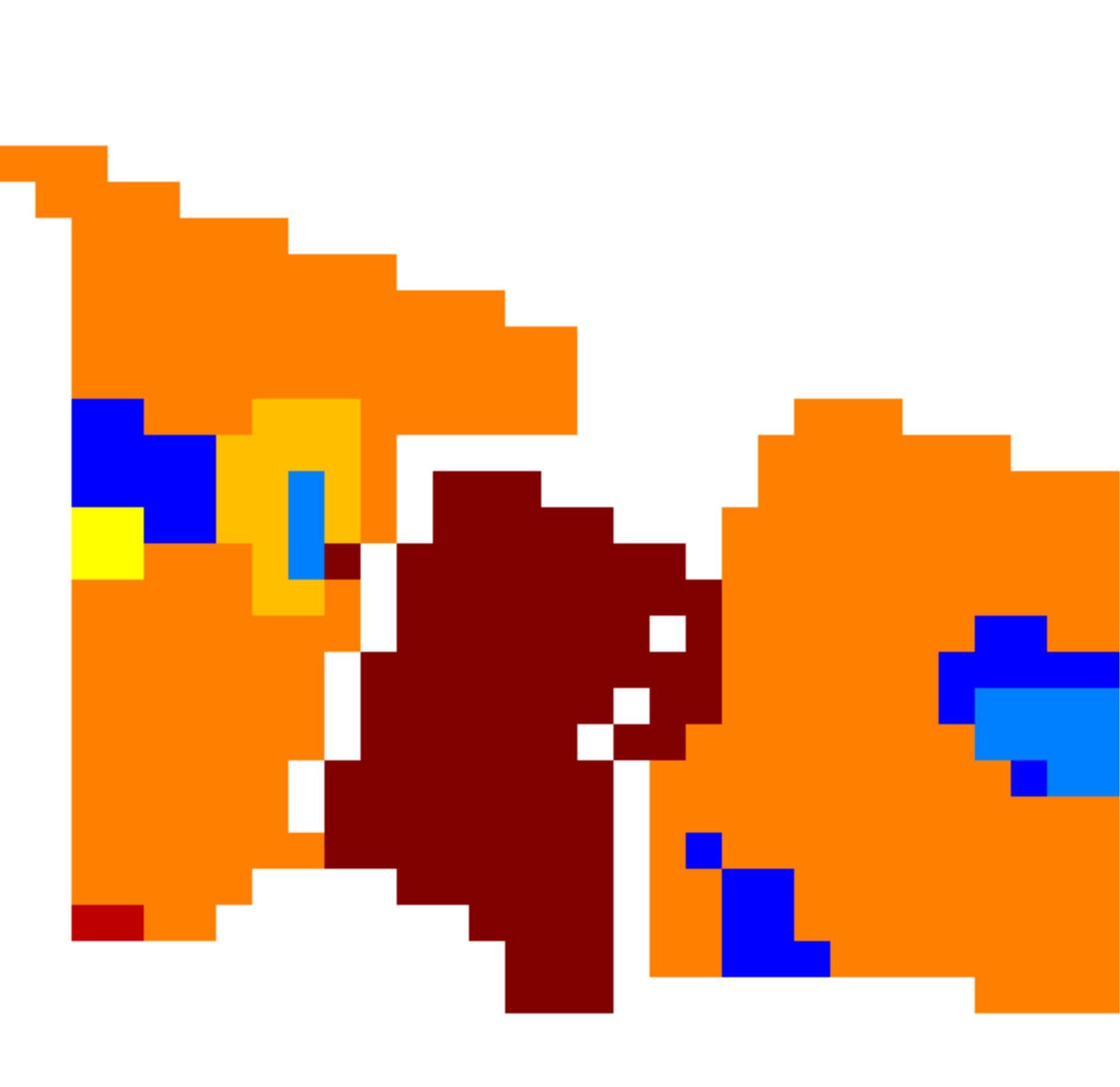}}}\hspace{2pt}
		\subfigure[]{%
			\resizebox*{3cm}{!}{\includegraphics{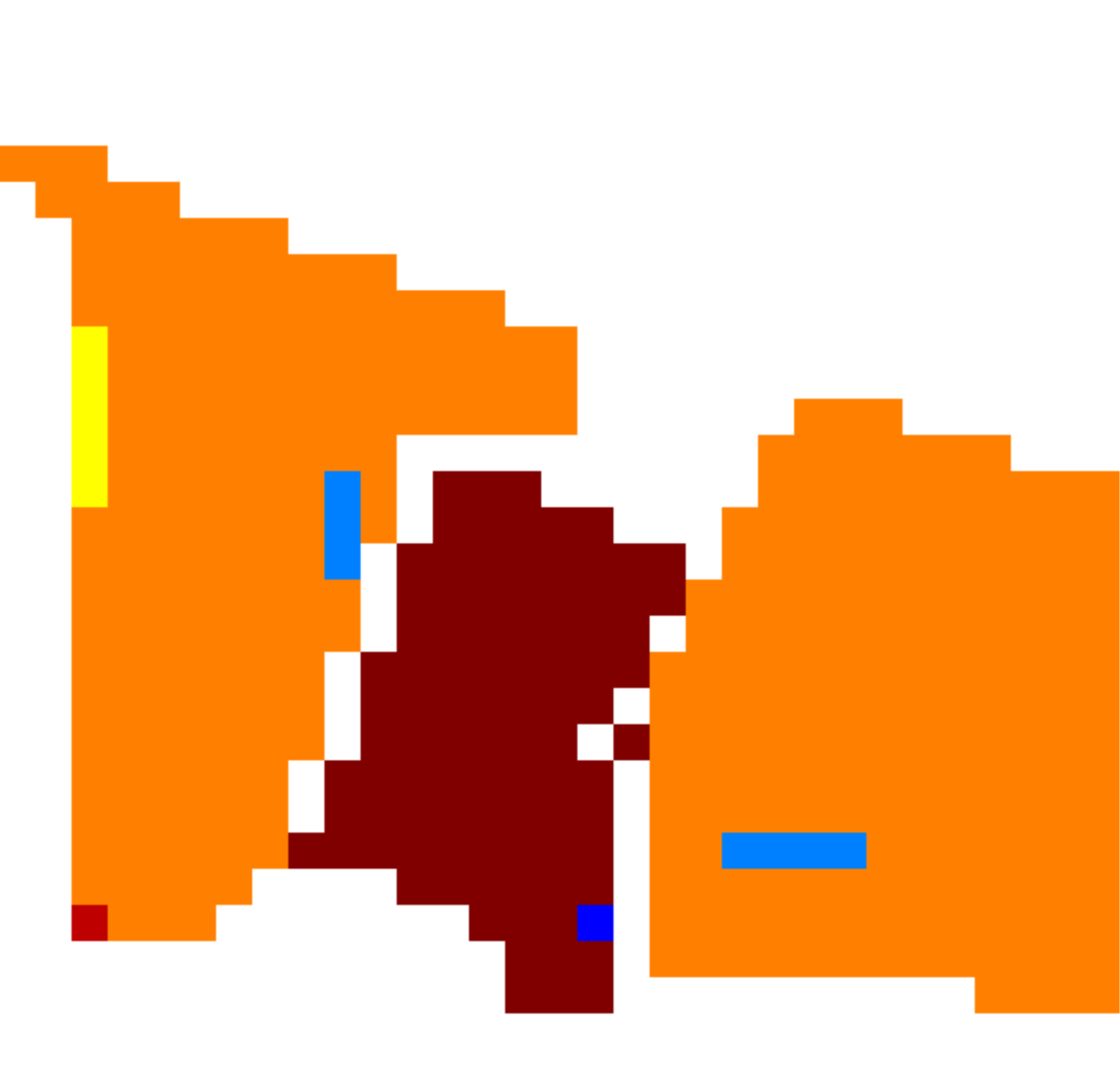}}}\hspace{2pt}			
		\subfigure[]{%
			\resizebox*{3cm}{!}{\includegraphics{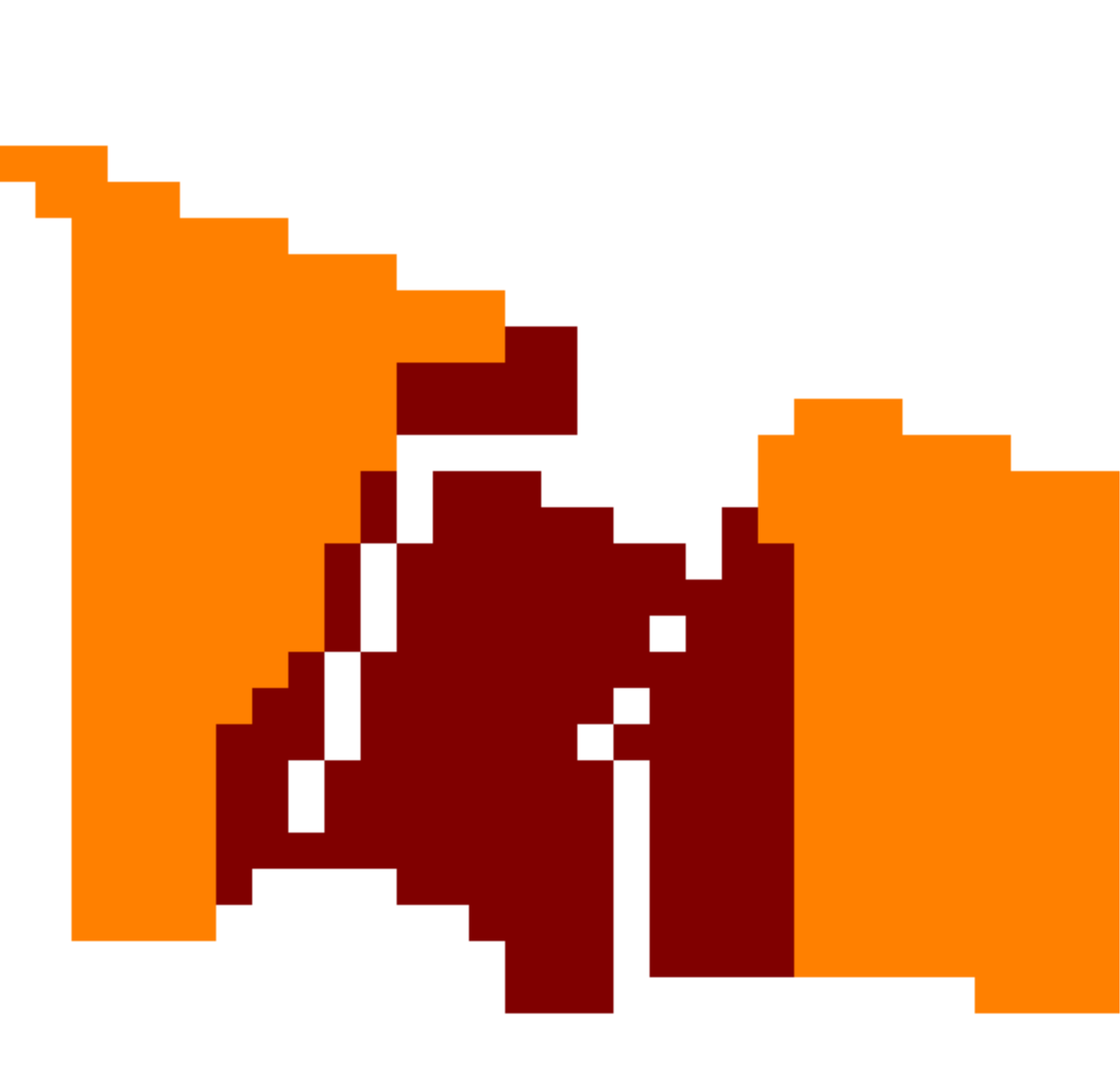}}}\hspace{2pt}	
		\subfigure[]{%
			\label{bd_MDGCN}
			\resizebox*{3cm}{!}{\includegraphics{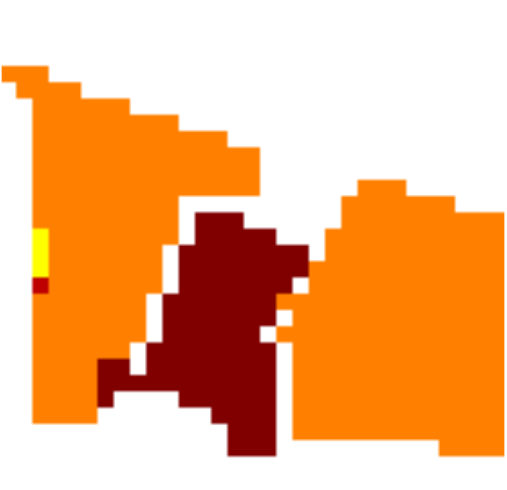}}}\hspace{2pt}	
		
		\caption{Classification maps obtained by different methods regarding a boundary region in Indian Pines dataset. (a) The studied boundary region; (b) Ground-truth map; (c) GCN; (d) S$^{2}$GCN; (e) R-2D-CNN; (f) DR-CNN; (g) MDA; (h) HiFi; (i) JSDF; (j) MDGCN.} 
		\label{IPClassificationMapsOfTheBoundaryRegion}
	\end{figure*}
	
	One of the defects in traditional CNN-based methods is that the weights of each convolution kernel are identical when convolving all image patches, which may produce misclassifications in boundary regions. Different from the coarse convolution of traditional CNN-based methods with fixed size and weights, the graph convolution of the proposed MDGCN can be flexibly applied to irregular image patches and thus will not significantly `erase' the boundaries of objects during the convolution process. Therefore, the boundary information will be preserved and our MDGCN will perform better than the CNN-based methods in boundary regions. To reveal this advantage, in Fig.~\ref{IPClassificationMapsOfTheBoundaryRegion}, we show the classification maps of a boundary region in the Indian Pines dataset obtained by different methods. The investigated boundary region is indicated by a black box in Fig.~\ref{bd_gt}. Note that the results near the class boundaries are quite confusing and inaccurate in the classification maps of GCN, S$^{2}$GCN, R-2D-CNN, DR-CNN, MDA, HiFi, and JSDF, since the spatial information is very limited to distinguish the pixels around class boundaries. In contrast, the classification map of the proposed MDGCN (see Fig.~\ref{bd_MDGCN}) is more compact and accurate than those of other methods.

	\subsection{Running Time}

	Table~\ref{RunningTime} reports the running time of deep models including GCN, S$^{2}$GCN, R-2D-CNN, DR-CNN, and the proposed MDGCN on the three datasets adopted above. The codes for all methods are written in Python, and the running time is reported on a desktop computer with a 2.20-GHz Intel Core i5 CPU with 12 GB of RAM and a RTX 2080 GPU. As can be observed, the CNN-based methods always consume massive computation time since they require deep layers and multiple types of convolution kernels, which results in a large number of parameters, and thus significantly increasing the computational complexity. In contrast, GCN and S$^{2}$GCN employ a fixed graph for convolution, and thus the number of parameters is greatly reduced. As a result, they need less computational time than the CNN-based methods. Due to the utilization of superpixel technique, the size of the graphs used in MDGCN is far smaller than that in GCN and S$^{2}$GCN. Consequently, the time consumption of the proposed MDGCN is the lowest among the five methods, even though MDGCN employs multiple graphs at different neighborhood scales.
	
	\begin{table}[!t]
		\centering
		\caption{Running Time Comparison (In Seconds) of Different Methods. `IP' Denotes Indian Pines Dataset, `PaviaU' Denotes University of Pavia Dataset, and `KSC' Denotes Kennedy Space Center Dataset}
		\begin{tabular}{cccccc}
			\toprule
			& GCN   & S$^{2}$GCN & R-2D-CNN & DR-CNN & MDGCN \\
			\midrule
			IP    & 656   & 1176  & 2178  & 5258  & 74 \\
			PaviaU & -     & -     & 2313  & 8790  & 258 \\
			KSC   & 147   & 287   & 1527  & 4528  & 20 \\
			\bottomrule
		\end{tabular}%
		\label{RunningTime}%
	\end{table}%

	\section{Conclusion}
	\label{Conclusions}
	
	In this paper, we propose a novel Multi-scale Dynamic Graph Convolutional Network (MDGCN) for hyperspectral image classification. Different from prior works that depend on a fixed input graph for convolution, the proposed MDGCN critically employs dynamic graphs which are gradually refined during the convolution process. Therefore, the graphs can faithfully encode the intrinsic similarity among image regions and help to find accurate region representations. Meanwhile, multiple graphs with different neighborhood scales are constructed to fully exploit the multi-scale information, which comprehensively discover the hidden spatial context carried by different scales. The experimental results on three widely-used hyperspectral image datasets demonstrate that the proposed MDGCN is able to yield better performance when compared with the state-of-the-art methods.

	\ifCLASSOPTIONcaptionsoff
	\newpage
	\fi

	
	
	%

	\bibliographystyle{IEEEtran}
	\bibliography{IEEEabrv,IEEEexample}

\end{document}